\documentclass[review]{elsarticle}
\usepackage{indentfirst}
\usepackage{bm}
\usepackage{threeparttable}
\usepackage{multirow}
\usepackage{amsmath}
\usepackage{array}
\usepackage{amssymb}
\usepackage{url}
\usepackage{hyperref}
\usepackage{booktabs}
\usepackage{color}

\usepackage{graphicx, subfigure}
\usepackage{array, threeparttable}

\begin{document}



\title{An ALE-type discrete unified gas kinetic scheme for low-speed continuum and rarefied flow simulations with moving boundaries}


\author{Yong Wang}
\ead{wangyong19890513@mail.nwpu.edu.cn}

\author[]{Chengwen Zhong\corref{mycorrespondingauthor}}
\ead{zhongcw@nwpu.edu.cn}
\cortext[mycorrespondingauthor]{Corresponding author}
\address{National Key Laboratory of Science and Technology on Aerodynamic Design and Research, Northwestern Polytechnical University, Xi'an, Shaanxi 710072, China}


\date{\today}

\begin{abstract}
    In this paper, the original discrete unified gas kinetic scheme (DUGKS) is extended to arbitrary Lagrangian-Eulerian (ALE) framework for simulating the low-speed continuum and rarefied flows with moving boundaries. For ALE method, the mesh moving velocity is introduced into the Boltzmann-BGK equation. The remapping-free scheme is adopted to develop the present ALE-type DUGKS, which avoids the complex rezoning and remapping process in traditional ALE method. As in some application areas, the large discretization errors will be introduced into the simulation if the geometric conservation is not guaranteed. Three compliant approaches of the geometric conservation law (GCL) are discussed and a uniform flow test case is conducted to validate these schemes. To illustrate the performance of present ALE-type DUGKS, four test cases are carried out. Two of them are the continuum flow cases, which are the flows around the oscillating circular cylinder and the pitching NACA0012 airfoil, respectively. Others are the rarefied flow cases, one is the moving piston driven by the rarefied gas, another is the flow caused by the plate oscillating in its normal direction. The results of all test cases are in good agreement with the other numerical and/or experimental results, demonstrating the capability of present ALE-type DUGKS to cope with the moving boundary problems at different flow regimes.
\end{abstract}


\maketitle

\section{Introduction}
Moving boundary problems can be found in various scientific and engineering fields. For example, in aerospace area, the store separation from the aircraft body, the motion of landing gear during the take-off and landing, etc. \cite{zhang2012applications}. For micro air vehicles, the design of rotary wings and flapping wings which include lots of moving boundary problems will be another application aera \cite{whitney2012conceptual}. In the above examples, usually the reference lengths of obstacle are much larger than the mean-free-path of gas molecule. According to the definition of Knudsen number \cite{hsue1946superaerodynamics}, the flows in above cases are the continuum flows. Furthermore, the moving boundary problems are also encountered in the application of rarefied flow regime. Such as, the sound wave generated by the oscillating plate, the motion of vanes for the Crookes radiometer, etc. \cite{tsuji2013moving}. Generally speaking, the macro-methods based on the N-S equations \cite{shyy1998computational} can cope with the corresponding problems at continuum flow regime, and the prevailing direct simulation Monte Carlo (DSMC) method \cite{shrestha2015numerical} can deal with that at rarefied flow regime. But in some applications, both macro-methods and DSMC will encounter obstacle as the flow regime out of its computational range, such as the flow in transition regime. Though some hybrid methods \cite{whitney2012conceptual, patronis2013hybrid} can be used, due to the different temporal and spatial scales, this kind of hybrid method also encounters great difficulties. Consequently, developing and improving a method which can simulating the moving boundary problems at all flow regimes will have enormous values for engineering application.

Recently, the discrete unified gas kinetic scheme (DUGKS) proposed by Guo et al. \cite{guo2013discrete} is a promising method which can handle flows at all regimes \cite{Pan2017An}. For this method, it combines the advantages of both lattice Boltzmann method \cite{succi2001lattice} (LBM) and unified gas kinetic scheme \cite{xu2010unified} (UGKS), where the flux at cell interface is easy to calculate like LBM, and the computational cost is declined compared with the UGKS. Some details can be found in Ref.~\cite{guo2013discrete, guo2015discrete, wang2015comparative, zhu2016discrete}. At current stage, the DUGKS is implemented under the stationary mesh, so the purpose of this paper is further extends the application range of original DUGKS, that is can deal with moving boundary problems.

Nowadays, there are many methods can handle the moving boundary problems. Based on the mesh systems used in the numerical computation, these methods simply can be divided into two categories: Eulerian method and Lagrangian method. For Eulerian method, the mesh is fixed at each iteration computation. The immersed boundary method (IBM) is one of the representation \cite{mittal2005immersed, Yuan2015An, yuan2018immersed}. In this method, the uniform Cartesian grids are used near the region of wall at present stage, the moving boundary will be regarded as a set of Lagrangian nodes, and the influence of wall (no-slipping condition) to the nearby nodes of the Cartesian grid is considered with some interpolation methods. Besides the applications in the continuum flow regime, the IBM coupled the UGKS now can also deal with the rarefied flow moving boundary problems \cite{ragta2017unified}. The primary disadvantage of IBM is in some applications, such as high Reynolds number flows, the amount of mesh is intolerable. The static mesh movement method \cite{stockie2001moving} will be another representation of Eulerian method to cope the moving boundary problems. During the numerical simulations, after an Eulerian step, a new mesh is generated according to certain requirement. In general, regenerating a new mesh is time-consuming for most applications. Besides. some interpolation methods also needed to transfer the flow variables from the old to the new mesh. For Lagrangian method, the moving velocity of mesh is equal to the local fluid flow velocity, so the mesh distortion and tangling is unavoidable for most cases.

To combine the advantages of both Eulerian and Lagrangian method, a famous method, that is arbitrary Lagrangian-Eulerian \cite{hirt1974arbitrary} (ALE) technique is developed and improved during the last few decades. Nowadays, the ALE method also can be divided into two types. In traditional procedure, three steps, that are the explicit Lagrangian phase, the rezoning phase, and the remapping phase, will be implemented. Similar to the static mesh movement method in pure Eulerian method, the mesh regeneration or modification, and flows variables transfer will be the two critical steps for this ALE type. In other words, if the quality of mesh can maintain very well during the moving process, this type ALE will degenerated into the pure Lagrangian method. For another type of ALE, the mesh velocity will be introduced into the governing equations (convective terms) to modify the net flux of cell interface. And the mesh moving technique will be introduced to maintain the mesh quality. Besides, the mesh moving velocity also will be constructed based on the old and the new meshes. In aerospace area, such as the aeroelastic analysis, this type of ALE method is usually used. Generally speaking, the time-consuming for the implement of mesh moving technique is less than that of regenerating a new mesh, and rezoning and remapping phases can be discarded for this type ALE, so the remapping-free ALE technique will be used in this paper to improve the original DUGKS.

For the methods of dealing with the moving boundary problems, one source of numerical error is violates the geometric conservation law (GCL). In some applications, it will yields erroneous results \cite{chang2015further}. Following the previous works, in this paper, GCL compliance schemes are considered to exclude this part of numerical errors.

The rest of this paper is organized as follows. In Sec.~\ref{NumericalMethods}, original DUGKS is introduced briefly, ALE-type DUGKS and several GCL schemes are detailed illustrated. In Sec.~\ref{Cases}, one case to verify the GCL schemes, and four test cases to validate the capacity of present method are conducted. Finally, a short conclusion is summarized in the Sec.~\ref{Conclusion}.

\section{Numerical Methods}\label{NumericalMethods}

\subsection{The sketch of original discrete unified gas kinetic scheme}
In this section, the original DUGKS proposed by Guo et al. \cite{guo2013discrete} is introduced briefly. The starting point of DUGKS is the Boltzmann-BGK equation, which can be expressed as
\begin{equation}\label{Maxwell}
    \frac{\partial{f}}{\partial{t}}+\bm{\xi}\cdot\nabla{f}=\Omega=-\frac{1}{\tau}[f-f^{eq}],
\end{equation}
where $f=f(\bm{x},\bm{\xi},\bm{\eta},\bm{\zeta}, t)$ is the velocity distribution function for particles moving in $D$-dimensional velocity space with $\bm{\xi}=(\xi_1,\dots,\xi_D)$ at position $\bm{x}=(x_1,\dots,x_D)$ and time $t$. $\bm{\eta}=(\xi_{D+1},\dots,\xi_3)$ is the rest components of the particle velocity with the length $L=3-D$. $\bm{\zeta}$ is a vector with $K$ dimension which represent the internal degree of freedom of molecules. $\tau$ is the relaxation time relating to the fluid dynamics viscosity $\mu$ and pressure $p$ with $\tau = \mu/p$. And $f^{eq}$ is the  Maxwellian equilibrium distribution function, which is given by
\begin{equation}
	f^{eq}=\frac{\rho}{(2\pi{R}T)^{(3+K)/2}}exp(-\frac{c^2+\eta^2+\zeta^2}{2RT}),
\end{equation}
where $R$ is the gas constant, $T$ is the fluid temperature, $\rho$ is the density of fluid, and $\bm{c}=(\bm{\xi}-\bm{u})$ is the peculiar velocity with $\bm{u}$ being the macroscopic flow velocity.

In order to remove the dependence of distribution function on the internal variable $\bm{\eta}$ and $\bm{\zeta}$, usually, two reduced distributions \cite{yang1995rarefied} can be introduced in practical computation, and respectively, given by
\begin{equation}
    g(\bm{x},\bm{\xi},t)=\int{f(\bm{x},\bm{\xi},\bm{\eta},\bm{\zeta},t)}d\bm{\eta}d\bm{\zeta},
\end{equation}
and
\begin{equation}
	h(\bm{x},\bm{\xi},t)=\int{(\eta^2+\zeta^2)f(\bm{x},\bm{\xi},\bm{\eta},\bm{\zeta},t)}d\bm{\eta}d\bm{\zeta}.
\end{equation}
With Eq.~(\ref{Maxwell}), the evolution equations for $g$ and $h$ can be expressed as
\begin{equation}\label{equ_g}
    \frac{\partial{g}}{\partial{t}}+\bm{\xi}\cdot\nabla{g}=\Omega_g=-\frac{1}{\tau}[g-g^{eq}],
\end{equation}
\begin{equation}\label{equ_h}
    \frac{\partial{h}}{\partial{t}}+\bm{\xi}\cdot\nabla{h}=\Omega_h=-\frac{1}{\tau}[h-h^{eq}],
\end{equation}
respectively, where the equilibrium distribution functions for $g^{eq}$ and $h^{eq}$ are given by
\begin{equation}
	g^{eq}=\int{f^{eq}}d\bm{\eta}d\bm{\zeta}=\frac{\rho}{(2\pi{R}T)^{D/2}}exp(-\frac{(\bm{\xi}-\bm{u})^2}{2RT}),
\end{equation}
\begin{equation}
	h^{eq}=(K+3-D)RTg^{eq}.
\end{equation}
The macro-quantities can be calculated with
\begin{equation}\label{dugks_marco}
	\rho=\int{g}d\bm{\xi}, \rho\bm{u}=\int{\bm{\xi}g}d\bm{\xi}, \rho{E}=\frac{1}{2}\int{(\xi^2g+h)}d\bm{\xi},
\end{equation}
and with ideal gas law, $p=\rho{R}T$, the pressure can be obtained. In additional, the relationship between dynamics viscosity $\mu$ and temperature $T$, based on the hard-sphere (HS) or variables hard-sphere (VHS) molecules, can be given by
\begin{equation}
	\mu=\mu_{ref}\left(\frac{T}{T_{ref}}\right)^\omega,
\end{equation}
where $\omega$ is the index related to the HS or VHS model, $\mu_{ref}$ is viscosity at the reference temperature $T_{ref}$.

As Eq.~(\ref{equ_g}) and Eq.~(\ref{equ_h}) are exactly the same, which can be rewritten as
\begin{equation}\label{equ_phi}
	\frac{\partial{\phi}}{\partial{t}}+\bm{\xi}\cdot\nabla{\phi}=\Omega=-\frac{1}{\tau}[\phi-\phi^{eq}],
\end{equation}
with $\phi$ represent $g$ or $h$. In DUGKS, Eq.~(\ref{equ_phi}) will be solved with finite volume method. And the discretization of this equation can be divided into the two steps: velocity-space discretization and physical-space discretization.

For particle velocity-space discretization, usually, a finite set of discretized micro-velocities will be used \cite{guo2013discrete}, and $\bm{\xi}_i$ represents the $i$-th discretized velocity. As the macro-quantities depending on the particle micro-velocities (Eq.~(\ref{dugks_marco})), the choose of the values of micro-velocity can be set coincide with the abscissas of quadrature rule. For low-speed continuum flows, the discretized velocities, weights, and corresponding equilibrium distribution functions developed in LBM, such as D2Q9 model \cite{qian1992lattice}, can be used in DUGKS, which also build the connection between LBM and DUGKS. For rarefied flows, usually, the Gauss-Hermit and Newton-Cotes quadrature rules will be used to integral the macro-quantities, and corresponding set of abscissas will be used as the set of discretized velocities.

\begin{figure}[!htp]
	\centering
	\subfigure[]{
		\includegraphics[width=0.35 \textwidth]{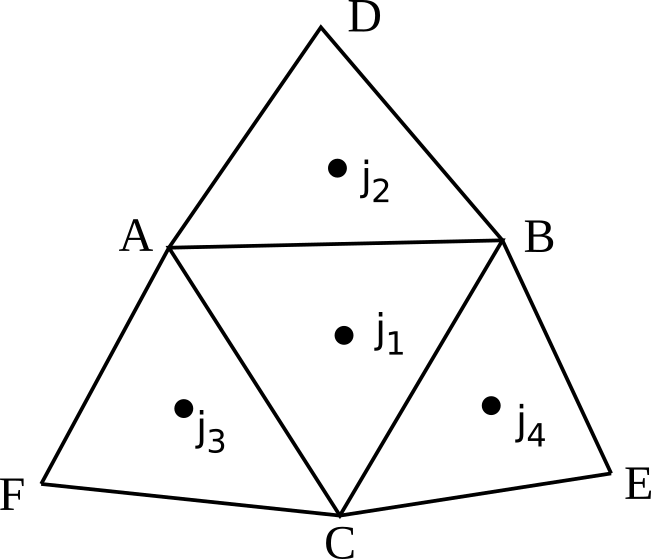}
		\label{structure1}
	}
	\subfigure[]{
		\includegraphics[width=0.35 \textwidth]{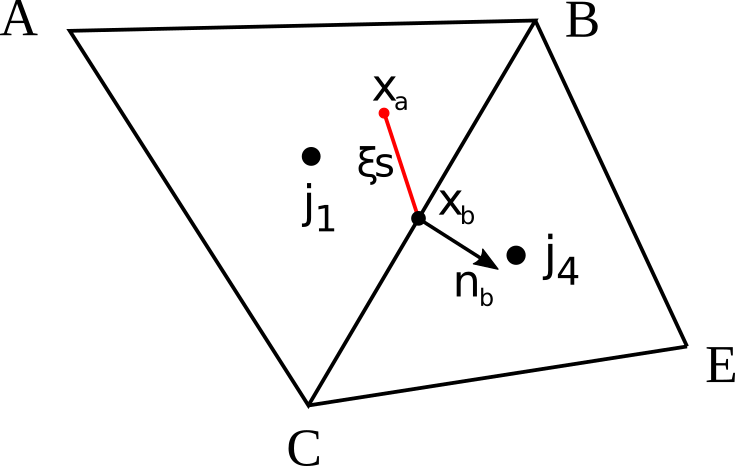}
		\label{structure2}
	}
	\caption{\label{structure_mesh} Schematics of (a) unstructured mesh used in DUGKS, (b) flux calculation for cell interface.}
\end{figure}

For physical-space discretization, in this paper, the finite volume method based on unstructured mesh is used. Fig.~\ref{structure1} shows the schematic of unstructured mesh, $j$ is the  center of triangle cell $ABC$, and subscript represent the index number of cell. If $\phi_j$ and $\Omega_j$ are the average values of $\phi$ and $\Omega$ in cell $ABC$, $\bigtriangleup{t}=t^{n+1}-t^n$ is the time step, and the mid-point rule is used for the time integration of the convection term and the trapezoidal rule for the collision term, Eq.~(\ref{equ_phi}) can be rewritten as
\begin{equation}\label{semi_equ}
	\phi_j^{n+1}(\bm{\xi})-\phi_j^n(\bm{\xi})+\frac{\bigtriangleup{t}}{\left|V_j\right|}\bm{F}^{n+1/2}(\bm{\xi})=\frac{\bigtriangleup{t}}{2}[\Omega_j^{n+1}(\bm{\xi})+\Omega_j^n(\bm{\xi})],
\end{equation}
where $V_j$ represent the volume of cell $ABC$. The $\bm{F}^{n+1/2}(\bm{\xi})$ is the flux of cell surface and given by
\begin{equation}\label{flux}
	\bm{F}^{n+1/2}(\bm{\xi})=\int_{\partial{V_j}}(\bm{\xi}\cdot{\bm{n}})\phi(\bm{x},\bm{\xi},t_{n+1/2})dS,
\end{equation}
where $\partial{V_j}$ is the cell surface, $\bm{x}$ is the center of cell interface, and $\bm{n}$ is the outward unit vector normal to the surface. To remove the implicit collision term, two new distribution functions will be introduced:
\begin{equation}\label{tilde_phi}
	\tilde{\phi}=\phi-\frac{\bigtriangleup{t}}{2}\Omega=\frac{2\tau+\bigtriangleup{t}}{2\tau}\phi-\frac{\bigtriangleup{t}}{2\tau}\phi^{eq},
\end{equation}
\begin{equation}\label{tilde_phi_plus}
	\tilde{\phi}^+=\phi+\frac{\bigtriangleup{t}}{2}\Omega=\frac{2\tau-\bigtriangleup{t}}{2\tau+\bigtriangleup{t}}\tilde{\phi}+\frac{2\bigtriangleup{t}}{2\tau+\bigtriangleup{t}}\phi^{eq}.
\end{equation}
Then Eq.~(\ref{equ_phi}) will be further rewritten as
\begin{equation}\label{govern_phi}
	\tilde{\phi}_j^{n+1}=\tilde{\phi}_j^{+,n}-\frac{\bigtriangleup{t}}{\left|V_j\right|}\bm{F}^{n+1/2}(\bm{\xi}).
\end{equation}
Due to the conservative property of the collision term, in practical computation, $\tilde{\phi}$ will be solved instead of the $\phi$. The Eq.~(\ref{dugks_marco}) for computing the macro-quantities also will be rewritten as
\begin{equation}\label{macro_phi}
	\rho=\int{\tilde{g}}d\bm{\xi}, \rho\bm{u}=\int{\bm{\xi}\tilde{g}}d\bm{\xi}, \rho{E}=\frac{1}{2}\int{(\xi^2\tilde{g}+\tilde{h})}d\bm{\xi}.
\end{equation}

For the calculation of interface flux, Fig.~\ref{structure2} shows the schematic of cell interface. $\bm{x}_b$ is the center of $BC$, and $\bm{n}_b$ is unit vector normal of $BC$ with the direction pointing from cell $ABC$ to $BEC$. Integral the Eq.~(\ref{equ_phi}) along the characteristic line within a half time step $s=\bigtriangleup{t}/2$, the original distribution function $\phi(\bm{x}_b,\bm{\xi},t_n+s)$ in Eq.~(\ref{flux}) can be updated with
\begin{equation}\label{equ_phi_plus}
	\phi(\bm{x}_b,\bm{\xi},t_n+s)-\phi(\bm{x}_b-\bm{\xi}s,\bm{\xi},t_n)=\frac{s}{2}[\Omega(\bm{x}_b,\bm{\xi},t_n+s)+\Omega(\bm{x}_b-\bm{\xi}s,\bm{\xi},t_n)].
\end{equation}
Similar the treatment to $\tilde{\phi}$, another two new distribution functions will be introduced and given by
\begin{equation}\label{bar_phi}
    \bar{\phi}=\phi-\frac{s}{2}\Omega=\frac{2\tau+s}{2\tau}\phi-\frac{s}{2\tau}\phi^{eq},
\end{equation}
\begin{equation}\label{bar_phi_plus}
    \bar{\phi}^+=\phi+\frac{s}{2}\Omega=\frac{2\tau-s}{2\tau+s}\bar{\phi}+\frac{2s}{2\tau+s}\phi^{eq}.
\end{equation}
Then Eq.~(\ref{equ_phi_plus}) can be rewritten as
\begin{equation}\label{bar_phi_stream}
	\bar{\phi}(\bm{x}_b,\bm{\xi},t_{n+1/2})=\bar{\phi}^+(\bm{x}_b-\bm{\xi}s,\bm{\xi},t_n).
\end{equation}
If we instead the $\tilde{\phi}$ in Eq.~(\ref{macro_phi}) by $\bar{\phi}$, the macro-quantities at cell interface can be calculated. And with Eq.~(\ref{bar_phi}), the original distribution function at $(\bm{x}_b,t_{n+1/2})$ is given by
\begin{equation}
  \phi(\bm{x}_b,\bm{\xi},t_{n+1/2})=\frac{2\tau}{2\tau+s}\bar{\phi}(\bm{x}_b,\bm{\xi},t_n+s)+\frac{s}{2\tau+s}\phi^{eq}(\bm{x}_b,\bm{\xi},t_n+s).
\end{equation}
As illustrated by Eq.~(\ref{bar_phi_stream}) and Fig.~\ref{structure2}, for given $\bar{\phi}^+$ at $\bm{x}_a=\bm{x}_b-\bm{\xi}s$, the $\phi^{n+1/2}$ at cell interface can be obtained. With Eq.~(\ref{tilde_phi}), (\ref{bar_phi}) and (\ref{bar_phi_plus}), we can build the relationship between $\tilde{\phi}$ and $\bar{\phi}^+$:
\begin{equation}\label{bar_phi_plus_a}
	\bar{\phi}^+=\frac{2\tau-s}{2\tau+\bigtriangleup{t}}\tilde{\phi}+\frac{3s}{2\tau+\bigtriangleup{t}}\phi^{eq}.
\end{equation}
So, with $\tilde{\phi}$ and corresponding equilibrium distribution function $\phi^{eq}$ stored at $\bm{x}_j$. According to Taylor expansion, we can calculate the $\phi^+(\bm{x}_a,t_n)$ with
\begin{equation}
	\bar{\phi}^+(\bm{x}_a,t_n)=\bar{\phi}^+(\bm{x}_j,t_n)+(\bm{x}_a-\bm{x}_j)\cdot\nabla\bar{\phi}^+(\bm{x}_j,t_n),
\end{equation}
where $\nabla\bar{\phi}^+$ is the gradient of $\bar{\phi}^+$. For the calculation of $\nabla\bar{\phi}^+$ under the unstructured mesh, in this paper, the least square method will be used:
\begin{equation}\label{leastsquare}
\min_{\nabla{\bar{\phi}^+_j}}\sum_{n}w_{j,n}[\bar{\phi}^+_{j,n} - \bar{\phi}^+_j - \nabla{\bar{\phi}^+_j} \cdot (\bm{x}_{j,n} - \bm{x}_j)]^2,
\end{equation}
where $w_{j,n} = 1/(\bm{x}_{j,n} - \bm{x}_j)^2$ is the geometrical weighting factor, $n$ is the total number of cell neighbor. Besides, in practical computation, when $\bar{\phi}^+$ has been calculated, with Eq.~(\ref{tilde_phi_plus}) and (\ref{bar_phi_plus_a}), the $\tilde{\phi}^+$ in Eq.~(\ref{govern_phi}) can be updated with
\begin{equation}
	\tilde{\phi}^+=\frac{4}{3}\bar{\phi}^+-\frac{1}{3}\tilde{\phi}.
\end{equation}

\subsection{The ALE-type discrete unified gas kinetic scheme}

\subsubsection{The discretized method for ALE-type DUGKS}\label{dugks}
In this section, the discretized formulation of ALE-type DUGKS will be introduced detailed. Under the ALE framework, during the simulation, the geometry information of the cell, such as volume, the location of cell center, the length of cell interface, etc., will be changed from time to time. Following the handling method used in macro numerical method based on the N-S equations and UGKS which also based on the Boltzmann-BGK equation \cite{chen2012unified}, the mesh moving velocity $\bm{v}$ that modify the net flux of cell interface is introduced, and Eq.~(\ref{Maxwell}) is rewritten as
\begin{equation}\label{Maxwell_moving}
\frac{\partial{f}}{\partial{t}}+(\bm{\xi}-\bm{v})\cdot\nabla{f}=\Omega=-\frac{1}{\tau}[f-f^{eq}].
\end{equation}
The discretization scheme that presented in original DUGKS will also be used in ALE-type DUGKS, then Eq.~(\ref{semi_equ}) and (\ref{flux}) also will be modified as
\begin{equation}\label{semi_equ_ale}
   \phi_j^{n+1}(\bm{\xi})\left|V_j^{n+1,*}\right|-\phi_j^n(\bm{\xi})\left|V_j^{n,*}\right|+\bigtriangleup{t}\bm{F}_{ALE}^{n+1/2}(\bm{\xi})=\frac{\bigtriangleup{t}}{2}[\Omega_j^{n+1}(\bm{\xi})\left|V_j^{n+1,*}\right|+\Omega_j^n(\bm{\xi})\left|V_j^{n,*}\right|],
\end{equation}
and
\begin{equation}\label{flux_ale}
   \bm{F}_{ALE}^{n+1/2}(\bm{\xi})=\int_{\partial{V_j}}(\bm{\xi}-\bm{v})\cdot{\bm{n}}\phi(\bm{x},\bm{\xi},t_{n+1/2})dS=\sum_{k}(\bm{\xi}-\bm{v}_{b,k}^{n+1/2})\cdot\bm{n}_{b,k}^*\phi^{n+1/2}(\bm{x}_{b,k},\bm{\xi})S_k^*,
\end{equation}
respectively. Where $V^{n+1,*}$ and $V^{n,*}$ are the cell volumes at $n+1$ and $n$ time steps, superscript $^*$ means that the value of volume at corresponding time maybe not equal to the true value of volume at that time and will be illustrated at next section. $k$ is total number of cell interface. $\bm{v}_b^{n+1/2}$ is the moving velocity of cell interface at $n+1/2$ time. $\bm{n}_b^*$ and $S_b^*$ are the outward unit normal vector and area of cell interface, respectively. The computational method for these three variables also will be illustrated at next section.

Similar to the original DUGKS, in ALE-type DUGKS, two new distribution functions, $\tilde{\phi}$ and $\tilde{\phi}^+$ also will be introduced, and Eq.~(\ref{semi_equ_ale}) can be modified as
\begin{equation}\label{equ_phi_ale}
	\tilde{\phi}_j^{n+1}=\left|\frac{V_j^{n,*}}{V_j^{n+1,*}}\right|\tilde{\phi}_j^{+,n}-\frac{\bigtriangleup{t}}{\left|V_j^{n+1,*}\right|}\bm{F}_{ALE}^{n+1/2}(\bm{\xi})=\left|V_{j,mod}\right|\tilde{\phi}_j^{+,n}-\frac{\bigtriangleup{t}}{\left|V_j^{n+1,*}\right|}\bm{F}_{ALE}^{n+1/2}(\bm{\xi}),
\end{equation}
where the formulations of $\tilde{\phi}_j^{n+1}$ and $\tilde{\phi}_j^{+,n}$ are same to that of original DUGKS. The scheme for reconstructing the $\phi^{n+1/2}$ at cell interface also same to that of original DUGKS, and $\bm{x}_a=\bm{x}_b^n-\bm{\xi}s$ is used to compute the location of interpolating point. Besides, the sign of $\bm{\xi}\bm{n}^n$ is used to judge the upwind direction. Finally, from the perspective of programming, if three modifications are added into the framework of original DUGKS solver, that are
\begin{description}
\item[(1)] Add the independent module of mesh moving and geometry information updated part into the old solver,
\item[(2)] Modify the $\tilde{\phi}_j^{+,n}$ in Eq.~(\ref{govern_phi}) with volume modified coefficient $\left|V_{mod}\right|$,
\item[(3)] Modify the vector product in flux computation formulation from $\bm{\xi\cdot\bm{n}}$ to $(\bm{\xi}-\bm{v})\cdot\bm{n}$.
\end{description}
It will become the ALE-type DUGKS solver, and has ability to deal with both continuum and rarefied flow problems with moving boundary.

Additionally, Laplace smoothing equations for mesh deformation \cite{lohner1996improved} is solved to update the unstructured mesh under moving boundary.

\subsubsection{Geometric Conservation Law}\label{dgcltheory}
The concept of GCL was firstly well-defined by Thomas and Lambard in 1979 \cite{thomas1978geometric}. Generally speaking, for uniform flow, if scheme based on the moving mesh is GCL complaint, any distribution must not introduced into flow domain at any time. `Free Stream Preservation Property' is the fundamental condition for any time-integral schemes on moving mesh \cite{chang2015further}. Following the  theoretical derivation used in macro numerical method, the Boltzmann-BGK equation is considered here. If integral the Eq.~(\ref{Maxwell}) on the control volume, and semi-discrete form is used, we have
\begin{equation}\label{equ_dgcl_f}
     \frac{d}{dt}\int{f}dV+\int_{\partial{V}}(\bm{\xi}-\bm{v}_b)\cdot\bm{n}_bf_bdS=-\frac{1}{\tau}\int[f-f^{eq}]dV.
\end{equation}

If flow is uniform, that is $f=const$ in each cell, and with $\int\bm{n}_bdS=0$, Eq.~(\ref{equ_dgcl_f}) can be simplified as
\begin{equation}\label{gover_dgcl}
	\frac{d}{dt}V_j=\int_{\partial{V}}\bm{v}_b\cdot\bm{n}_bdS=\sum_k{\bm{v}_{b,k}\bm{n}_{b,k}S_{b,k}}=\sum_k{\bm{v}_{b,k}\bm{S}_{b,k}},
\end{equation}
which is the governing equation of GCL, and means that the variation of volume of a moving cell equals to the integration of the volume-flux (or ``sweeping volume'') of all the surfaces (named as ``face'' in the following context) surrounding the control volume \cite{chang2015further}.

In this paper, the mesh velocity $\bm{v}_b$ of cell interface in Eq.~(\ref{flux_ale}) is given as
\begin{equation}
	\bm{v}_b^{n+1/2}=\frac{\bm{x}_b^{n+1}-\bm{x}_b^n}{\bigtriangleup{t}}.
\end{equation}

Based on the mesh velocity and GCL governing equation, three discretized geometric conservation law (DGCL) compliant schemes which decide the $V_j^{n+1,*}$, $V_j^{n,*}$, $\bm{S}_b^*$ will be discussed.

(1) DGCL scheme1:

If we set:
\begin{equation}\label{dgcl1_v}
	V_j^{n+1,*}=V_j^{n+1}, V_j^{n,*}=V_j^{n},
\end{equation}
that are the true values of volume at corresponding times will be used. And following the idea used in Ref.~\cite{lesoinne1996geometric}, the $\bm{S}_b^*$ can be calculated with
\begin{equation}\label{dgcl1_s}
	\bm{S}_b^*=\bm{S}_b^{n+\frac{1}{2}}=\frac{\bm{S}_b^{n+1}+\bm{S}_b^n}{2}.
\end{equation}
Through the simple mathematical derivation, Eq.~(\ref{dgcl1_v}) and (\ref{dgcl1_s}) will automatically satisfy the Eq.~(\ref{gover_dgcl}) during the mesh moving process.

(2) DGCL scheme2:

If we set:
\begin{equation}\label{dgcl2_s}
	V_j^{n,*}=V_j^n, \bm{S}_b^*=\bm{S}_b^n,
\end{equation}
to satisfy the DGCL, the volume $V^{*,n+1}$ must be modified \cite{chang2015further}. With the first-order Euler time discretized scheme, Eq.~(\ref{gover_dgcl}) can be rewritten as
\begin{equation}
    \frac{V_j^{n+1}-V_j^n}{\bigtriangleup{t}}=\sum_k{\bm{v}_{b,k}^{n+1/2}\bm{S}_{b,k}^n},
\end{equation}
and $V_j^{n+1,*}$ can be modified with
\begin{equation}\label{dgcl2_v}
	V_j^{n+1,*}=V_j^n+\bigtriangleup{t}\sum_k{\bm{v}_{b,k}^{n+1/2}\bm{S}_{b,k}^{n}}.
\end{equation}

(3) DGCL scheme3:

If we set:
\begin{equation}\label{dgcl3_s}
V_j^{n+1,*}=V_j^{n+1}, \bm{S}_b^*=\bm{S}_b^{n+1},
\end{equation}
similar to the DGCL scheme2, the $V_j^n$ must modified and can be calculated with
\begin{equation}\label{dgcl3_v}
	V_j^{n,*}=V_j^{n+1}-\bigtriangleup{t}\sum_k{\bm{v}_{b,k}^{n+1/2}\bm{S}_{b,k}^{n+1}}.
\end{equation}

As the DGCL scheme1 is much easy to implement, and only one time level ($\bm{F}^{n+1/2}$) is needed to calculate the flux at cell interface, it can natural couple with current ALE-type DUGKS. Though DGCL scheme2 and scheme3 are much complex, further improved ALE-type DUGKS framework such as high-order DUGKS \cite{wu2018third} or multi-time-level implicit method \cite{chang2015further} which the geometry information at intermediate time-levels are much difficult to define \cite{koobus1999second}, these two schemes are the good choice. Besides, the scheme2 and scheme3 are the volume-constrained scheme, the face-constrained scheme \cite{chang2015further} also can be used but not considered in this paper.

\subsubsection{Boundary conditions}\label{BoundaryCondition}
In this section, the boundary conditions used in this paper will be illustrated in detail.

For the wall boundary condition, depended on the flow conditions, that are continuum or rarefied flows, the non-equilibrium extrapolation \cite{guo2002extrapolation} or diffuse-scattering rule \cite{guo2013discrete} will be used. For continuum flow, the original distribution functions at $n+1/2$ can be given by
\begin{equation}
    f_w^{n+1/2}(\bm{\xi}_i)=f_w^{eq}(\bm{\xi}_i;\rho_w,\bm{u}_w)+f_j^{n+1/2}-f_j^{eq,n+1/2}(\bm{\xi}_i),
\end{equation}
where subscript $w$ represent the wall boundary, $j$ is the neighbor cell of wall interface, $\rho_w$ and $\bm{u}_w$ are the density and velocity at wall, respectively. For rarefied flow, the particles which the direction reflect from the wall can be calculated as
\begin{equation}
    f_w^{n+1/2}(\bm{\xi}_i)=f^{eq}(\bm{\xi}_i;\rho_w,\bm{u}_w), \bm{\xi}_i\cdot{\bm{n}_w}>0,
\end{equation}
where $\bm{n}_b$ represent the unit vector with the direction normal to the wall pointing to the cell. And the density at wall $\rho_w$ is determined by the condition that no particles can go through the wall:
\begin{equation}
    \sum_{\bm{\xi}_i\cdot\bm{n}_w>0}(\bm{\xi}_i-\bm{u}_w)\cdot\bm{n}_wf^{eq,n+1/2}_w(\bm{\xi}_i;\rho_w,\bm{u}_w)+\sum_{\bm{\xi}_i\cdot\bm{n}_w<0}(\bm{\xi}_i-\bm{u}_w)\cdot\bm{n}_wf_w^{n+1/2}=0,
\end{equation}
then
\begin{equation}
    \rho_w=-\frac{\sum_{\bm{\xi}_i\cdot\bm{n}_w<0}(\bm{\xi}_i-\bm{u}_w)\cdot\bm{n}_wf_w^{n+1/2}}{\sum_{\bm{\xi}_i\cdot\bm{n}_w>0}(\bm{\xi}_i-\bm{u}_w)\cdot\bm{n}_wf^{eq,n+1/2}_w(\bm{\xi}_i;1,\bm{u}_w)},
\end{equation}
where the distribution functions $f_w^{n+1/2}$ with direction $\bm{\xi}_i\cdot{\bm{n}_w}<0$ can be constructed following the procedure described in Sec.\ref{dugks}.

For the test cases of flow around the obstacle, in this paper, the far-field boundary condition will be used. Similar the treatment to the wall boundary, the directions reflect from the boundary can be calculated as
\begin{equation}
   f^{n+1/2}(\bm{\xi}_i)=f^{eq}(\bm{\xi}_i;\rho_0,\bm{u}_0), \bm{\xi}_i\cdot{\bm{n}}>0,
\end{equation}
where the $\rho_0$ and $\bm{u}_0$ are the density and velocity of free-stream, respectively.

\section{Numerical results and discussions}\label{Cases}
In this section, several test cases are set up to validate the proposed ALE-type DUGKS in this paper. The first case is the GCL compliance test and shows why its important the GCL be implemented. The second and third cases are low-speed continuum flows around the oscillating circular cylinder and pitching NACA0012 airfoil, respectively, which show the method presented in this paper also can get goods results compared with macroscopic method. The fourth and last cases are low-speed rarefied flows. One is the moving piston driven by the rarefied gas. Another is the flow caused by a plate oscillating in its normal direction. This case is a typical problem in MEMS devices, but currently not finish the systematic research \cite{tsuji2014gas}. For continuum cases, only $g$ distribution function governing equation is solved, and three-points Gauss-Hermite quadrature \cite{guo2013discrete} is used to calculate the macro-quantities. For rarefied cases, both $g$ and $h$ distribution functions will be solved, and the quadrature rules are the Gauss-Hermit and Newton-Cotes formulations \cite{zhu2016discrete}. Besides, the continuum flow problems are 2D flows and rarefied flow test cases are set as 1D flows.

ALE-type DUGKS formulation has been coded with the help of Code\underline{ }Saturne \cite{archambeau2004code}, an open-source computational fluid dynamics software of Electricite De France (EDF), France (\url{http://code-saturne.org/cms/}). We appreciate the development team of  Code\underline{ }Saturne for their great works.

\subsection{The uniform flow for GCL compliance test}
Uniform flow usually used as the basic case for GCL compliance test, some additional test cases for GCL complaint scheme can be found in Ref.~\cite{chang2015further}. For this problem, the computational domain is a square with the size equal to 20, and the number of cells is 40$\times$40. The time steps $\bigtriangleup$t$=0.2$. The initial velocity $u=U_0=0.1$, $v=0$, and the initial density $\rho_0=1.0$. Furthermore, the far-field boundary condition presented in Sec.~\ref{BoundaryCondition} will be used in this case.

\begin{figure}[!htp]
	\centering
	\subfigure[]{
		\includegraphics[width=0.4 \textwidth]{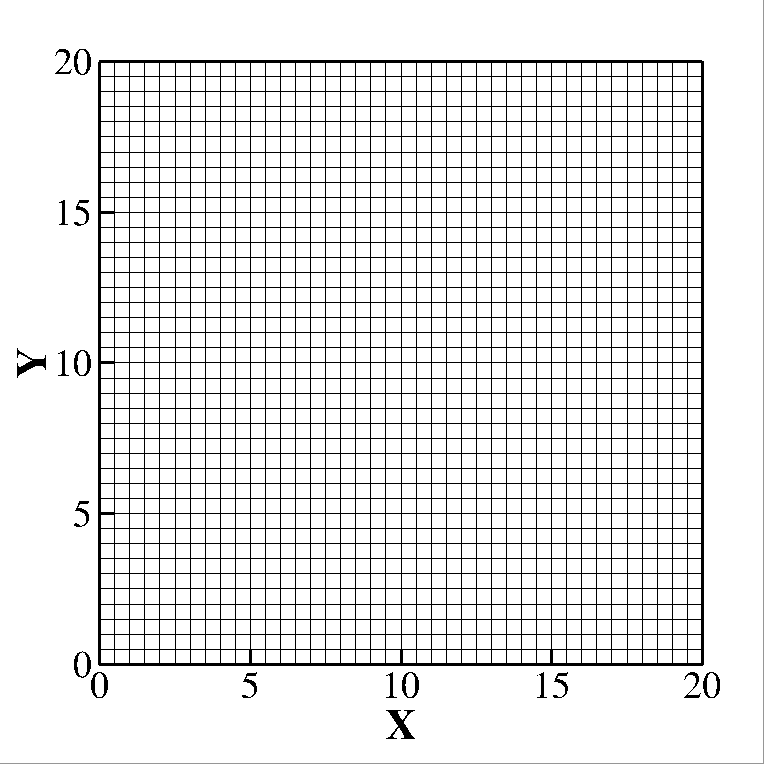}
		\label{dgclmesh1}
	}
	\subfigure[]{
		\includegraphics[width=0.4 \textwidth]{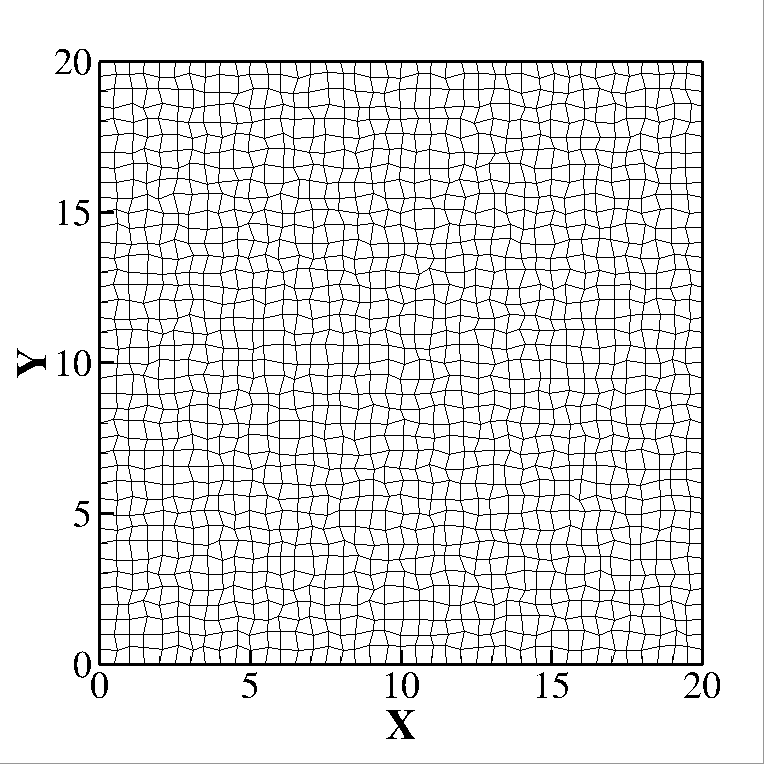}
		\label{dgclmesh2}
	}
	\caption{\label{dgclmesh} Mesh for uniform flow. (a) initial mesh, (b) instantaneous mesh.}
\end{figure}

\begin{figure}[!htp]
	\centering
	\subfigure[]{
		\includegraphics[width=0.4 \textwidth]{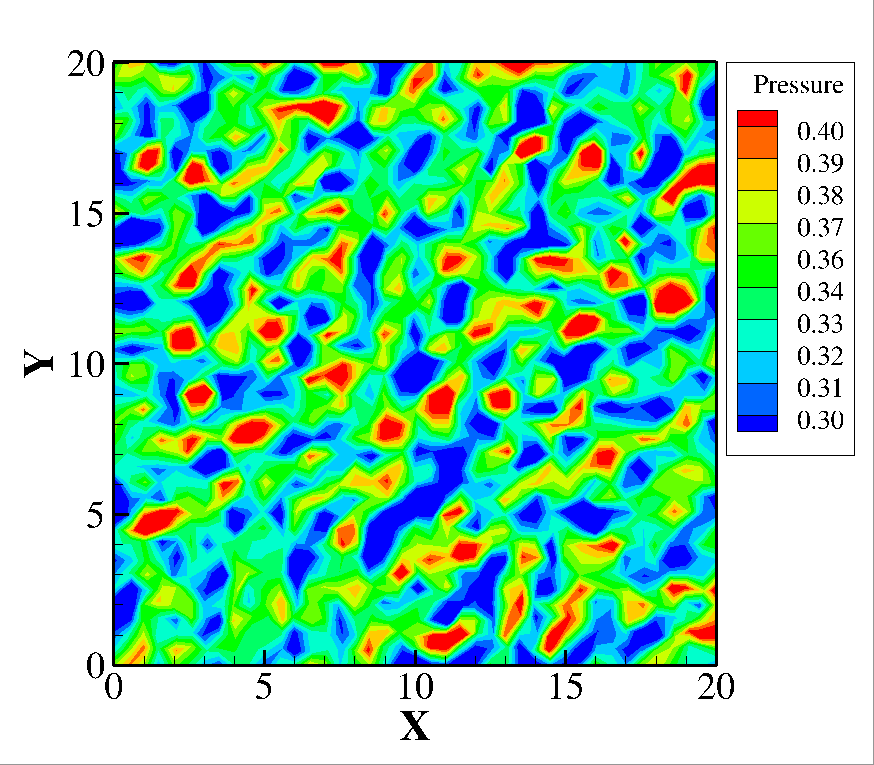}
		\label{dgclpressure}
	}
	\subfigure[]{
		\includegraphics[width=0.4 \textwidth]{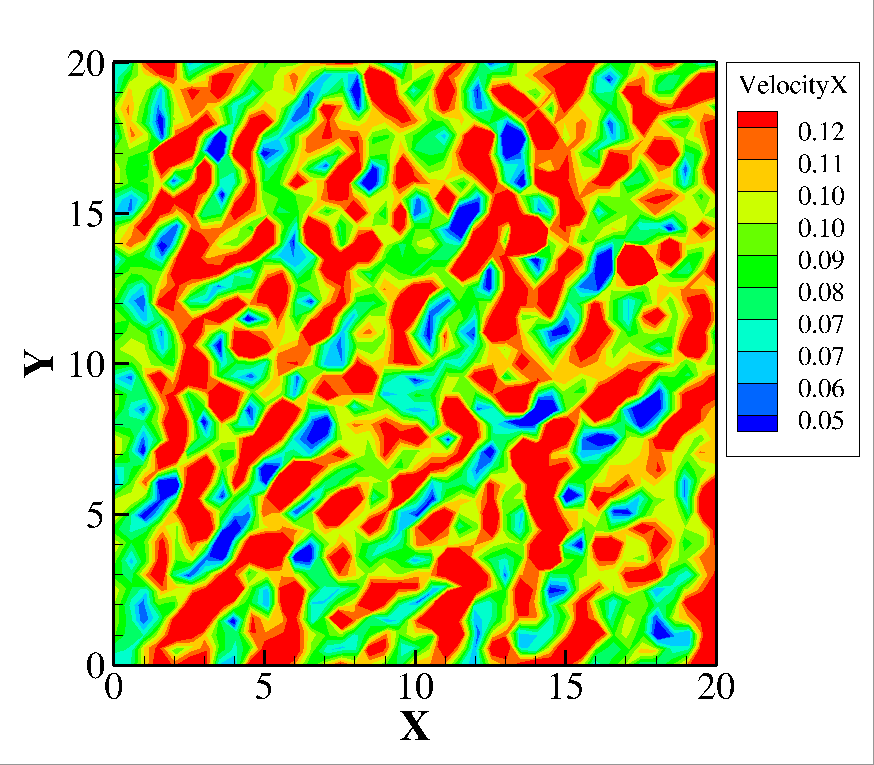}
		\label{dgclvel}
	}
	\caption{\label{dgclvelpressure} (a) Pressure and (b) Velocity contours of uniform flow when DGCL is violated.}
\end{figure}

Firstly, the important of GCL will be illustrated. For the mode of mesh moving, the grid nodes will be oscillating randomly at its original positions with amplitude of $\pm0.5\bigtriangleup$$x$ ($\bigtriangleup x$ is the size of a cell). Fig.~\ref{dgclmesh} shows the mesh, respectively, at initial time and instantaneous time when grid nodes start moving. It has been demonstrated that in Sec.~\ref{dgcltheory}, when update the distribution function, if Eq.~(\ref{dgcl1_v}) and (\ref{dgcl1_s}) are used, the DGCL will be satisfied automatically. Otherwise, if Eq.~(\ref{dgcl2_s}) or (\ref{dgcl3_s}) are used, and without volume-constrained, the DGCL will be violated. Fig.~\ref{dgclvelpressure} shows the pressure and velocity contours at 10 time iterations when DGCL violated scheme be implemented. It is clear that the pressure and velocity have been polluted seriously, and simulation can not be continued. So, if the grid nodes moving with some extreme ways, the GCL error is much larger and must be eliminated.

Secondly, the performance of DGCL schemes used in ALE-type DUGKS will be checked. In this part, the grid nodes are also oscillating randomly at its original positions, and  spatial errors for pressure and velocity will be calculated at four different mesh sizes (the numbers of cells are equal to $20\times20$, $40\times40$, $80\times80$, and $160\times160$, respectively). In Table \ref{tab:dgclerror}, after 1000 iterations, the $L_2$ and $L_\infty$ errors for all schemes maintain the significantly small values, it means that all the DGCL compliance schemes can eliminate the GCL error very well.

As the amplitude of random oscillating for previous numerical tests is little larger, the mesh quantity during the mesh moving is much terrible, so the GCL error is greatly larger. From our further tests, if this amplitude drop to 0.1$\bigtriangleup$$x$, as the mesh quality improved, the GCL error of DGCL violated scheme much declined. So, if the modes of mesh moving are quite regular, the GCL error will be  a small value compared with other numerical errors. But, in some areas like aero-elasticity, it has been reported that the GCL error will yields erroneous results \cite{guillard2000significance}. So, following the suggests proposed in Ref.~\cite{chang2015further}, the DGCL compliance scheme of ALE-type DUGKS will be ulteriorly studied in the further works. Besides, the DGCL scheme1 will be used in this paper as it is easy to implement. The DGCL scheme2 and scheme3 maybe useful to develop other ALE-type DUGKS like high-order DUGKS \cite{wu2018third} or multi-steps implicit DUGKS, where the geometry information of cell at middle time level are not easy to defined.

\begin{table}
	\centering
	\caption{\label{tab:dgclerror} Spatial errors of numerical solutions for uniform flow}
	\begin{tabular}{cccccccc}
		\hline
		\multicolumn{1}{c}{\multirow{2}{*}{$\bigtriangleup$$x$(cell size)}} & \multicolumn{1}{c}{\multirow{2}{*}{DGCL scheme}} & \multicolumn{2}{c}{Pressure} & \multicolumn{2}{c}{U}  & \multicolumn{2}{c}{V} \\
		\cline{3-4}
		\cline{5-6}
		\cline{7-8}
		~&~& \multicolumn{1}{c}{$L_2$} & \multicolumn{1}{c}{$L_\infty$} & \multicolumn{1}{c}{$L_2$} & \multicolumn{1}{c}{$L_\infty$}  & \multicolumn{1}{c}{$L_2$} & \multicolumn{1}{c}{$L_\infty$} \\
		\hline
		\multicolumn{1}{c}{\multirow{3}{*}{1.000}} & Scheme1 & 7.44$e$-14 & 1.17$e$-14 & 2.29$e$-13 & 4.07$e$-14 & 3.38$e$-13 & 6.02$e$-14 \\
		~ & Scheme2 & 1.15$e$-13 & 2.14$e$-14 & 5.98$e$-13 & 1.19$e$-13 & 7.47$e$-13 & 1.21$e$-13 \\
		~ & Scheme3 & 8.07$e$-14 & 5.88$e$-15 & 1.21$e$-13 & 2.04$e$-14 & 9.86$e$-14 & 2.99$e$-14 \\
		\cline{1-8}
		\multicolumn{1}{c}{\multirow{3}{*}{0.500}} & Scheme1 & 3.00$e$-13 & 1.52$e$-14 & 2.95$e$-13 & 3.48$e$-14 & 2.56$e$-13 & 2.09$e$-14 \\
		~ & Scheme2 & 2.63$e$-13 & 9.55$e$-15 & 2.81$e$-13 & 3.29$e$-14 & 2.25$e$-13 & 1.85$e$-14 \\
		~ & Scheme3 & 2.95$e$-13 & 1.02$e$-14 & 2.27$e$-13 & 2.15$e$-14 & 1.90$e$-13 & 1.31$e$-14 \\
		\cline{1-8}
		\multicolumn{1}{c}{\multirow{3}{*}{0.250}} & Scheme1 & 1.10$e$-12 & 4.10$e$-14 & 7.55$e$-13 & 3.84$e$-14 & 6.79$e$-13 & 2.67$e$-14 \\
		~ & Scheme2 & 1.04$e$-12 & 1.64$e$-14 & 6.48$e$-13 & 2.96$e$-14 & 5.73$e$-13 & 1.80$e$-14 \\
		~ & Scheme3 & 9.63$e$-13 & 1.52$e$-14 & 6.23$e$-13 & 2.56$e$-14 & 5.48$e$-13 & 1.51$e$-14 \\
		\cline{1-8}
		\multicolumn{1}{c}{\multirow{3}{*}{0.125}} & Scheme1 & 3.33$e$-12 & 5.71$e$-14 & 1.96$e$-12 & 4.38$e$-14 & 1.91$e$-12 & 4.45$e$-14 \\
    	~ & Scheme2 & 2.64$e$-12 & 2.10$e$-14 & 1.45$e$-12 & 2.02$e$-14 & 1.47$e$-12 & 1.76$e$-14 \\
		~ & Scheme3 & 2.73$e$-12 & 2.14$e$-14 & 1.48$e$-12 & 1.94$e$-14 & 1.50$e$-12 & 1.82$e$-14 \\
		\lasthline
	\end{tabular}
\end{table}

\subsection{Continuum flow around an oscillating circular cylinder}\label{cylinder_case}
In this section, the continuum flow around the oscillating circular cylinder is simulated. For this case, the flow is incompressible, the cylinder oscillates sinusoidally in the horizontal direction, and the equation of motion can be expressed as

\begin{equation}
    x(t) = -Asin(2{\pi}ft),
\end{equation}
where $x$ is the displacement of cylinder at horizontal direction, $A$ is the amplitude and $f$ is oscillating frequency. Following the set up used in Ref.~\cite{dutsch1998low}, two key parameters will be defined, that are the Reynolds number, Re, and the Keulegan-Carpenter number, KC, respectively. These two parameters will dominate the pattern of this oscillating flow, and the expressions are
\begin{equation}
   Re = \frac{\rho{U_{max}}d}{\mu},
\end{equation}
\begin{equation}
	KC =\frac{U_{max}}{fd},
\end{equation}
respectively, where $d$ is diameter of the cylinder, $U_{max}$ is maximum velocity of cylinder in horizontal direction, $\rho$ is the fluid density, and $\mu$ is the fluid viscosity. In this simulation, the Re = 100, and KC = 5 will be considered.

\begin{figure}[!htp]
	\centering
	\subfigure[]{
		\includegraphics[width=0.4 \textwidth]{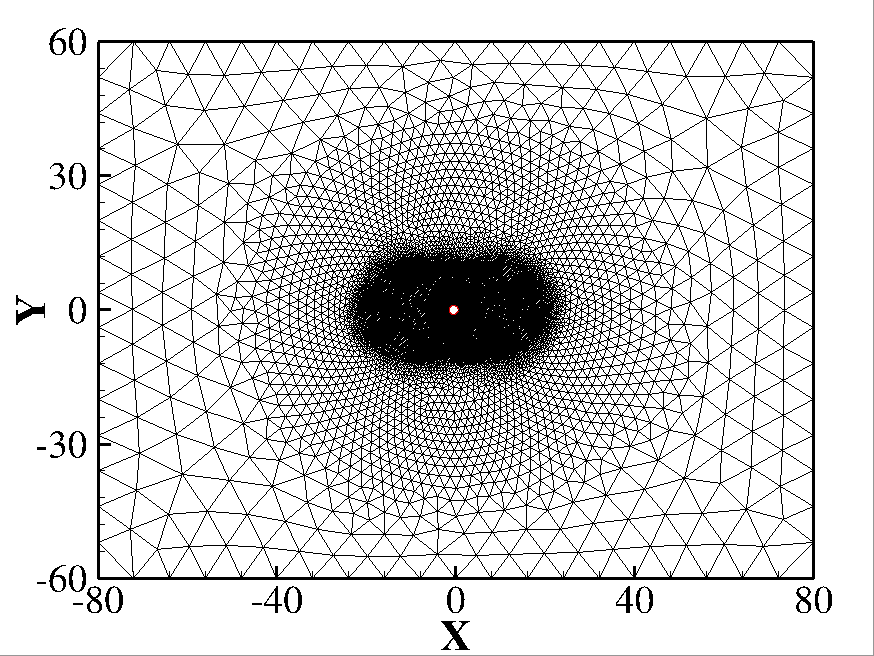}
		\label{cylinder_mesh1}
	}
	\subfigure[]{
		\includegraphics[width=0.4 \textwidth]{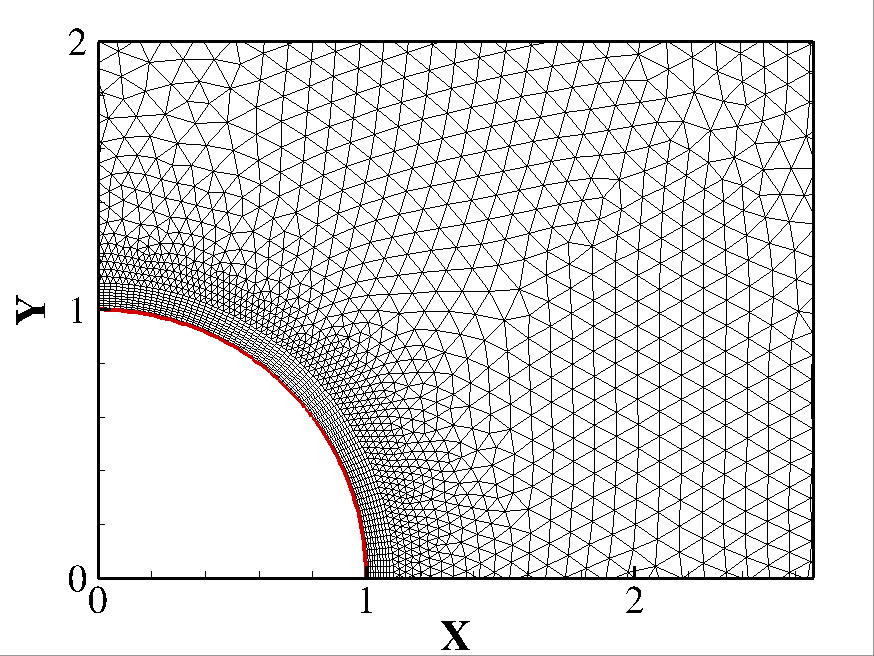}
		\label{cylinder_mesh2}
	}
	\caption{\label{cylinder_mesh} Mesh for flow around the oscillating circular cylinder: (a) Full domain and (b) near the circular cylinder surface.}
\end{figure}

Fig.~\ref{cylinder_mesh} shows the mesh used in this case. The size of computational domain is $[80d\times60d]$, it is very large so as to eliminate the influence of the far-field boundary condition. The total number of cells in the domain is about 49000, with 240 points along the cylinder surface. Quadrangular cells will be used to discretize the region near the cylinder surface with width equal to $0.1d$, and others are the triangles. At initial time, the cylinder is located at the center of computational domain, and $\rho=1.0$, $u=v=0.0$ will be used to initialize the distribution functions. $\rho=1.0$, $u=v=0.0$ also used as far-field boundary conditions.

\begin{figure}[!htp]
	\centering
    \subfigure[]{
    	\includegraphics[width=0.4 \textwidth]{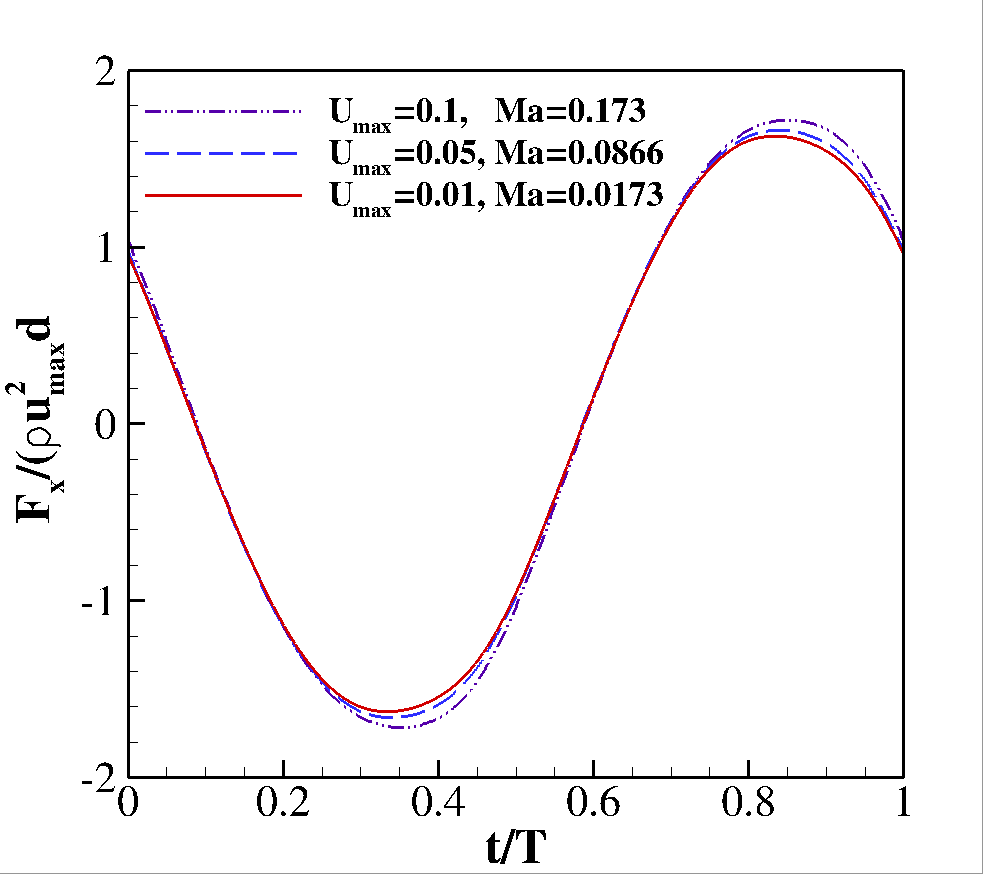}
    	\label{cylinder_ma_whole}
    }
    \subfigure[]{
    	\includegraphics[width=0.4 \textwidth]{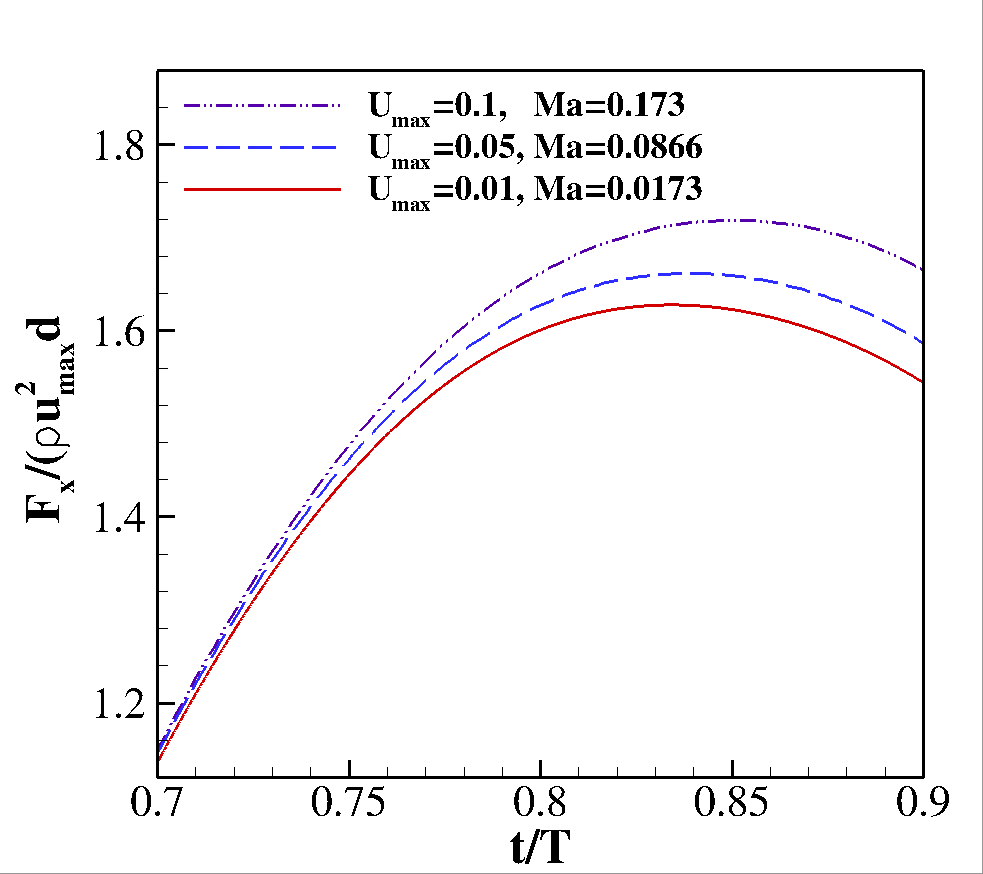}
    	\label{cylinder_ma_part}
    }
	\caption{\label{cylinder_ma_comp} The dimensionless inline force $F_x$ for flow around the oscillating circular cylinder at different $U_0$  ($\bigtriangleup$t$=0.00125$, $\bigtriangleup{x}=256/d$), (a) The full oscillating period $T$ and (b) The part enlarged region.}
\end{figure}

\begin{figure}[!htp]
	\centering
	\subfigure[]{
		\includegraphics[width=0.4 \textwidth]{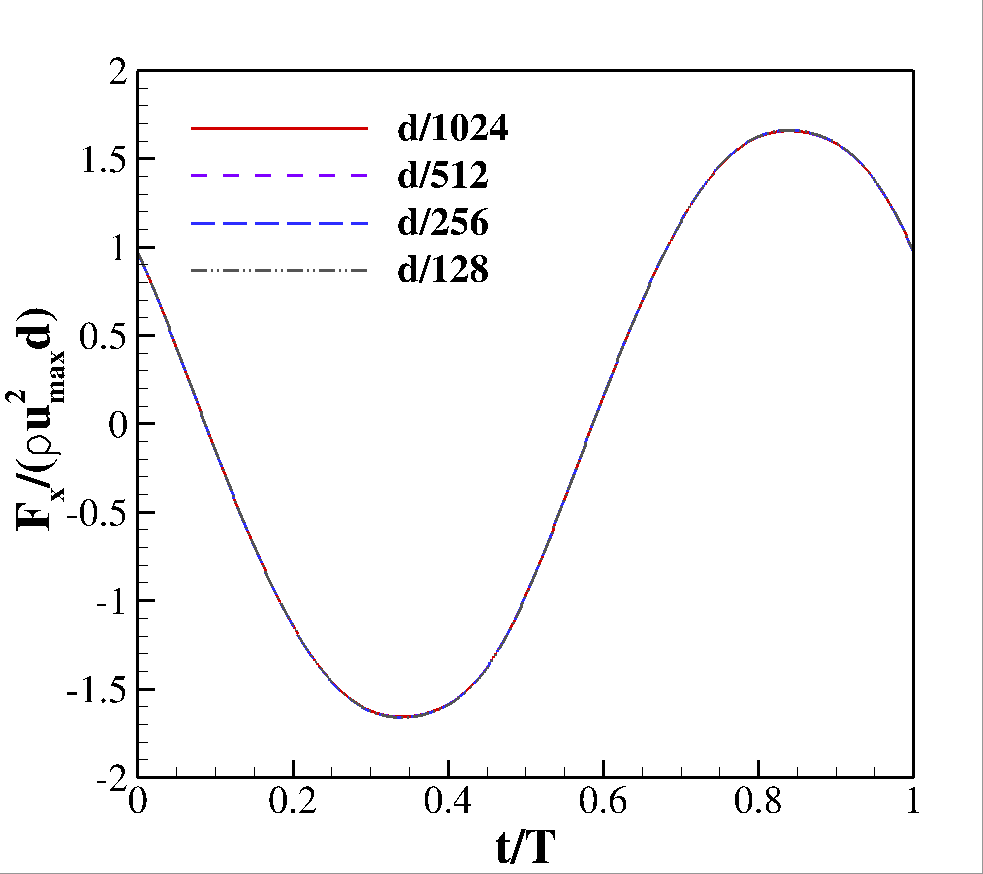}
		\label{cylinder_dx_whole}
	}
	\subfigure[]{
		\includegraphics[width=0.4 \textwidth]{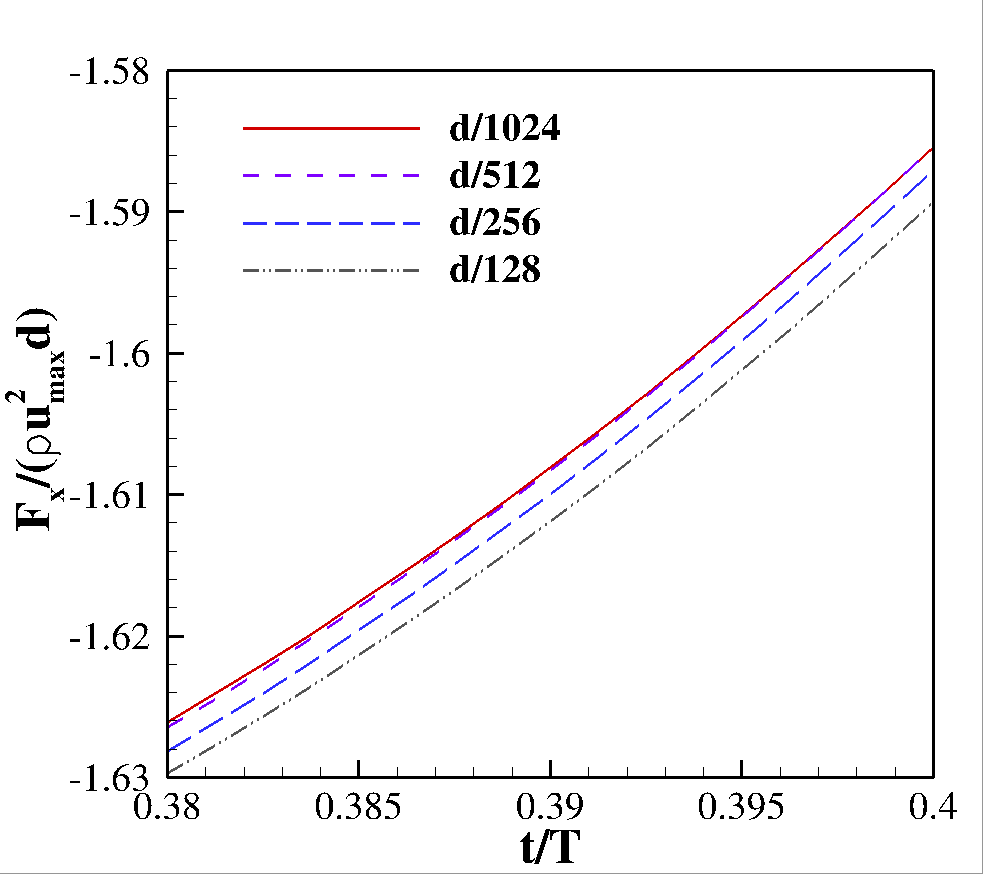}
		\label{cylinder_dx_part}
	}
	\caption{\label{cylinder_dx_comp} The dimensionless inline force $F_x$ for flow around the oscillating circular cylinder at different minimum grid size ($U_{max}=0.05$, $\bigtriangleup{t}=0.00125$), (a) The full oscillating period $T$ and (b) The part enlarged region.}
\end{figure}

\begin{figure}[!htp]
	\centering
	\subfigure[]{
		\includegraphics[width=0.4 \textwidth]{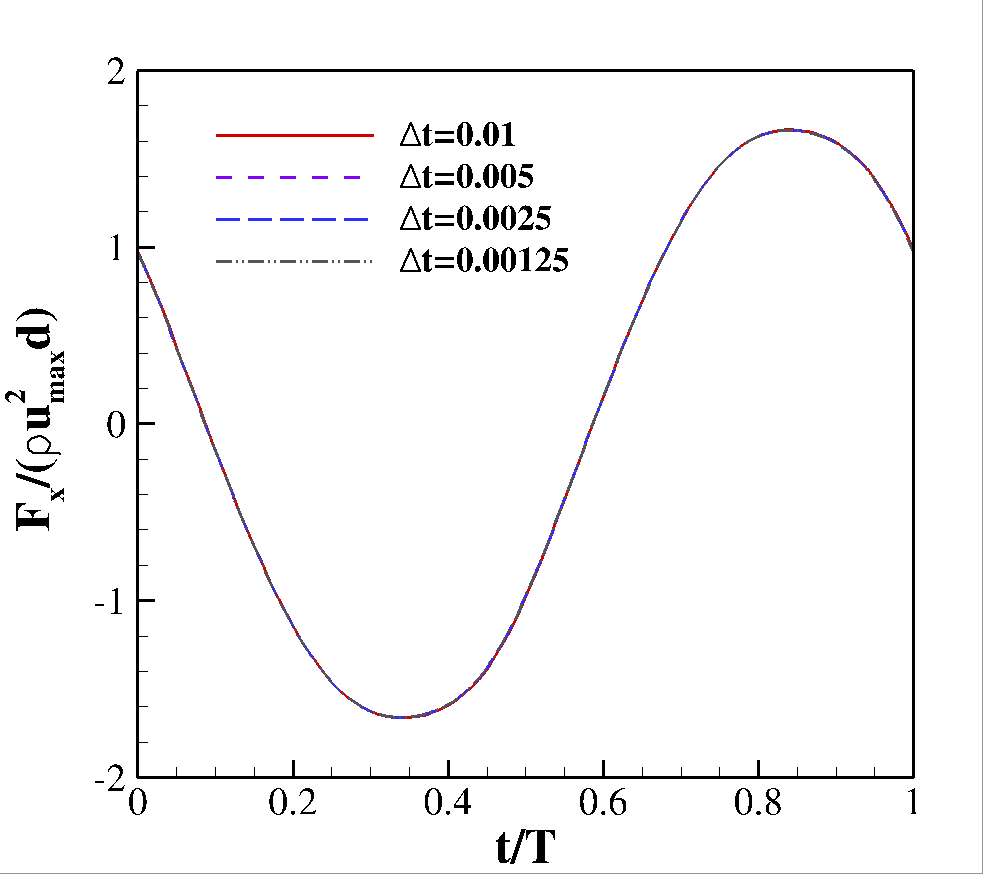}
		\label{cylinder_dt_whole}
	}
	\subfigure[]{
		\includegraphics[width=0.4 \textwidth]{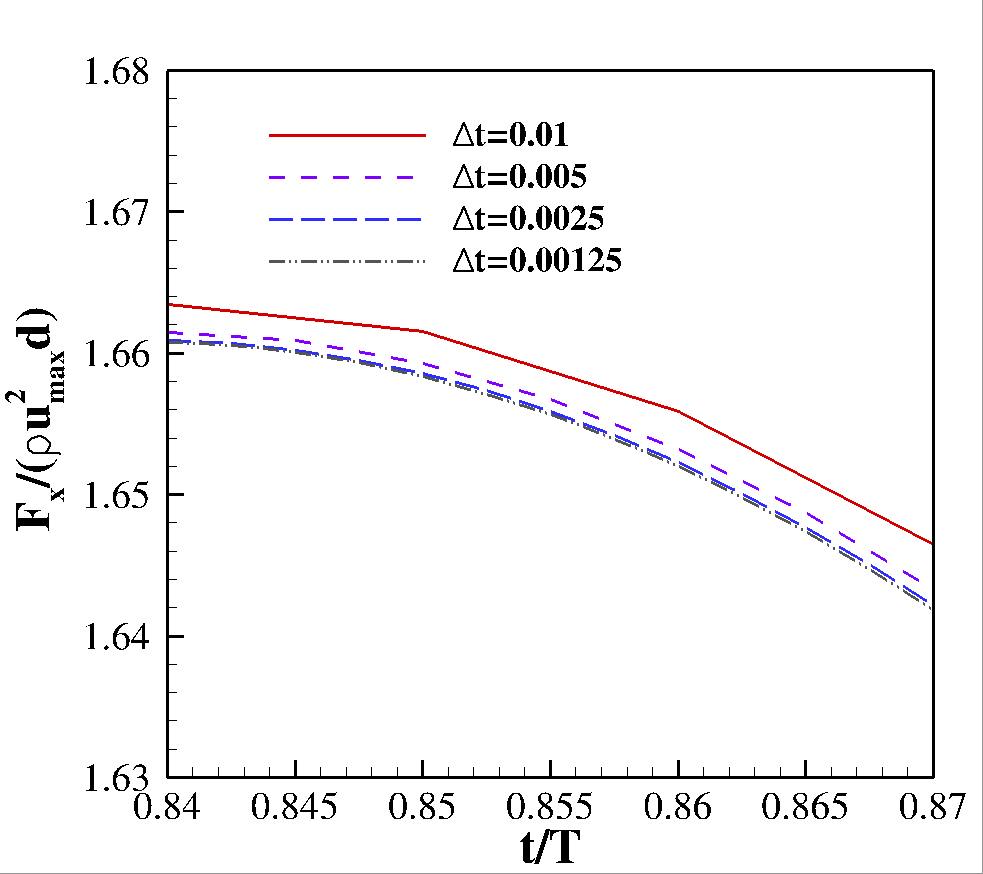}
		\label{cylinder_dt_part}
	}
	\caption{\label{cylinder_dt_comp} The dimensionless inline force $F_x$ for flow around the oscillating circular cylinder at different time step $\Delta$$t$ ($U_{max}=0.05$, $\bigtriangleup{x}=256/d$), (a) The full oscillating period $T$ and (b) The part enlarged region.}
\end{figure}

Firstly, the numerical convergency of inline force $F_x$ at different Mach number, Ma, time step, $\bigtriangleup$$t$, and minimum grid size, $\bigtriangleup$$x$ are tested. For the test based on the Ma, three values, which equal to 0.173 ($U_{max}=0.1$), 0.0866 ($U_{max}=0.05$), and 0.0173 ($U_{max}=0.01$) are considered (The sound speed equal to $1/\sqrt{3}$ in this case), respectively. As shown in Fig.~\ref{cylinder_ma_comp}, for one oscillating period $T$, it is clear that the inline force $F_x$  will much influenced by the different Ma, especially for the amplitude of $F_x$. Because in framework of LBM, the equilibrium distribution function used in this case will recover the compressible Navier-Stokes equations, so the compressible effect might lead to some undesirable errors in numerical simulations \cite{guo2000lattice}. For the tests based on the $\bigtriangleup{x}$ and the $\bigtriangleup{t}$, as illustrated in Fig.~\ref{cylinder_dx_comp} and Fig.~\ref{cylinder_dt_comp}, respectively, the influences to result are much declined compared with that of Ma. Consider the computational time, Ma=0.0866 ($U_{max}=0.05$), $\bigtriangleup{x}=256/d$ and $\bigtriangleup{t}=0.05$ will be used in the following simulations.

\begin{figure}[!htp]
	\centering
	\subfigure[]{
		\includegraphics[width=0.4 \textwidth]{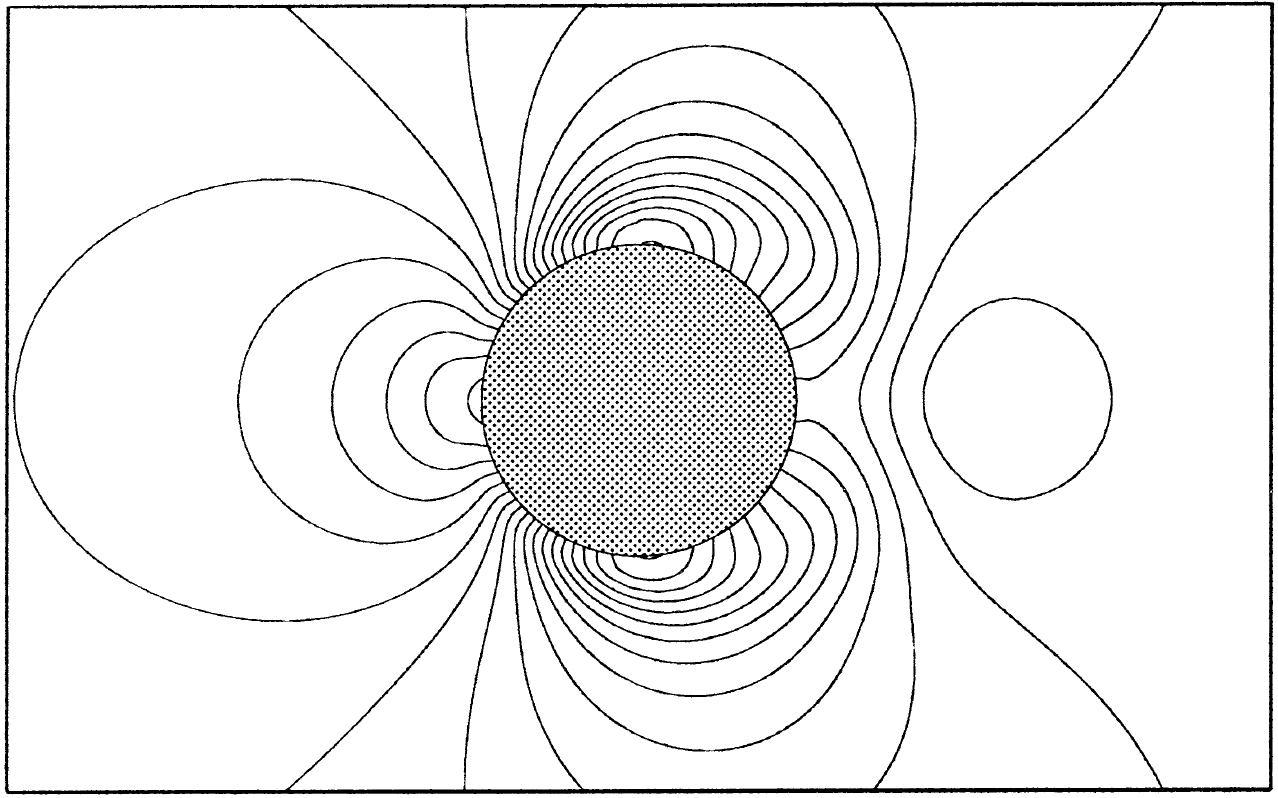}
		\label{cylinder_pressure0_ref}
		\includegraphics[width=0.407 \textwidth]{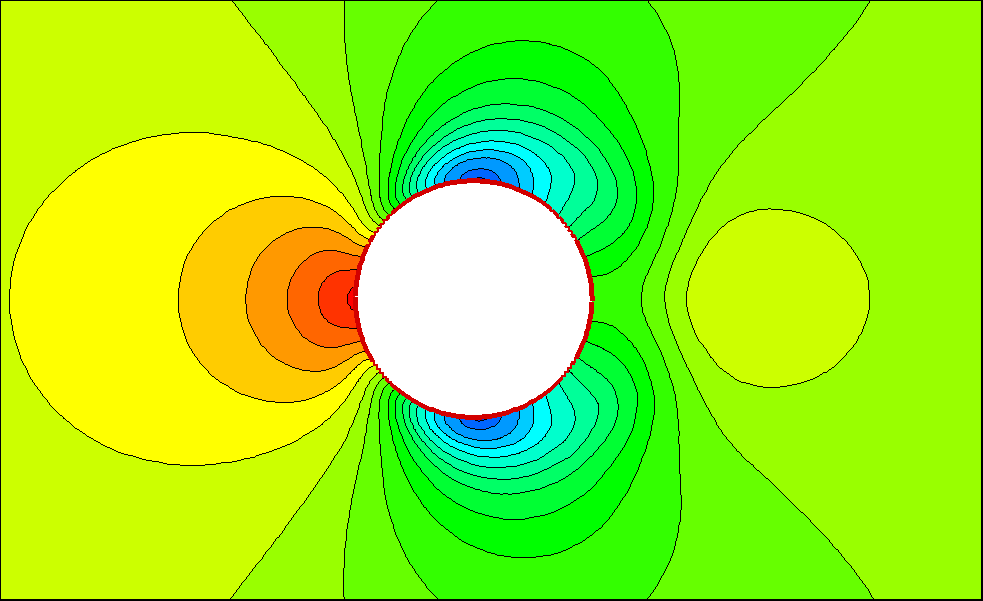}
		\label{cylinder_pressure0_pres}
	}
	\subfigure[]{
		\includegraphics[width=0.4 \textwidth]{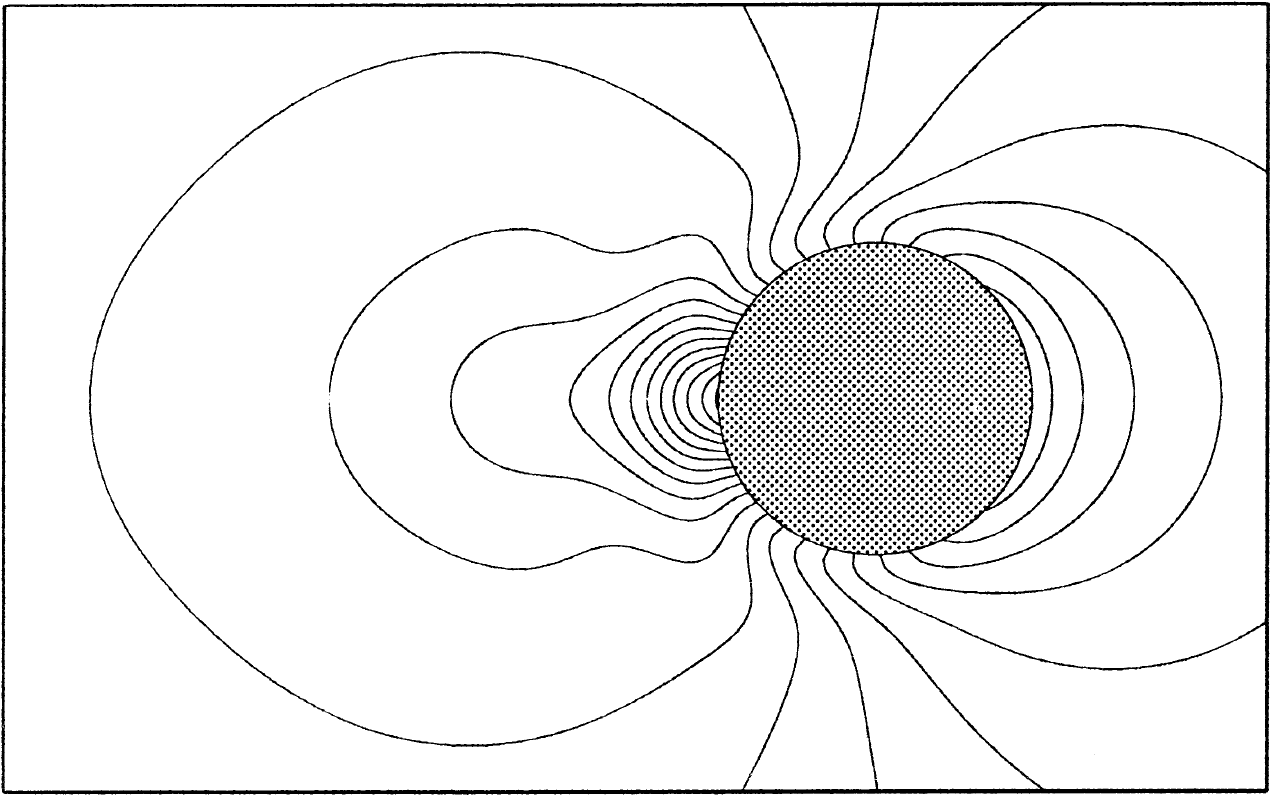}
		\label{cylinder_pressure288}
		\includegraphics[width=0.407 \textwidth]{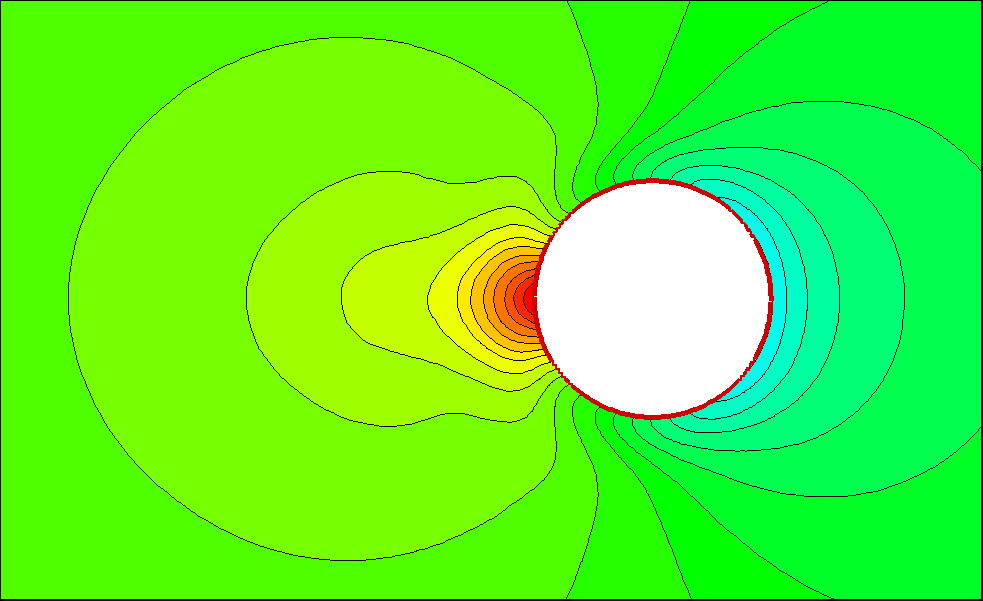}
		\label{cylinder_pressure288_pres}
	}
	\caption{\label{cylinder_pressue} Pressure contours for flow around an oscillating circular cylinder at phase (a) $0^\circ$ and (b) $288^\circ$, D{\"u}tsch et al. \cite{dutsch1998low} (left) and present (right) results.}
\end{figure}

\begin{figure}[!htp]
	\centering
	\subfigure[]{
		\includegraphics[width=0.4 \textwidth]{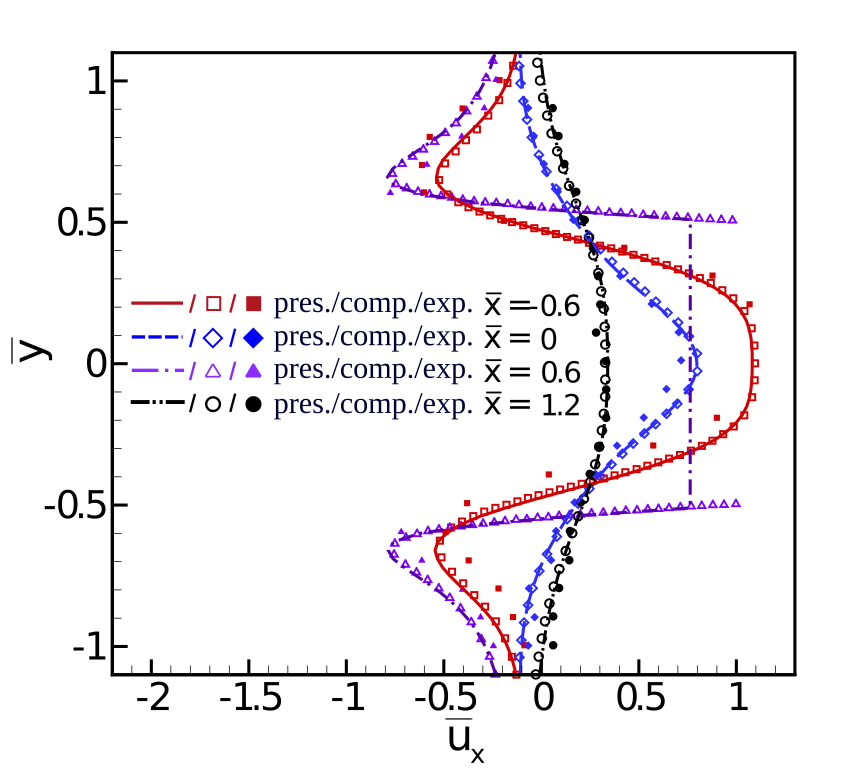}
		\label{cylinder_180u}
		\includegraphics[width=0.4 \textwidth]{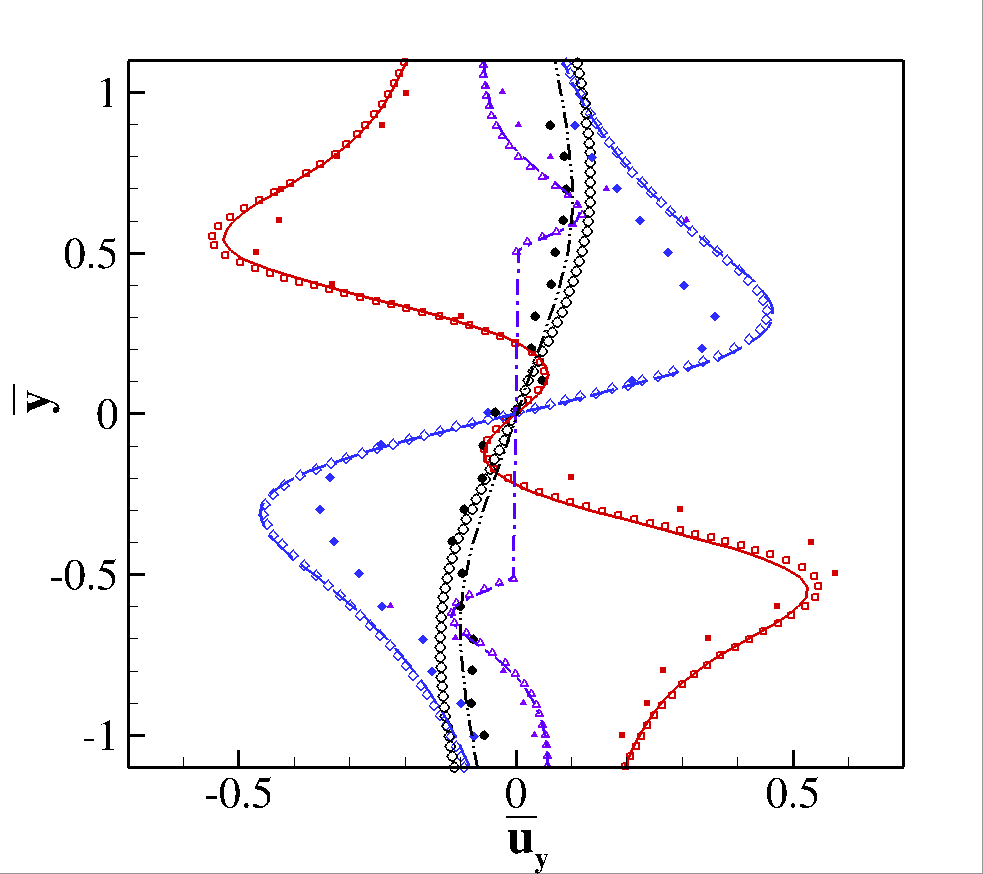}
		\label{cylinder_180v}
	}
	\subfigure[]{
	    \includegraphics[width=0.4 \textwidth]{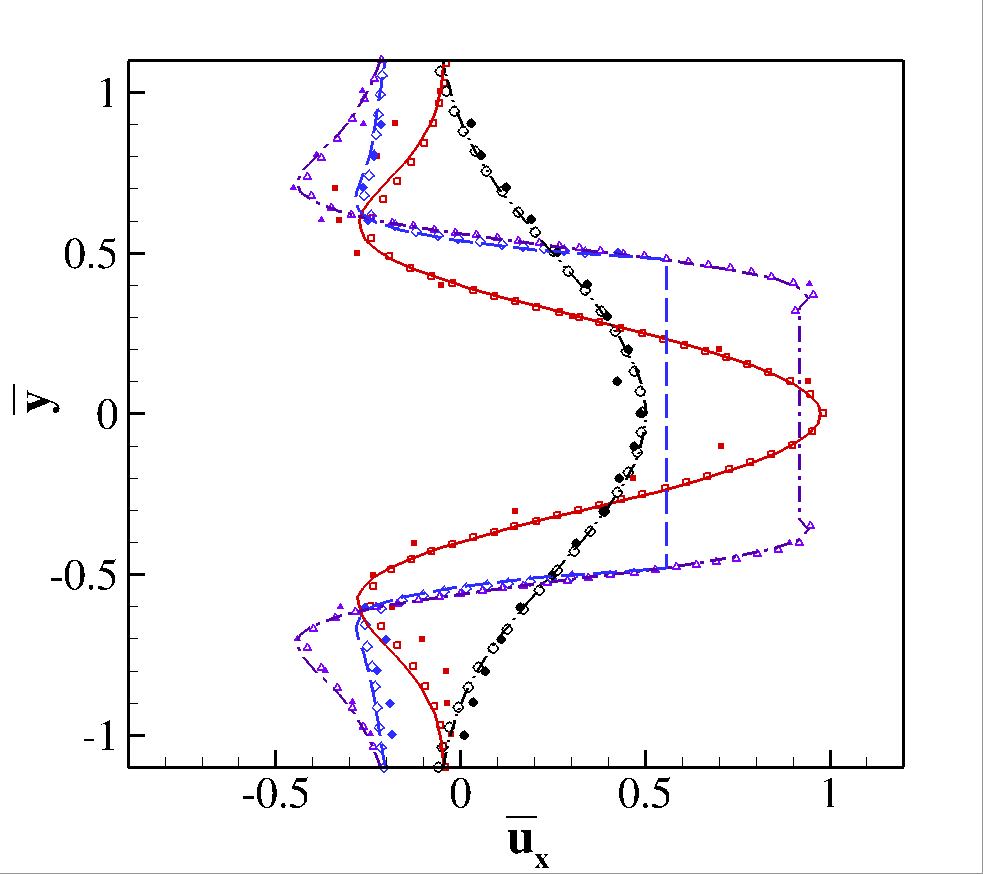}
	    \label{cylinder_210u}
	    \includegraphics[width=0.4 \textwidth]{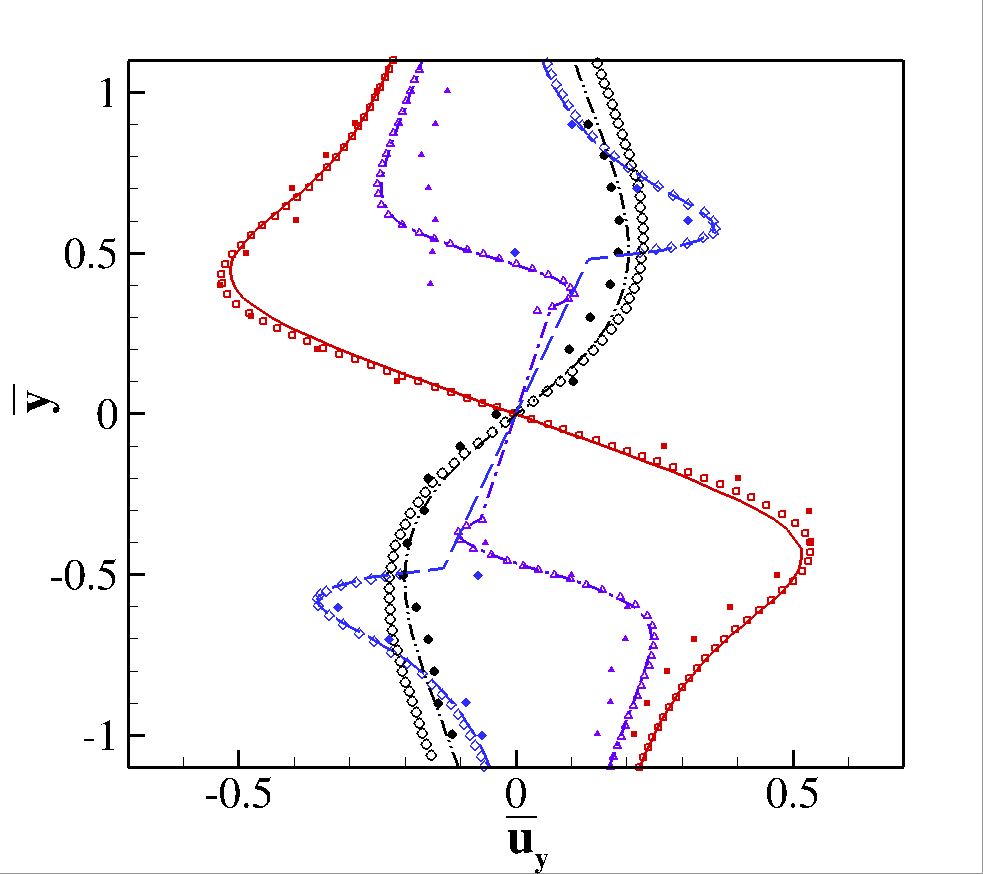}
	    \label{cylinder_210v}
    }
	\subfigure[]{
	    \includegraphics[width=0.4 \textwidth]{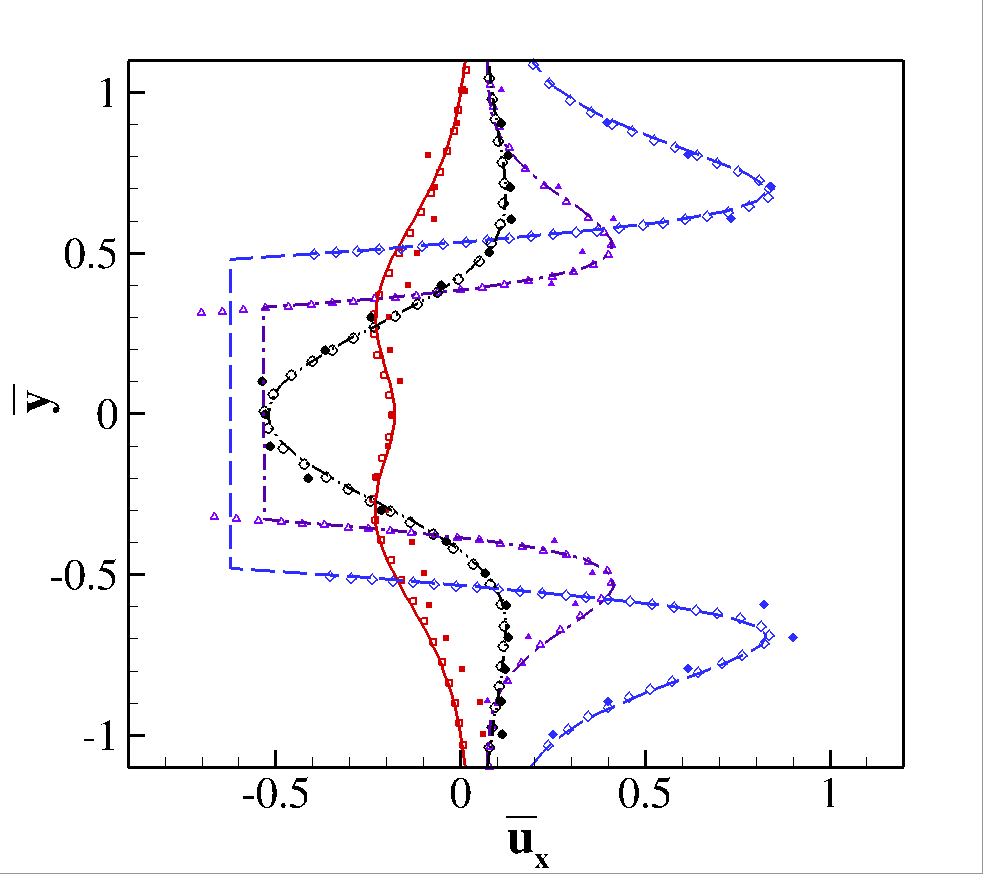}
	    \label{cylinder_330u}
	    \includegraphics[width=0.4 \textwidth]{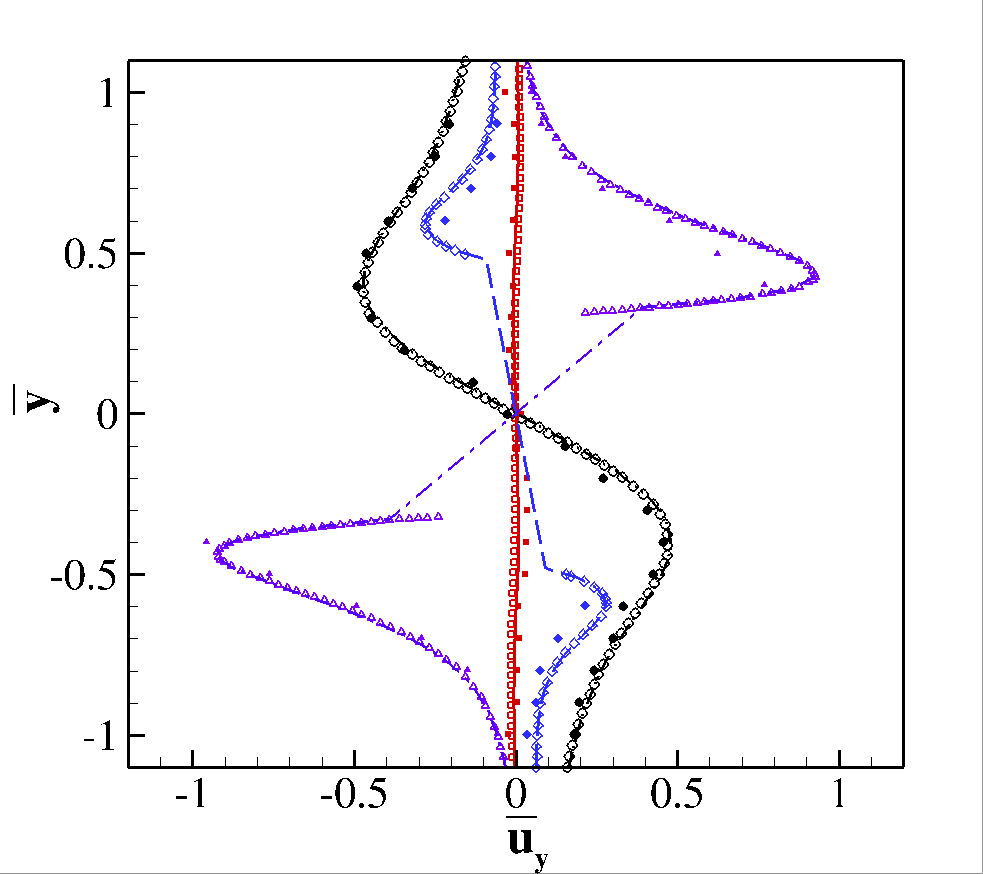}
        \label{cylinder_330v}	
    }
	\caption{\label{cylinder_profile} The velocity profiles of flow around the oscillating circular cylinder at four vertical cross sections $\bar{x} = -0.6, 0, 0.6, 1.2$ for phase position (a) $180^o$, (b) $210^o$ and (c) $330^o$. The present results are compared with the computational/experimental values of D{\"u}tsch et al. \cite{dutsch1998low}.}
\end{figure}

Secondly, some other results are compared to further validate our computation quantitatively. Pressure isolines at phase position $0^\circ$ and $288^\circ$ are shown in Fig.~\ref{cylinder_pressue}. These isolines are agree well with the numerical results of D{\"u}tsch et al. \cite{dutsch1998low}. Furthermore, the velocity profiles $\bar{u}_x$ and $\bar{u}_y$ at four vertical cross section $\bar{x} = -0.6, 0, 0.6, 1.2$ for three phase positions $180^\circ$, $210^\circ$ and $330^\circ$ are also compared with numerical and experiment results obtained by D{\"u}tsch et al. \cite{dutsch1998low}, where $\bar{u}_x$, $\bar{u}_y$, and $\bar{x}$ ($\bar{y}$ is the vertical distance for comparison) are defined as

\begin{equation}
	\bar{x}=\frac{x}{d}, \bar{y}=\frac{y}{d}, \bar{u}_x=\frac{u_x}{u_{max}}, \bar{u}_y=\frac{u_y}{u_{max}},
\end{equation}
$x$ and $y$ are the coordinates relative to the equilibrium position of cylinder, $u_x$ and $u_y$ are the velocity components in horizontal and vertical direction. As shown in Fig.~\ref{cylinder_profile}, generally, our results also agree well with the numerical and experiment results of D{\"u}tsch et al. \cite{dutsch1998low}.

\begin{figure}[!htp]
	\centering	
	\includegraphics[width=0.9 \textwidth]{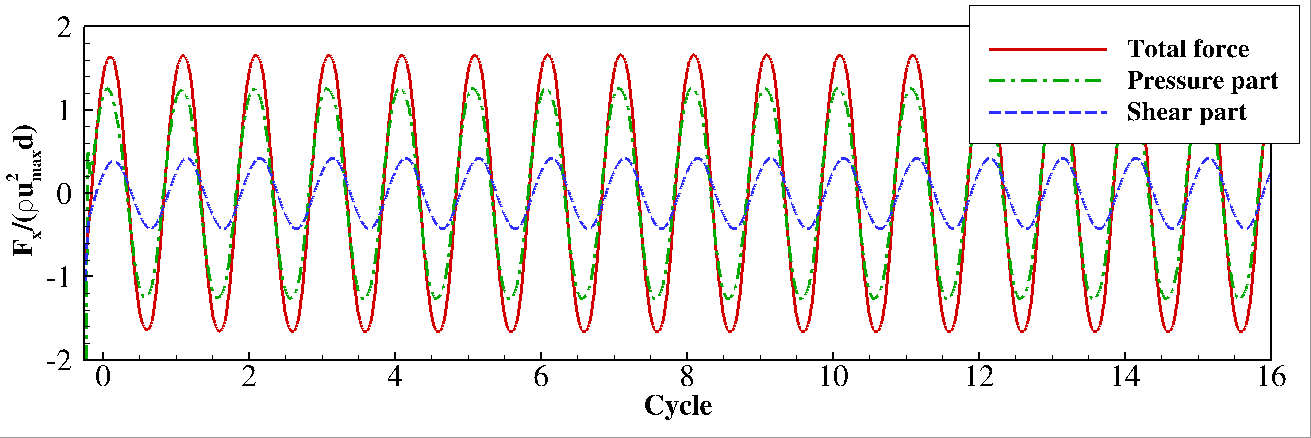}
	\caption{\label{cylinder_cycle} The history of dimensionless inline force $F_x$ for flow around the oscillating circular cylinder, and the contributions of pressure and shear to the total force.}
\end{figure}

\begin{figure}[!htp]
	\centering	
	\includegraphics[width=0.4 \textwidth]{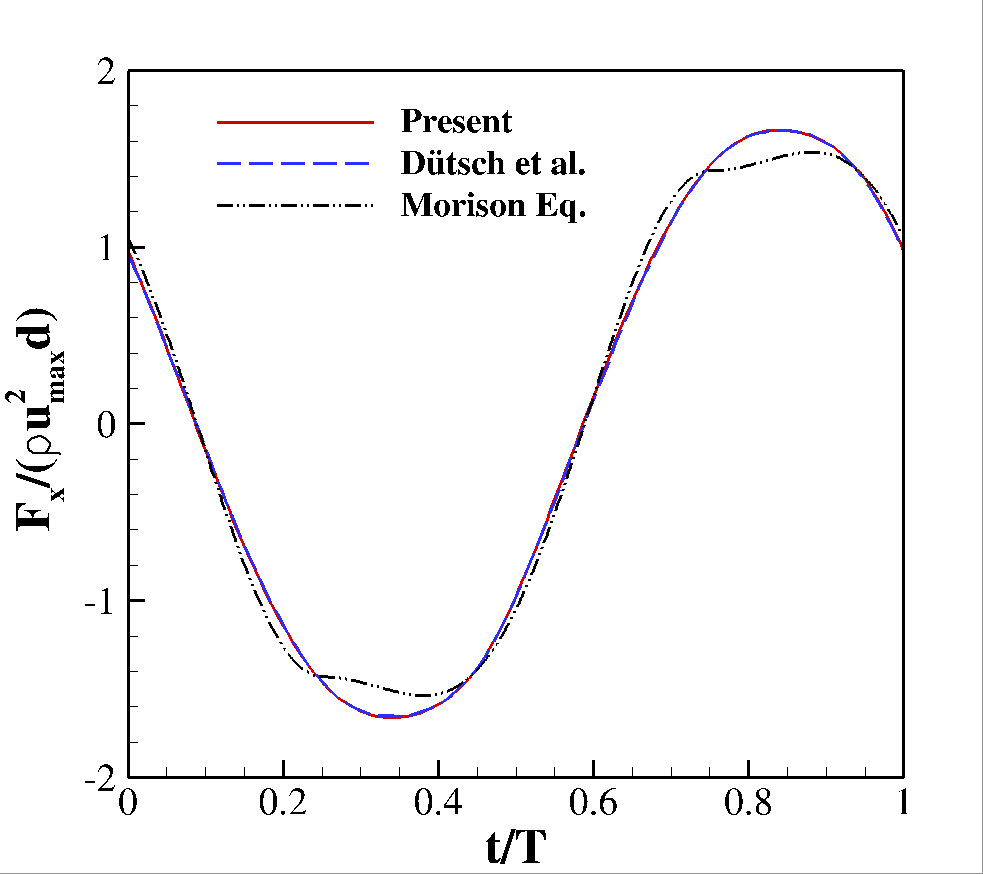}
	\caption{\label{cylinder_cd_comp} The compare of inline force $F_x$ for flow around the oscillating circular cylinder during an period $T$, compared with the results of D{\"u}tsch et al. \cite{dutsch1998low} and Morison equation \cite{morison1950force} (Eq.~(\ref{cylinder_empirical})).}
\end{figure}

For oscillating flow, the semi-empirical equation of Morison et al. \cite{morison1950force} are widely used to estimate the inline force $F_x$ on body. When the circular cylinder oscillating in the stationary fluids, the time-dependent inline force $F_x$ is expressed as

\begin{equation}\label{cylinder_empirical}
  F_x=-\frac{1}{2}\rho{d}{c_d}\dot{x}\left|\ddot{x}\right|-\frac{1}{4}\pi\rho{d^2}c_i\left|\ddot{x}\right|,
\end{equation}
where $x$ is the displacement of cylinder, $c_d$ and $c_i$ are the drag coefficient and the added mass coefficient, respectively. Integral the pressure and stress at the surface of cylinder, the $F_x$ can be calculated, then with the help of least-squares fitting or Fourier analysis, the $c_d$ and $c_i$ also can be evaluated. Fig.~\ref{cylinder_cycle} shows the time history of $F_x$, it is clear that the pressure is dominating contribution to the total force. Similar behavior also described by D{\"u}tsch et al. \cite{dutsch1998low}. The fitted $c_d$ and $c_i$, and some other numerical results are compared in Table \ref{tab:cylinder_cd}, though the $c_i$ is little higher, generally the present ALE-DUGKS can get good results. Given the $c_d$ and $c_i$, Eq.~(\ref{cylinder_empirical}) can evaluate the empirical values of $F_x$ at one oscillating period. In Fig.~\ref{cylinder_cd_comp}, our results are great well with that of D{\"u}tsch et al. \cite{dutsch1998low}, and rougher consistency with the empirical values of Morison et al. \cite{morison1950force}.

\begin{table}
	\centering
	\caption{\label{tab:cylinder_cd} Drag coefficient $c_d$ and added mass coefficient $c_i$ for the flow over an circular cylinder oscillating  in the stationary fluid.}
	\begin{tabular}{ccc}
		\hline
		\multicolumn{1}{c}{Work} & \multicolumn{1}{c}{$c_d$} & \multicolumn{1}{c}{$c_i$} \\
		\hline
		D{\"u}tsch et al. \cite{dutsch1998low}{$^a$} & 2.09 & 1.45   \\
		Uzuno{\u{g}}lu et al. \cite{uzunouglu2001low}{$^a$} & 2.10 & 1.45   \\
		Yuan et al. \cite{yuan2018immersed} & 2.10 & 1.47   \\
		Present & 2.10 & 1.48   \\
		\lasthline
	\end{tabular}
	\begin{tablenotes}
	   \centering
	   \item{Note: $a$, Data are based on the finest meshes in Ref.~\cite{dutsch1998low} and \cite{uzunouglu2001low}.}
    \end{tablenotes}
\end{table}

\subsection{Continuum flow around a pitching NACA0012 airfoil}
\begin{figure}[!htp]
	\centering
	\subfigure[]{
		\includegraphics[width=0.4 \textwidth]{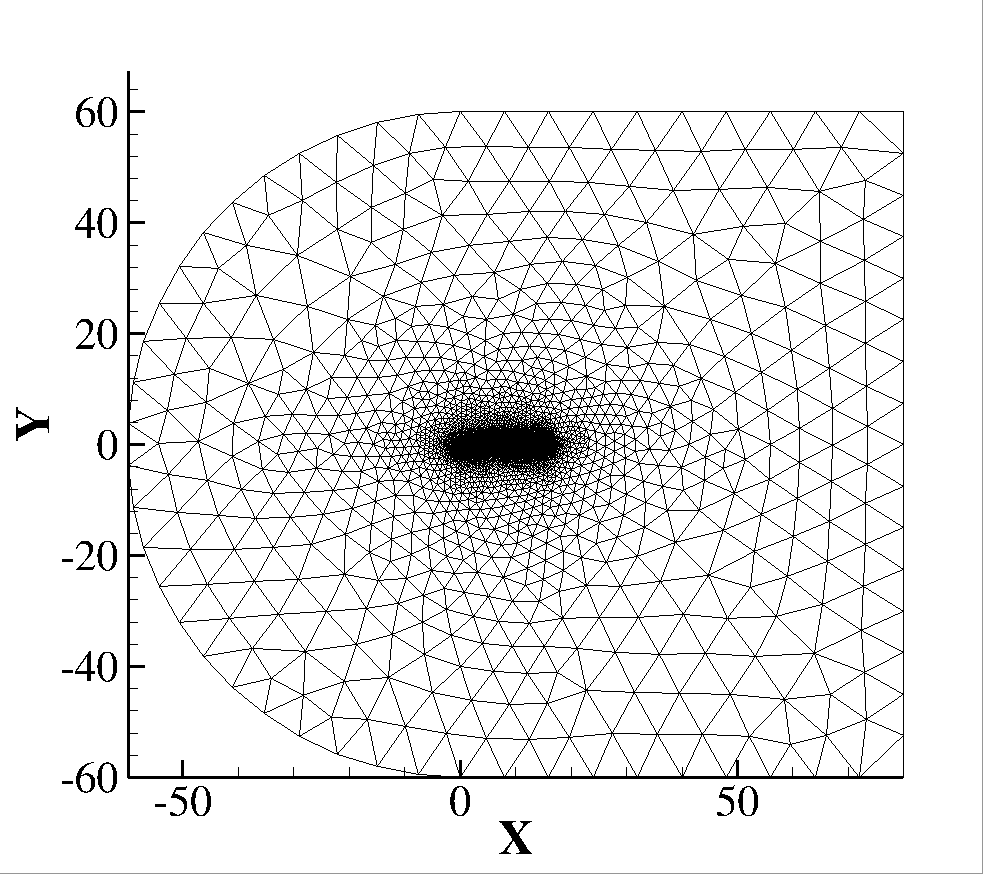}
		\label{pitch_mesh_whole}
	}
	\subfigure[]{
		\includegraphics[width=0.4 \textwidth]{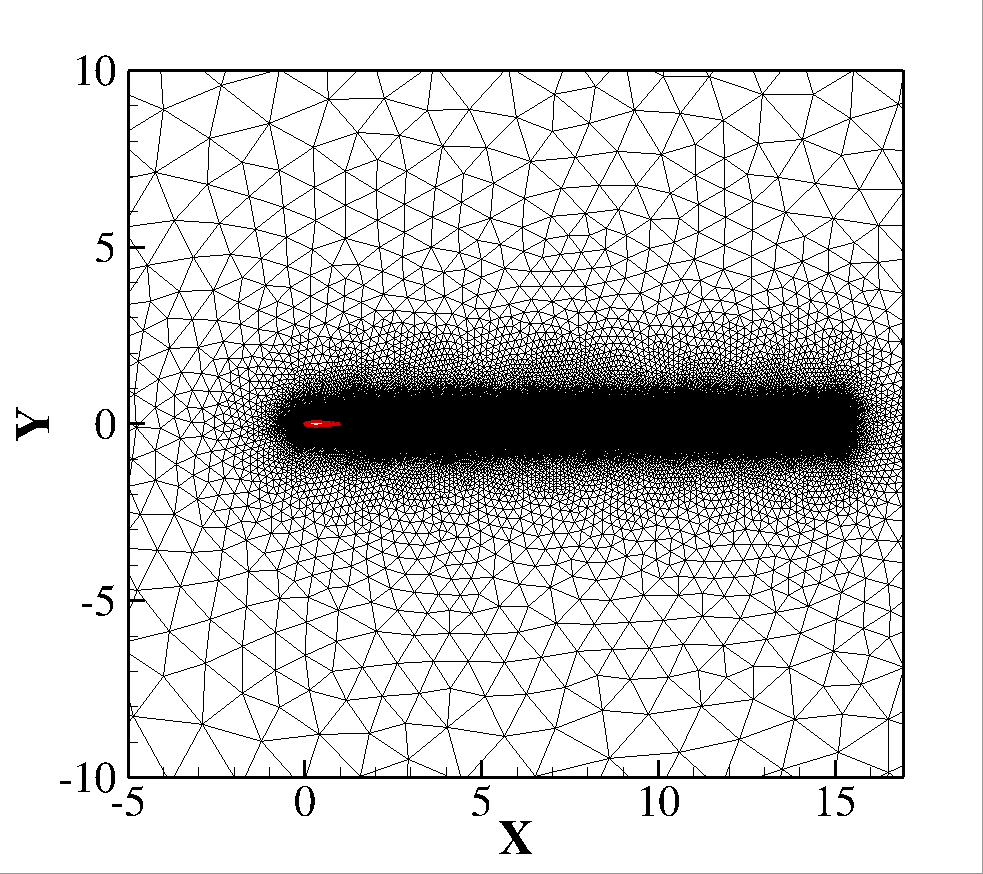}
		\label{pitch_mesh_part1}
	}
    \subfigure[]{
	    \includegraphics[width=0.4 \textwidth]{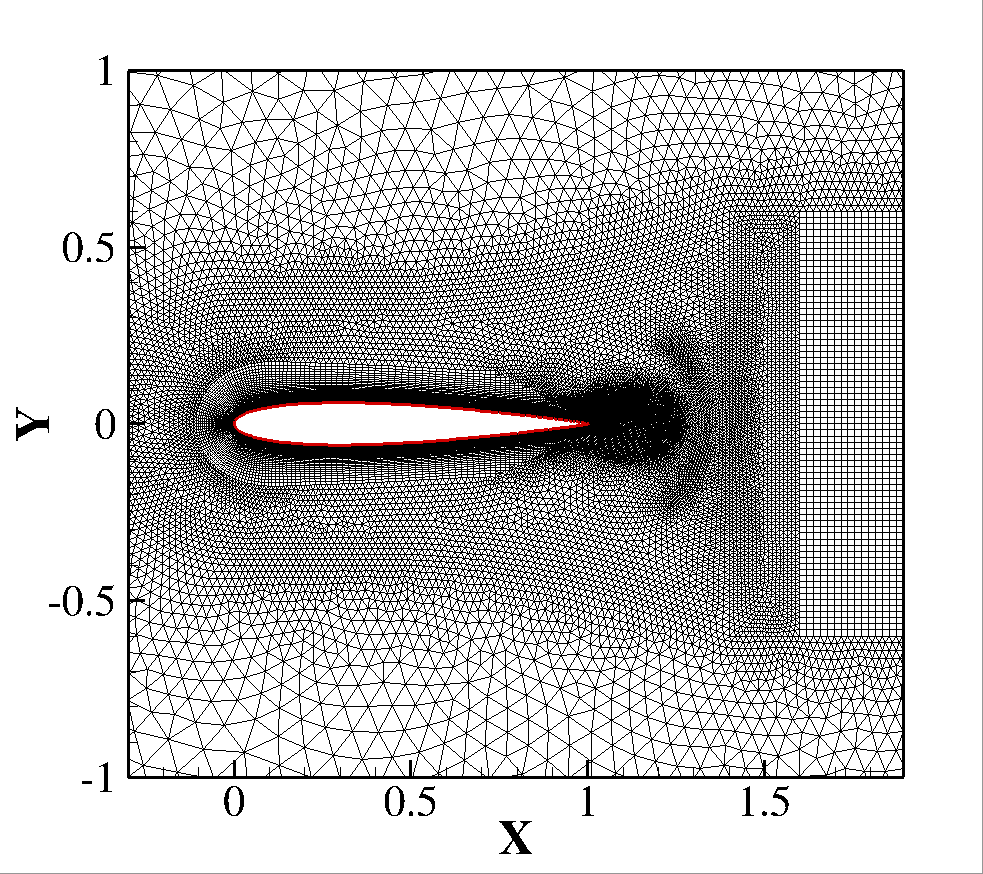}
	    \label{pitch_mesh_part2}
    }
    \subfigure[]{
	    \includegraphics[width=0.4 \textwidth]{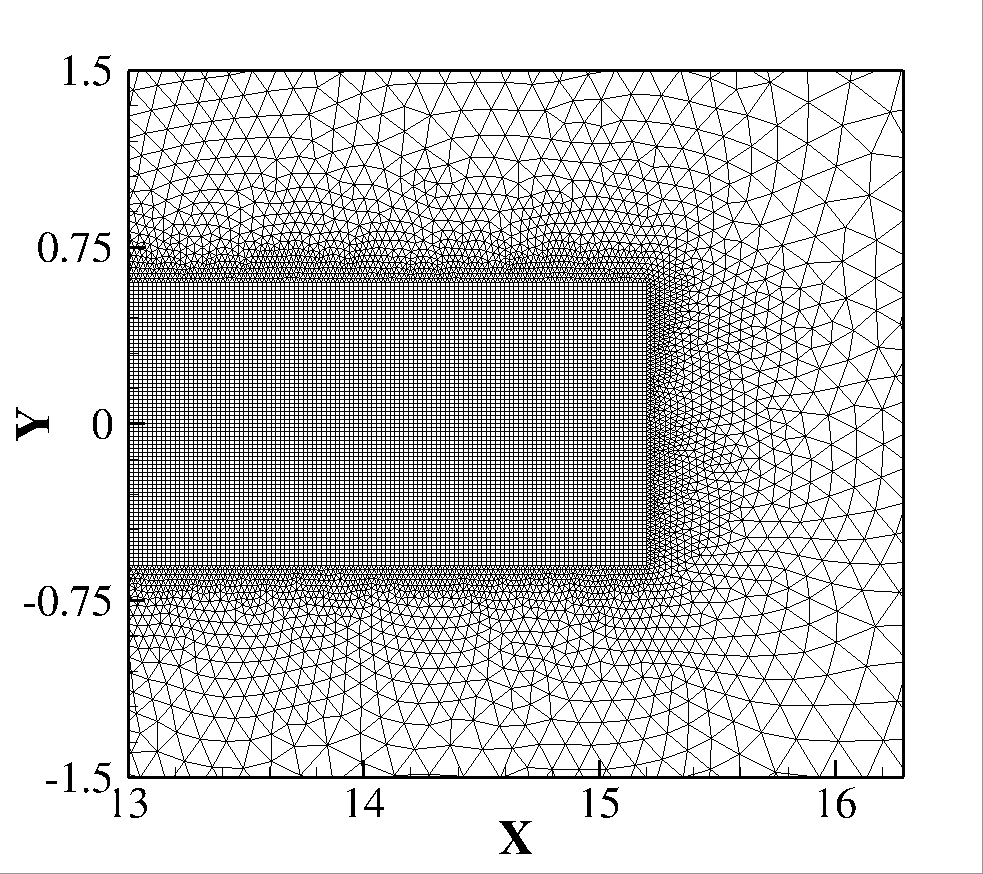}
	    \label{pitch_mesh_part3}
    }
	\caption{\label{pitch_mesh} Mesh for flow around the pitching NACA0012 airfoil, (a) Full domain and (b)(c)(d) Part enlarged regions near the airfoil.}
\end{figure}

When study the details of propulsion for insects, birds, or fishes, the flows around oscillating airfoil with pitching and/or heaving motions usually used as benchmark test cases \cite{young2004oscillation}. In this section, only motion of pitching is considered, which also can check the performance of present ALE-DUGKS to cope with the rotary moving boundary problem. The profile of airfoil is NACA0012, and pitching at its quarter-chord. The variation of attack of angle (AoA) $\alpha$ for pitching airfoil can be expressed as

\begin{equation}
   \alpha=dsin(2\pi{f}{t})
\end{equation}
where $d$ is amplitude of pitching, and $f$ is pitching frequency. Usually, for pitching airfoil, a new parameter, that is reduced frequency, $k$, will be defined. The relationship between $k$ and $f$ is $k=2\pi{f}c/U_0$, where $c$ is the length of airfoil chord ($c=1$), and $U_0$ is the velocity of free-stream. The Reynolds number, Re, based on the $c$, is set as 12000. Generally, flow under this Re can be treated as laminar flow \cite{liang2011high}. Fig.~\ref{pitch_mesh} shows the mesh used in this case. The total number of cells in the domain is $140,831$, along with 291 points at the airfoil surface. Region near the airfoil is discretized into rectangle cells, and the minimum size of cell is $1.0 \times 10^{-4}$. About 15$c$ length away from the airfoil trail, Cartesian grids are used to capture the vortical patterns. The number of cells is little larger in this case, if only consider some results like force coefficients, about $50\sim60\times10^3$ cells are enough to get good results. In addition, as shown in Sec.~\ref{cylinder_case}, a small value of $U_0$ ($U_0=0.05$) is used to eliminate the compressible effect.

\begin{figure}[!htp]
	\centering
	\subfigure[]{
		\includegraphics[width=0.4 \textwidth]{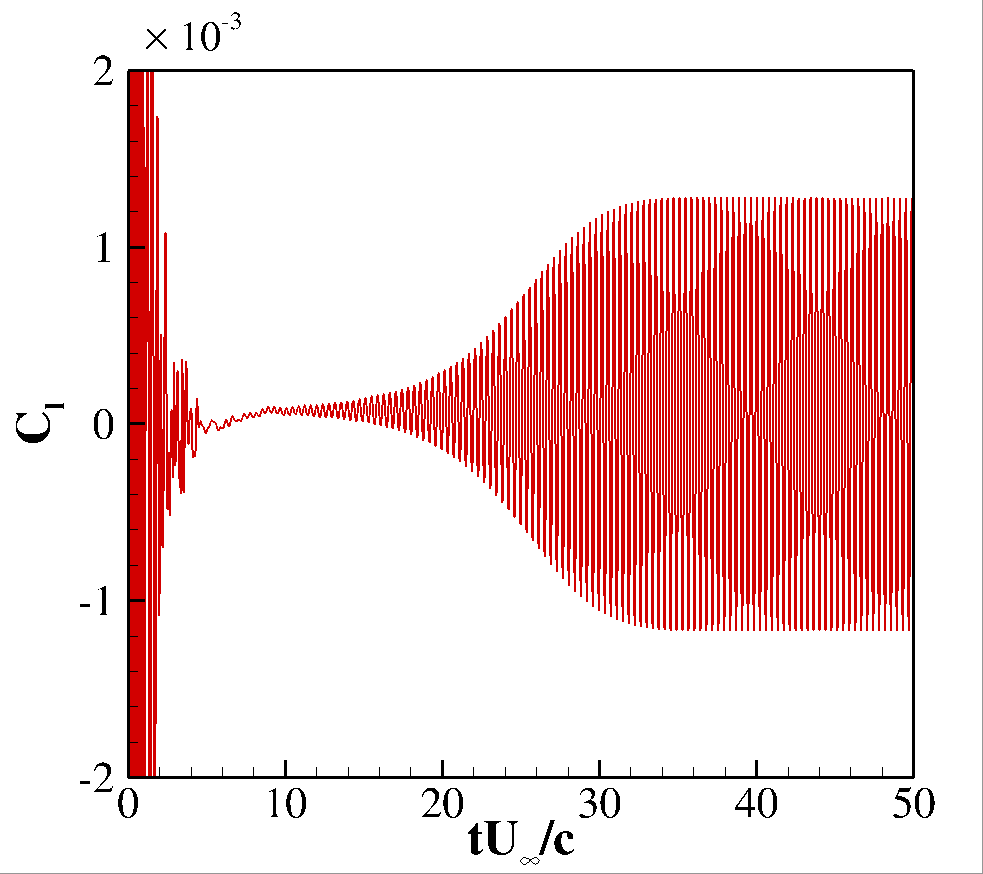}
		\label{pitch_stat_cl}
	}
	\subfigure[]{
		\includegraphics[width=0.4 \textwidth]{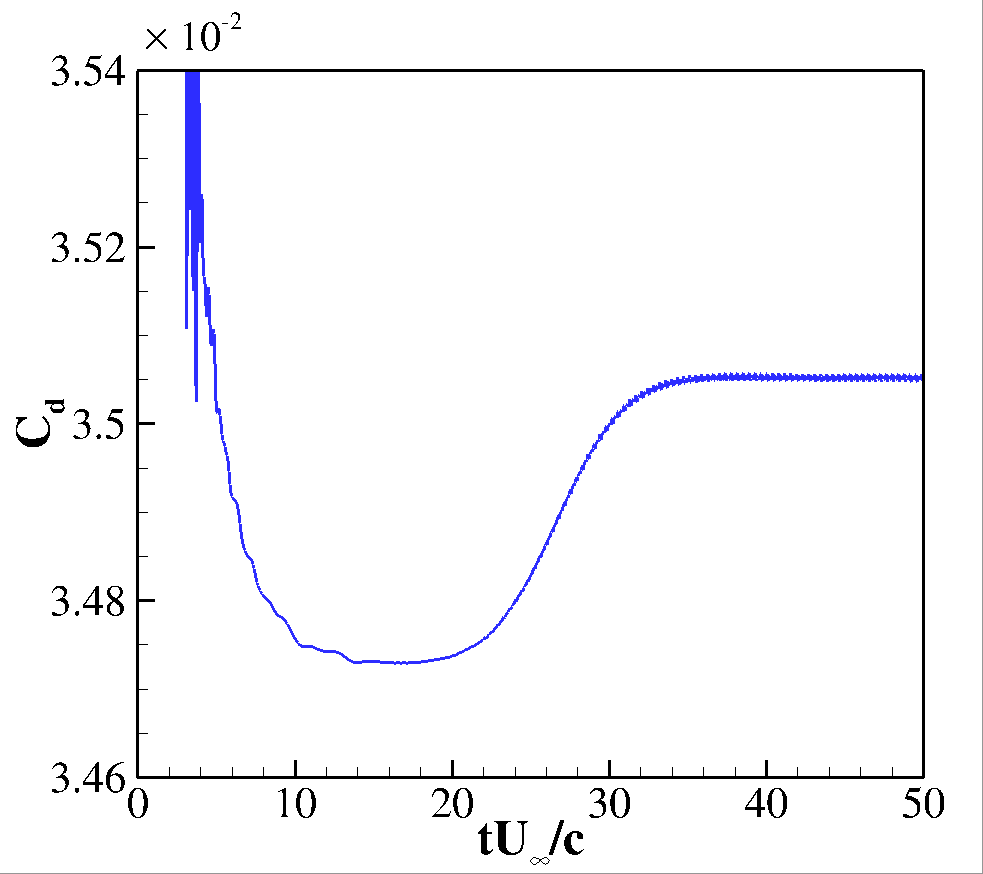}
		\label{pitch_stat_cd}
	}
	\caption{\label{pitch_stat_f} The histories of (a) lift and (b) drag coefficient for flow around the NACA0012 airfoil at reduced frequency $k=0$.}
\end{figure}

Firstly, the flow around stationary airfoil is simulated. Reported by Koochesfahani \cite{koochesfahani1989vortical} with experiment, under this Reynolds number, the phenomenon of vortex-shedding will generate, and the equivalent reduced frequency, $k_{equi.}$, based on the $f$, is 8.7. Fig.~\ref{pitch_stat_f} shows the time evolutions of lift coefficient $C_l$ and drag coefficient $C_d$, where $C_l$ and $C_d$ are defined as

\begin{equation}\label{pitch_force}
	C_l=\frac{F_x}{0.5\rho_0{U_0^2}c},	
	C_d=\frac{F_y}{0.5\rho_0{U_0^2}c},
\end{equation}
$F_x$ and $F_y$ are the components of force at $x$-direction and $y$-direction, and $\rho_0$ is density of free-stream ($\rho_0=1.0$ in this case). As the amplitude of $C_l$ is very small, and maybe due to the unstructured mesh is used, the maximum and minimum of $C_l$ is little different. From our test, the numerical result of $k_{equi.}$ is equal to 8.23, close to the experiment value. Besides, the mesh at the region about $0.5c$ width behind the airfoil trail must be refined. If the mesh at this region is too coarse, such as O-type mesh, the simulation will result a steady flow.

\begin{figure}[!htp]
	\centering
    \includegraphics[width=0.9 \textwidth]{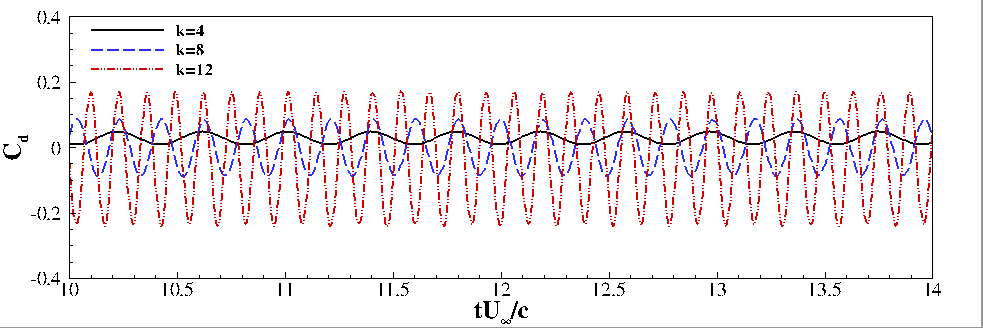}
	\caption{\label{pitch_cd_compare} The histories of drag coefficient for flow around the pitching NACA0012 airfoil at different reduced frequencies $k$, $d=2^\circ$.}
\end{figure}

\begin{figure}[!htp]
	\centering
	\subfigure[]{
		\includegraphics[width=0.4 \textwidth]{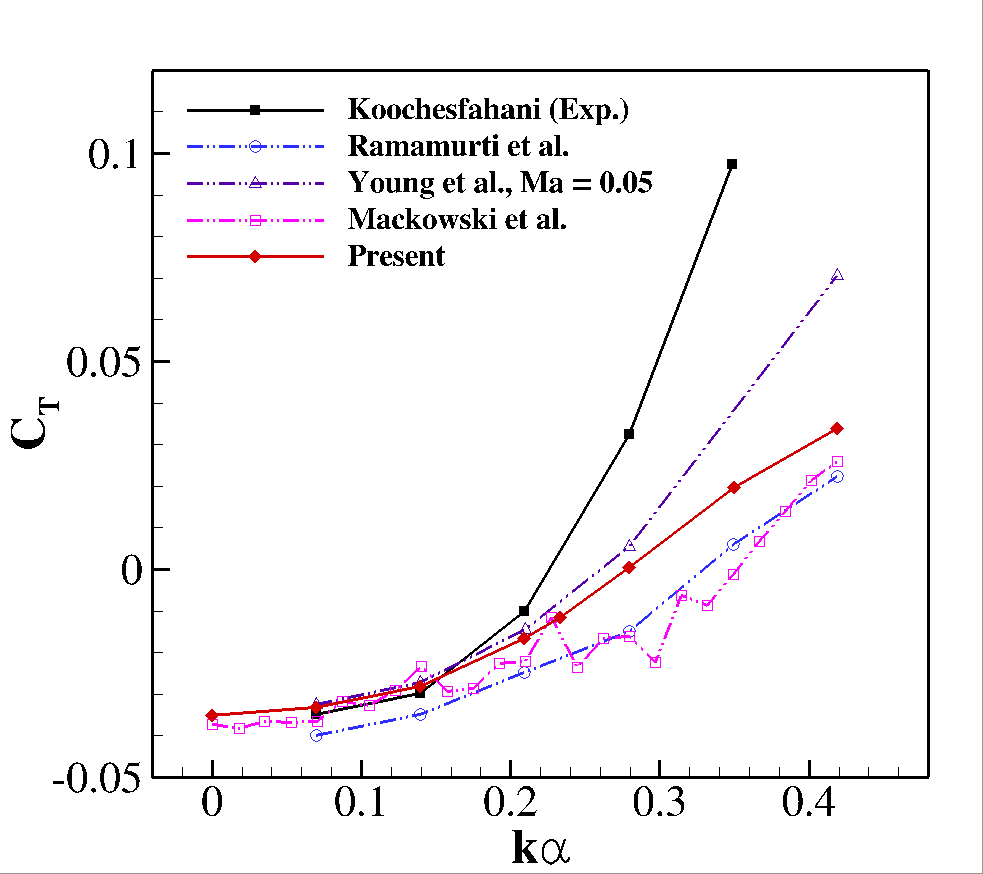}
		\label{pitch_cd_alf2}
	}
	\subfigure[]{
		\includegraphics[width=0.4 \textwidth]{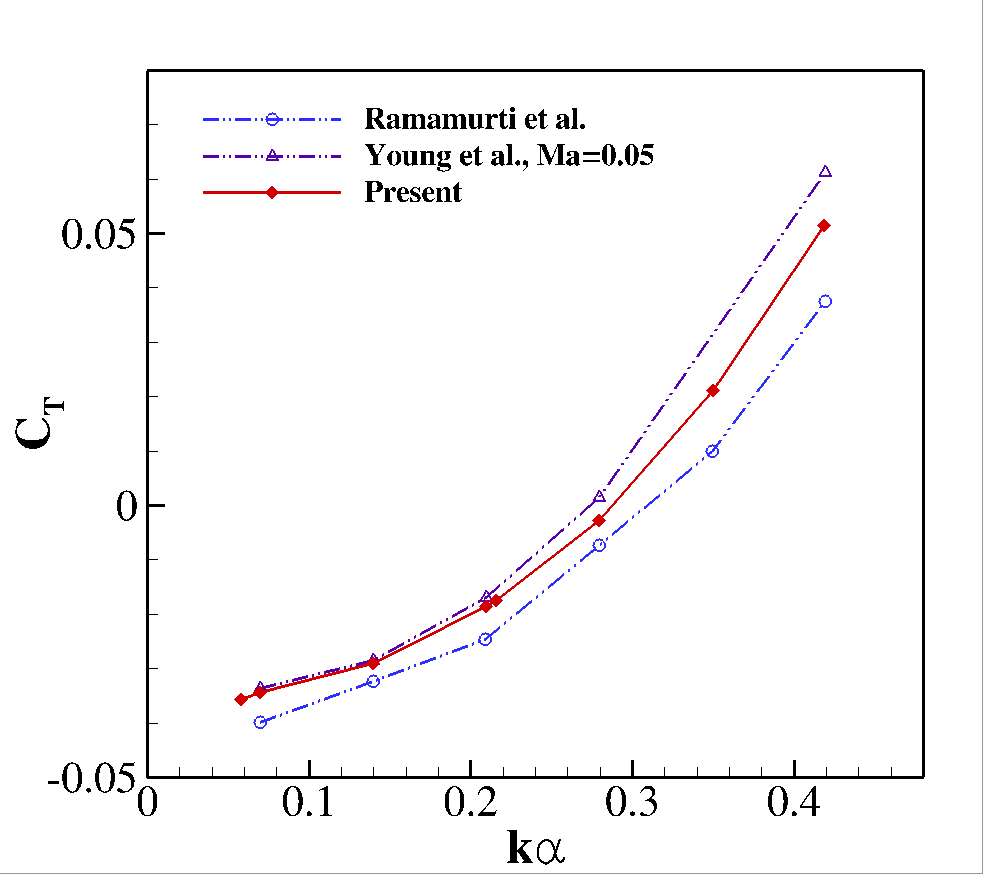}
		\label{pitch_cd_alf4}
	}
	\caption{\label{pitch_cd} The mean thrust coefficients $\bar{C}_T$ for flow around the pitching NACA0012 airfoil at (a) AoA=2$^\circ$ and (b) AoA = 4$^\circ$.}
\end{figure}

\begin{figure}[!htp]
	\centering
	\subfigure[]{
		\includegraphics[width=0.4 \textwidth]{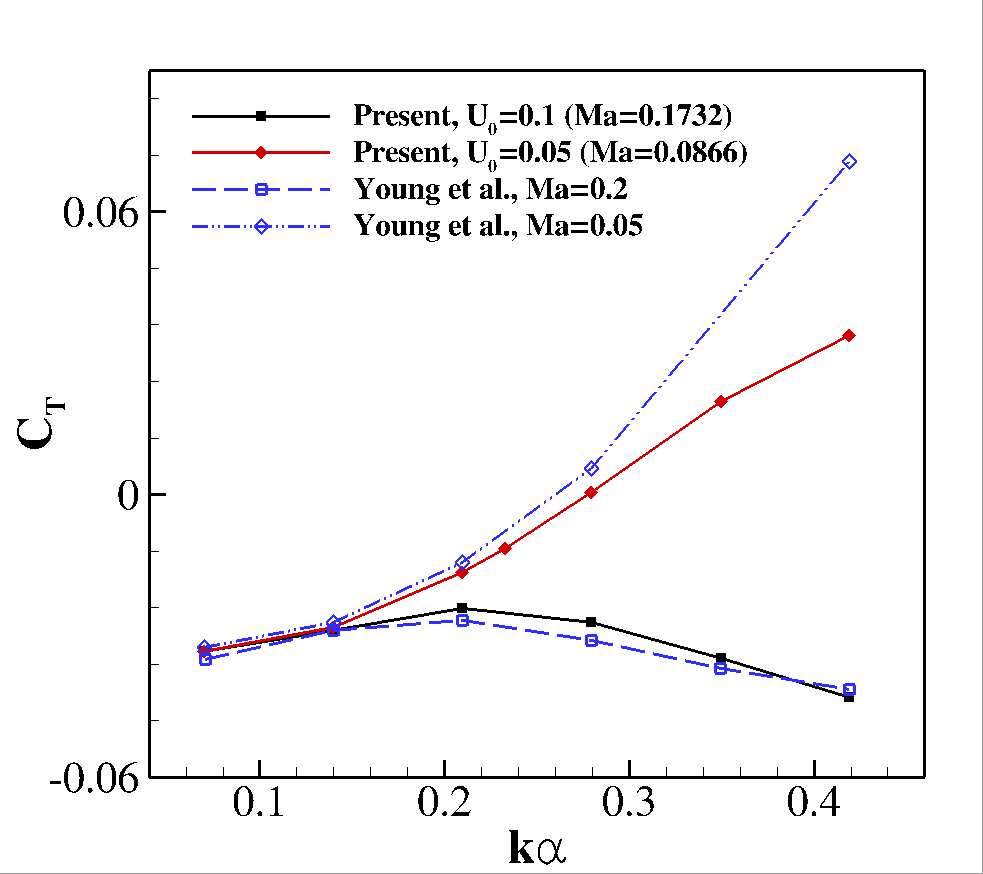}
		\label{pitch_ma_alf2}
	}
	\subfigure[]{
		\includegraphics[width=0.4 \textwidth]{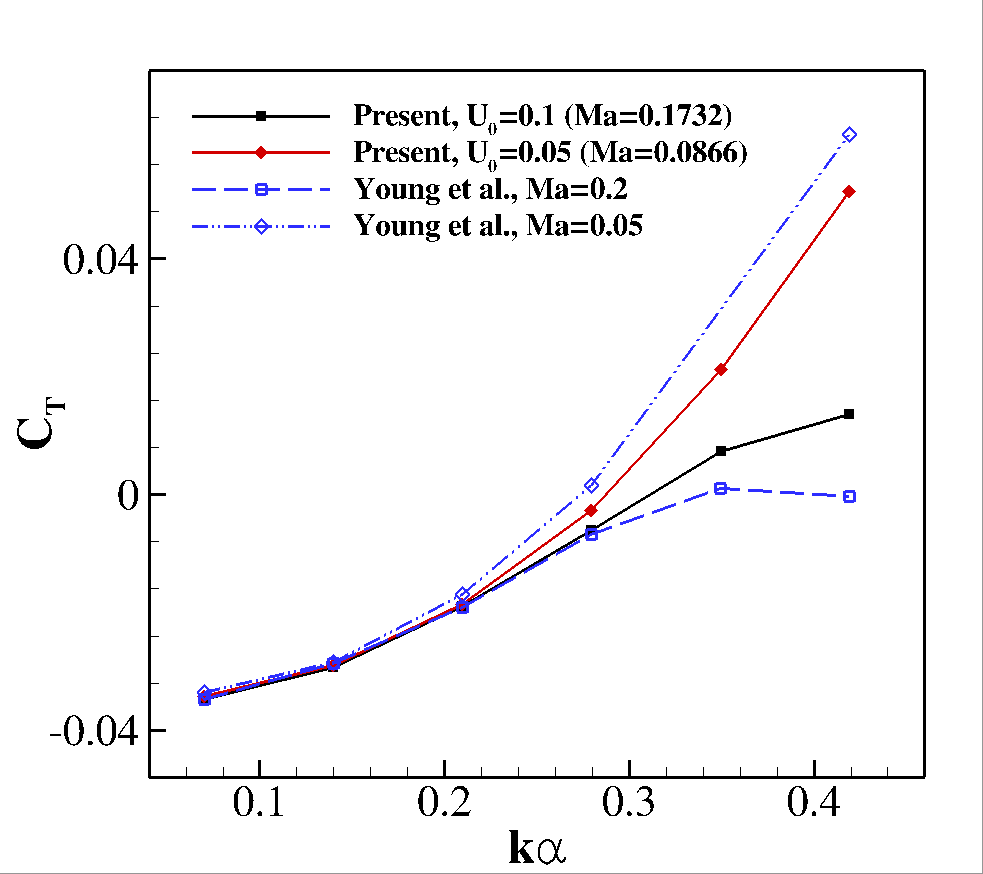}
		\label{pitch_ma_alf4}
	}
	\caption{\label{pitch_ma} The mean thrust coefficients $\bar{C}_T$ for flow around the pitching NACA0012 airfoil at (a) AoA=2$^\circ$ and (b) AoA = 4$^\circ$ at different Mach number, Ma.}
\end{figure}

\begin{figure}[!htp]
	\centering
	\subfigure[]{
		\includegraphics[width=0.4 \textwidth]{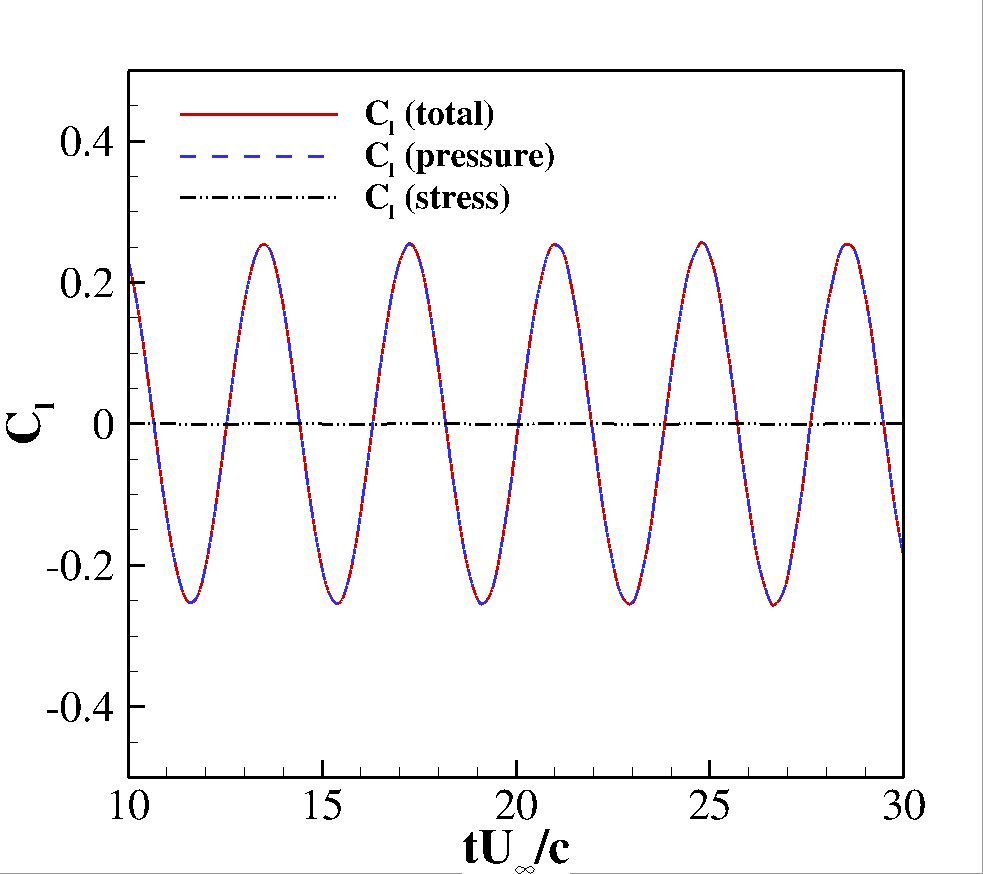}
		\label{pitch_cl_alf4k0p835}
	}
	\subfigure[]{
		\includegraphics[width=0.4 \textwidth]{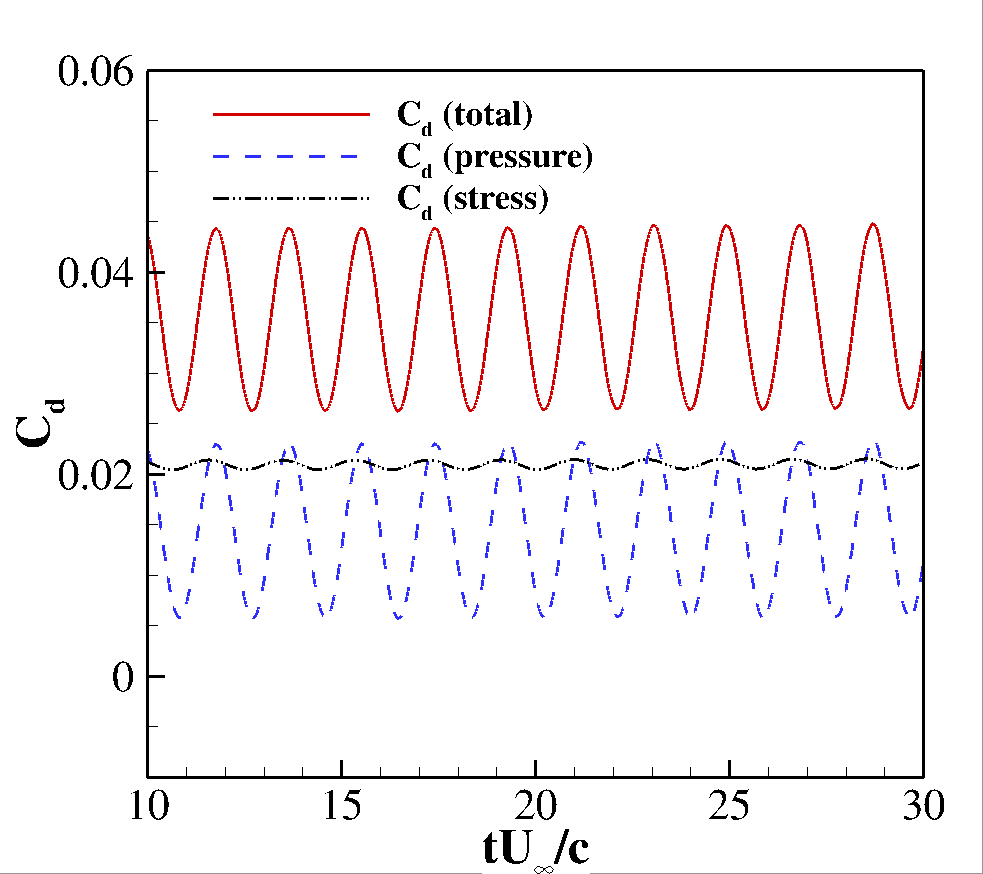}
		\label{pitch_cd_alf4k0p835}
	}
	\caption{\label{pitch_clcd_alf4k0p835} The histories of (a) lift and (b) drag coefficient for flow around the pitching NACA0012 airfoil at AoA=4$^\circ$ and $k=0.835$, and the contributions of pressure and stress to the total forces.}
\end{figure}

\begin{figure}[!htp]
	\centering
	\subfigure[]{
		\includegraphics[width=0.4 \textwidth]{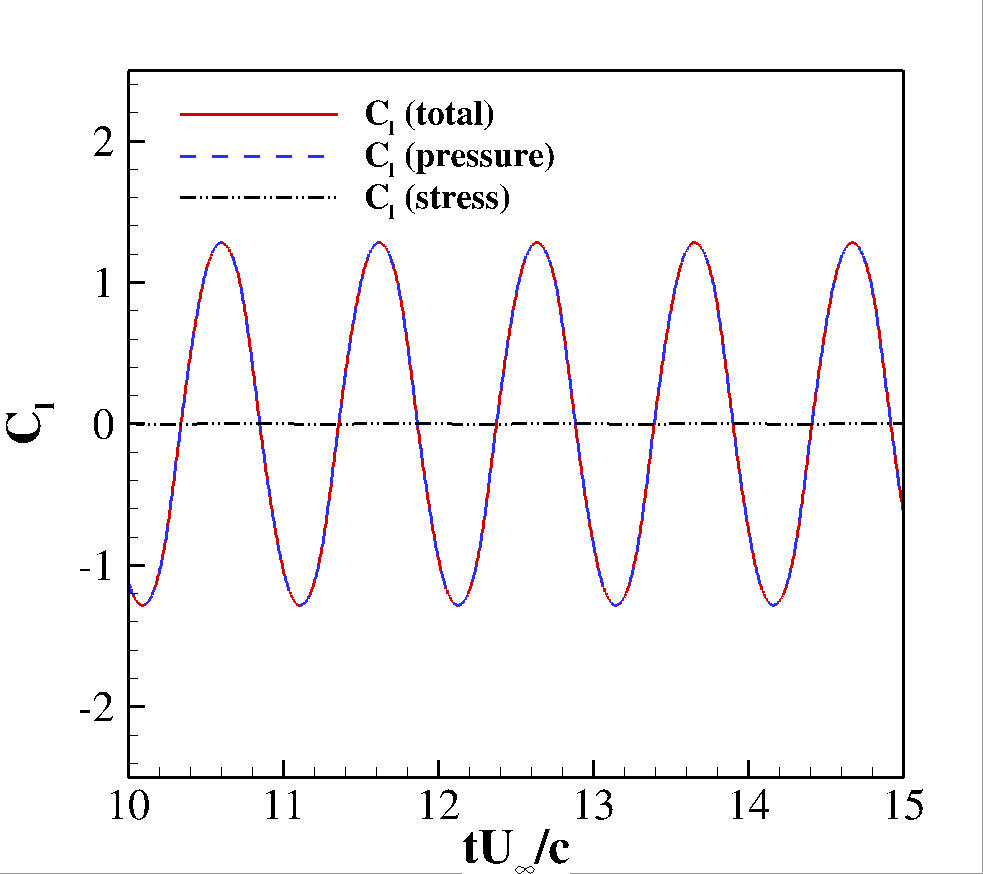}
		\label{pitch_cl_alf4k3p09}
	}
	\subfigure[]{
		\includegraphics[width=0.4 \textwidth]{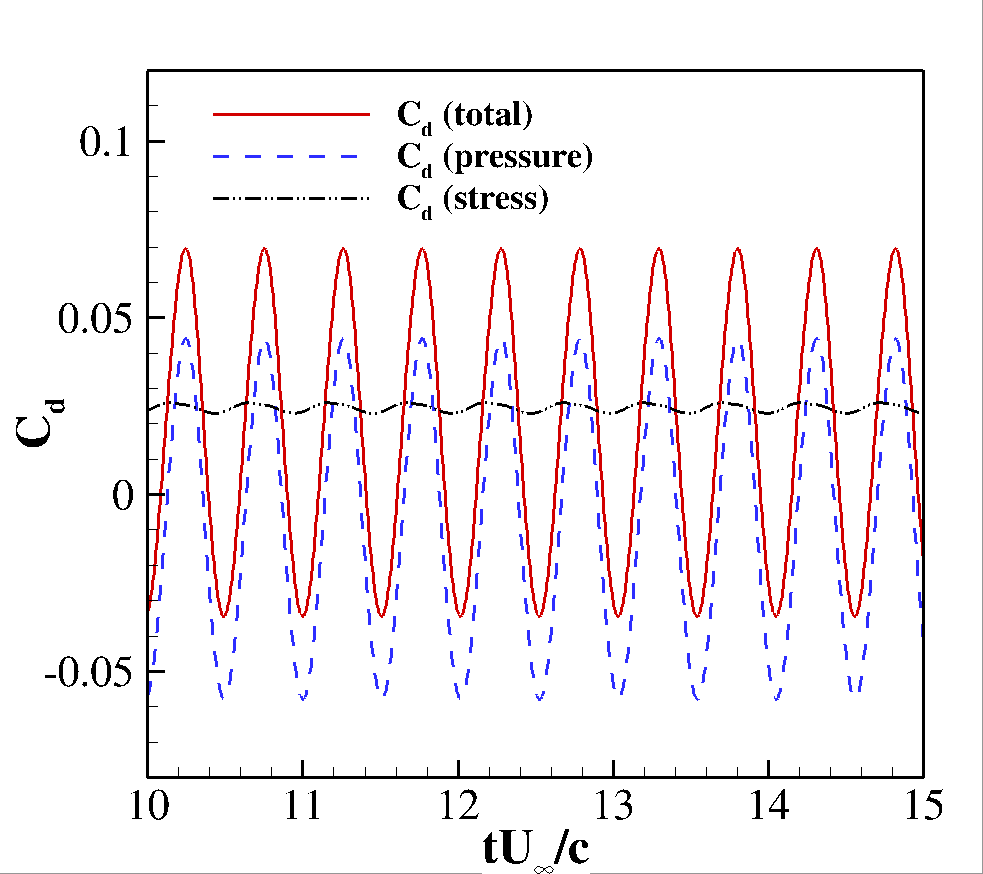}
		\label{pitch_cd_alf4k3p09}
	}
	\caption{\label{pitch_clcd_alf4k3p09} The histories of (a) lift and (b) drag coefficient for flow around the pitching NACA0012 airfoil at AoA=4$^\circ$ and $k=3.09$, and the contributions of pressure and stress to the total forces.}
\end{figure}

Secondly, a series of flow over the pitching airfoil are simulated at $d=2^\circ$ and $4^\circ$, respectively, with different reduced frequency $k$. Fig.~\ref{pitch_cd_compare} shows the time evolutions of $C_d$ at three values of $k$. The $C_d$ are also calculated with Eq.~(\ref{pitch_force}). At small value of $k$, both the maximum and the minimum of $C_l$ are larger than zero, so the flow will generate the drag force. When enhance the value of $k$, the minimum of $C_d$ will little than zero. Though the flow is still generate the drag force, the  magnitude is decreasing. If we continue enhance the value of $k$, finally, the absolute value of minimum of $C_d$ will larger than that of maximum of $C_d$, and flow will generate the thrust force. Here, we define a new force coefficient, $C_T$, represent the thrust coefficient, and $C_T=-C_d$. Fig.~\ref{pitch_cd} shows the mean thrust coefficient $\bar{C}_T$ at different $k\alpha$. Generally, we get the same tendency compared with other numerical results \cite{ramamurti2001simulation,young2004oscillation,mackowski2015direct} and experiment result \cite{koochesfahani1989vortical}. For $d=2^\circ$, when $k\alpha$ less than 0.2, the numerical results are consistency with each others, and close to the experiment values. But when $k\alpha$ larger than 0.2, the discrepancy becomes obvious. Some reasons described in Ref.~\cite{ramamurti2001simulation} maybe explain this discrepancy. Fig.~\ref{pitch_ma} shows the $\bar{C}_T$ at different Ma, the discrepancy at high values of $k\alpha$ is much larger. We demonstrate again that if compressible flow solver is used, the Ma of free-stream must set with a small value, at least for this pitching airfoil test case. The similar profiles also reported by Young et al. \cite{young2004oscillation}. Fig.~\ref{pitch_clcd_alf4k0p835} and Fig.~\ref{pitch_clcd_alf4k3p09} show the time evolutions of $C_l$ and $C_d$ at two different reduced frequencies. We reproduce the same phenomenons illustrated by Liang et al. \cite{liang2011high} at these two flow conditions. That is, for lift coefficient, though the amplitudes are very larger, the zero mean lift will acting on the airfoil. Stress force almost equal to zero, only the pressure is dominating contribution to the total lift force. For drag coefficient, the stress force almost keep the same values at these two flow conditions, but the instantaneous pressure declined, and will offset the stress force, consequently the magnitude of total drag force is reduced.

\begin{figure}[!htp]
	\centering
	\subfigure[$k=0$ ($k_{equi.}=8.23$)]{
		\includegraphics[width=0.8 \textwidth]{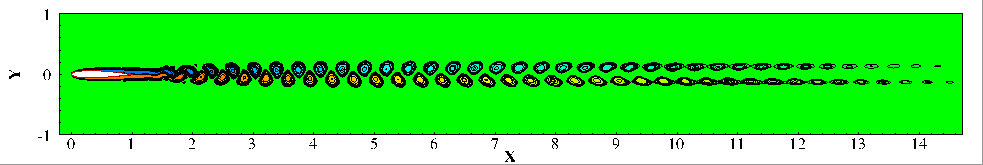}
		\label{pitch_vor_stat}
	}
    \subfigure[$k=2$ ($k<k_{equi.}$)]{
	    \includegraphics[width=0.8 \textwidth]{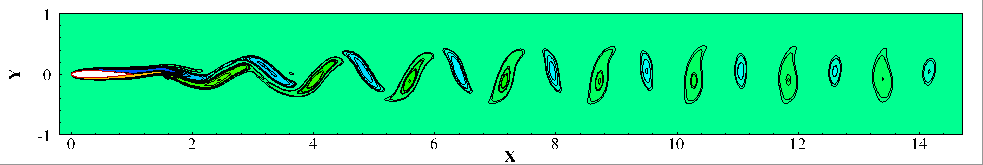}
	    \label{pitch_vor_alf2k2}
    }
    \subfigure[$k=4$ ($k<k_{equi.}$)]{
	    \includegraphics[width=0.8 \textwidth]{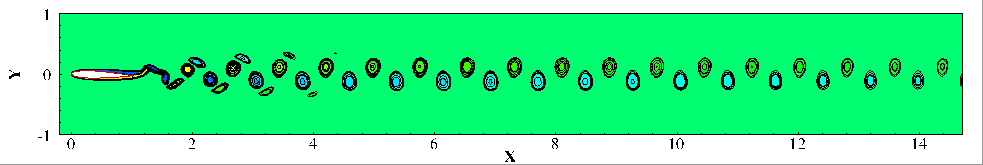}
	    \label{pitch_vor_alf2k4}
    }
    \subfigure[$k=6$ ($k<k_{equi.}$)]{
	    \includegraphics[width=0.8 \textwidth]{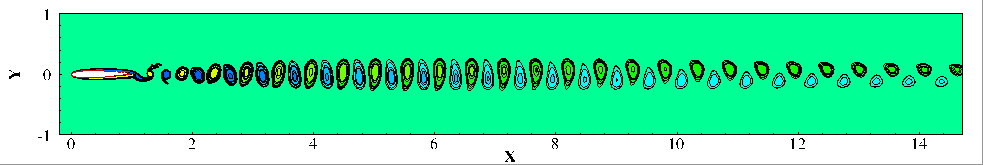}
	    \label{pitch_vor_alf2k6}
    }
    \subfigure[$k=8$ ($k<k_{equi.}$)]{
	    \includegraphics[width=0.8 \textwidth]{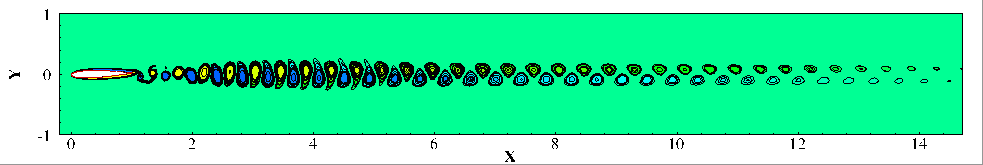}
	    \label{pitch_vor_alf2k8}
    }
    \subfigure[$k=10$ ($k>k_{equi.}$)]{
	    \includegraphics[width=0.8 \textwidth]{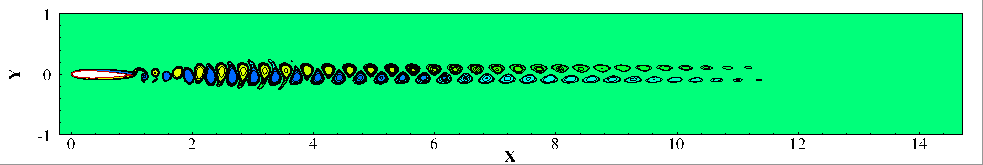}
    	\label{pitch_vor_alf2k10}
    }
    \subfigure[$k=12$ ($k>k_{equi.}$)]{
	    \includegraphics[width=0.8 \textwidth]{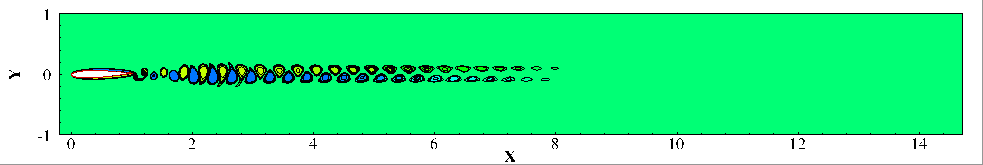}
	    \label{pitch_vor_alf2k12}
    }
	\caption{\label{pitch_vor_alf2} Vortical patterns for flow around the NACA0012 airfoil at different reduced frequency $k$ with AoA=2$^\circ$.}
\end{figure}

\begin{figure}[!htp]
	\centering
	\subfigure[$k=0$ ($k_{equi.}=8.23$)]{
	   \includegraphics[width=0.8 \textwidth]{pitch_vor_stat}
	}
    \subfigure[$k=0.835$ ($k<k_{equi.}$)]{
	   \includegraphics[width=0.8 \textwidth]{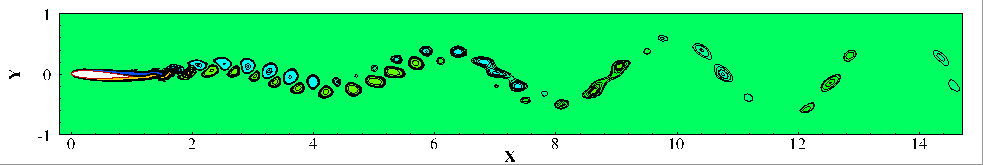}
	   \label{pitch_vor_alf4k0p835}	
    }
    \subfigure[$k=2$ ($k<k_{equi.}$)]{
	   \includegraphics[width=0.8 \textwidth]{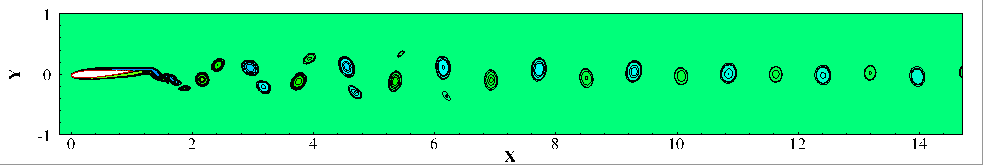}
	   \label{pitch_vor_alf4k2}
    }
    \subfigure[$k=3.09$ ($k<k_{equi.}$)]{
	   \includegraphics[width=0.8 \textwidth]{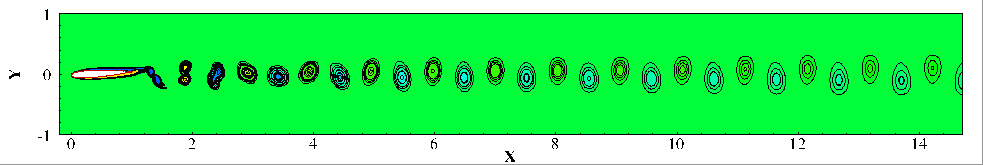}
	   \label{pitch_vor_alf4k3p09}
    }
    \subfigure[$k=4$ ($k<k_{equi.}$)]{
	   \includegraphics[width=0.8 \textwidth]{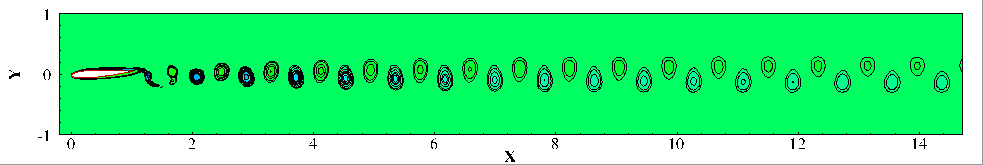}
	   \label{pitch_vor_alf4k4}
    }
    \subfigure[$k=5$ ($k<k_{equi.}$)]{
	   \includegraphics[width=0.8 \textwidth]{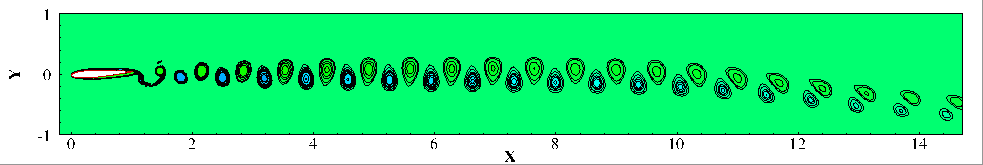}
	   \label{pitch_vor_alf4k5}
    }
    \subfigure[$k=6$ ($k<k_{equi.}$)]{
	   \includegraphics[width=0.8 \textwidth]{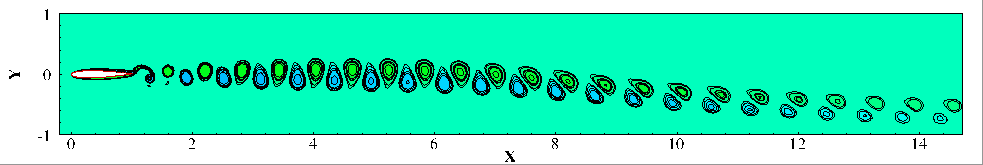}
	   \label{pitch_vor_alf4k6}
    }
\caption{\label{pitch_vor_alf4} Vortical patterns for flow around the NACA0012 airfoil at different reduced frequency $k$ with AoA=4$^\circ$.}
\end{figure}

Fig.~\ref{pitch_vor_alf2} and Fig.~\ref{pitch_vor_alf4} show the vortical patterns at $d=2^\circ$ and $d=4^\circ$, respectively. For comparison, the vortical pattern at $k=0$ (stationary airfoil) is also presented. For $d=2^\circ$, at small value of $k$, the much larger structures of vortex street will generate compared with that of stationary airfoil. When enhance the value of $k$, as the thrust force is generated, original two rows of vortex street seem merge into one row. Besides, if $k<k_{equi.}$, even though about 15$c$ length far away from the airfoil trail, the clear structures of vortex street are kept very well. Otherwise, these structures are quickly dissipated at high values of $k$. For $d=4^\circ$, when $k=0.835$, the wake assumes a form of undulating vortex sheet. And with $k=3.09$, double-vortex pattern is captured. But about $2c$ length behind the airfoil trail, this structure seems to dissipate and merge into one another. For $k=2$, that the value between 0.835 and 3.09, the wake pattern seems that the length of undulating vortex sheet is declined (about $5c$ behind the airfoil trail), and the double-vortex features also generate but not very clear. At higher values of $k$, double-vortex features are disappearing, and the vortex streets that approximate a straight line are deflected. Similar structures of vortex street also illustrated by Liang et al. \cite{liang2011high} with numerical method and by Koochesfahani \cite{koochesfahani1989vortical} with experiment.

\subsection{A moving piston driven by the rarefied gas}
\begin{figure}[!htp]
	\centering
	\subfigure[]{
		\includegraphics[width=0.4 \textwidth]{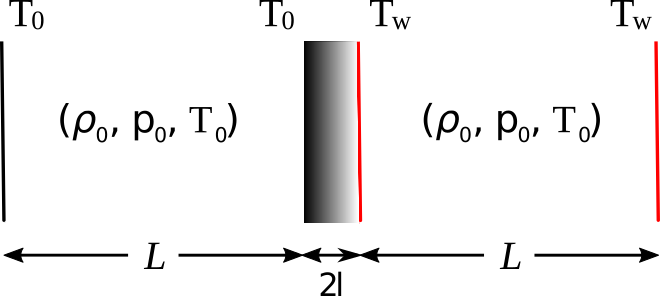}
		\label{piston1}
	}
	\subfigure[]{
		\includegraphics[width=0.4 \textwidth]{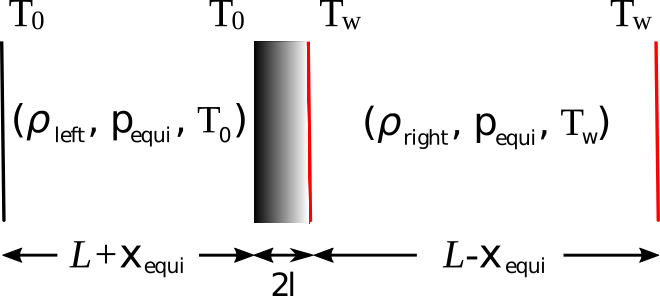}
		\label{piston2}
	}
	\caption{\label{piston} Schematics of the moving piston driven by the rarefied gas, (a) Initial stage and (b) Equilibrium stage.}
\end{figure}

In this section, a case that the rarefied gas driving a piston is simulated. This problem has been studied by Dechrist{\'e} et al. \cite{dechriste2012numerical} with a deterministic numerical scheme coupled with immersed boundary method, and Shrestha et al. \cite{shrestha2015numerical} with DSMC. Fig.~\ref{piston} shows the schematics of this problem. The one-dimensional computational domain will divided into two sub-domains by a piston. The length of one sub-domain is $L$, and the width of piston is $2l$. At initial time, these two sub-domains will be filled with the same gas, and the density $\rho$, pressure $p$, and temperature $T$ also set the same values. For right part of computational domain, as the temperature of wall higher than that of gas, the pressure will enhance, and push the piston moving from right to left. Finally, if the pressures at two walls of piston are same, the piston will stop moving. With the mass conservation and the state equation of gas for each part, we have \cite{dechriste2012numerical}:
\begin{equation}	
	\rho_0L_0=\rho_{left}(L_0+x_{equi}), \rho_0L_0=\rho_{right}(L_0-x_{equi}),
\end{equation}
and
\begin{equation}
	\rho_{left}RT_0=p_{equi}, \rho_{right}RT_w=p_{equi},
\end{equation}
where $R$ is the gas constant. So, the equilibrium location of piston is
\begin{equation}\label{piston_theory}
    x_{equi}=L_0\frac{1-T_w/T_0}{1+T_w/T_0},
\end{equation}
and, respectively, the density and pressure at equilibrium states for each part are
\begin{equation}
	\rho_{left}=\frac{p_{equi}}{RT_0}, \rho_{right}=\frac{p_{equi}}{RT_w}, p_{equi}=\frac{L_0}{L_0+x_{equi}}p_0.
\end{equation}

Following the set up described by Shrestha et al. \cite{shrestha2015numerical}, the gas is argon, the mass of atom is $m_g=6.63 \times 10^{-26}kg$, and the diameter of atom is $3.68 \times 10^{-10}m$ with the hard-sphere collision model. The initial values for $T_0$, $T_w$ and $p_0$ are equal to $270K$, $330K$, and $10Pa$, respectively. And the initial velocity for each part are set as zero. With ideal gas law, the initial density can be calculated. In our simulations, two computational geometries are considered, that are $L=0.1m$, $l=0.01m$, and $L=0.01m$, $l=0.001m$, respectively. Consequently, based on the width of piston ($2l$) and the given condition of gas, the flows under two Knudsen (Kn) numbers, $Kn=0.031$ and $Kn=0.31$ will be simulated. Besides, 400 cells are used for the total computational domain.

\begin{figure}[!htp]
	\centering
	\subfigure[]{
		\includegraphics[width=0.4 \textwidth]{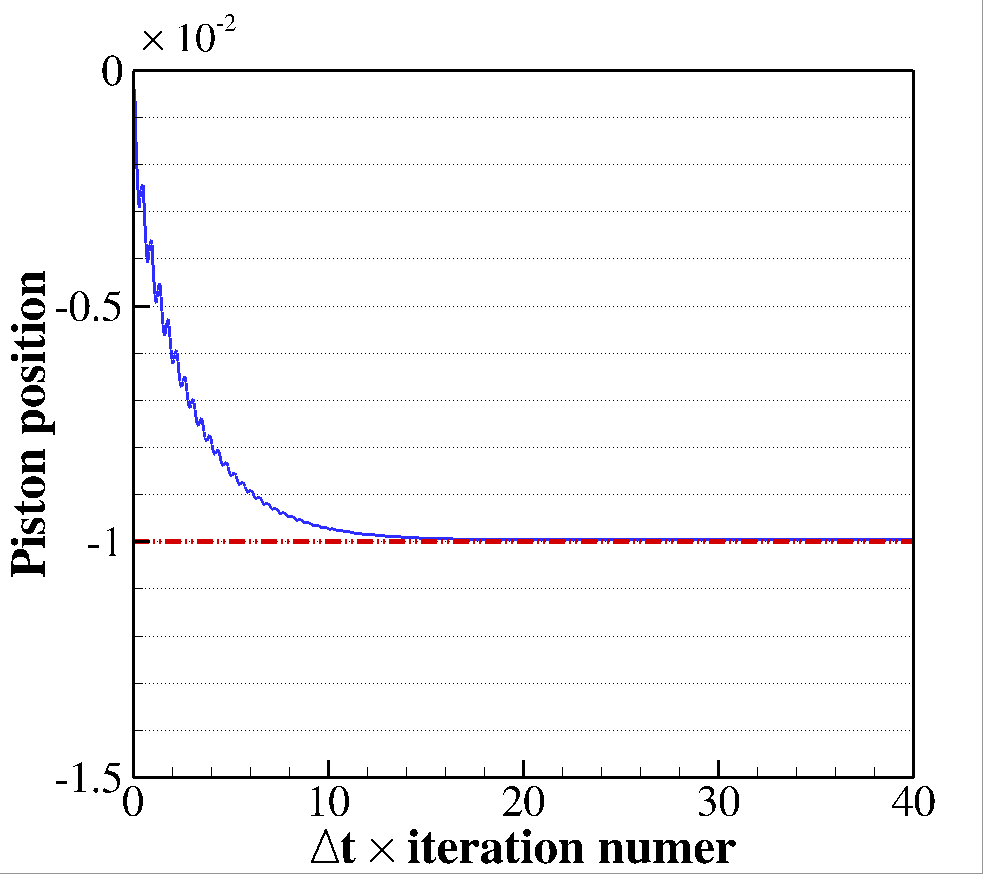}
		\label{piston_loc_kn0p031}
	}
	\subfigure[]{
		\includegraphics[width=0.4 \textwidth]{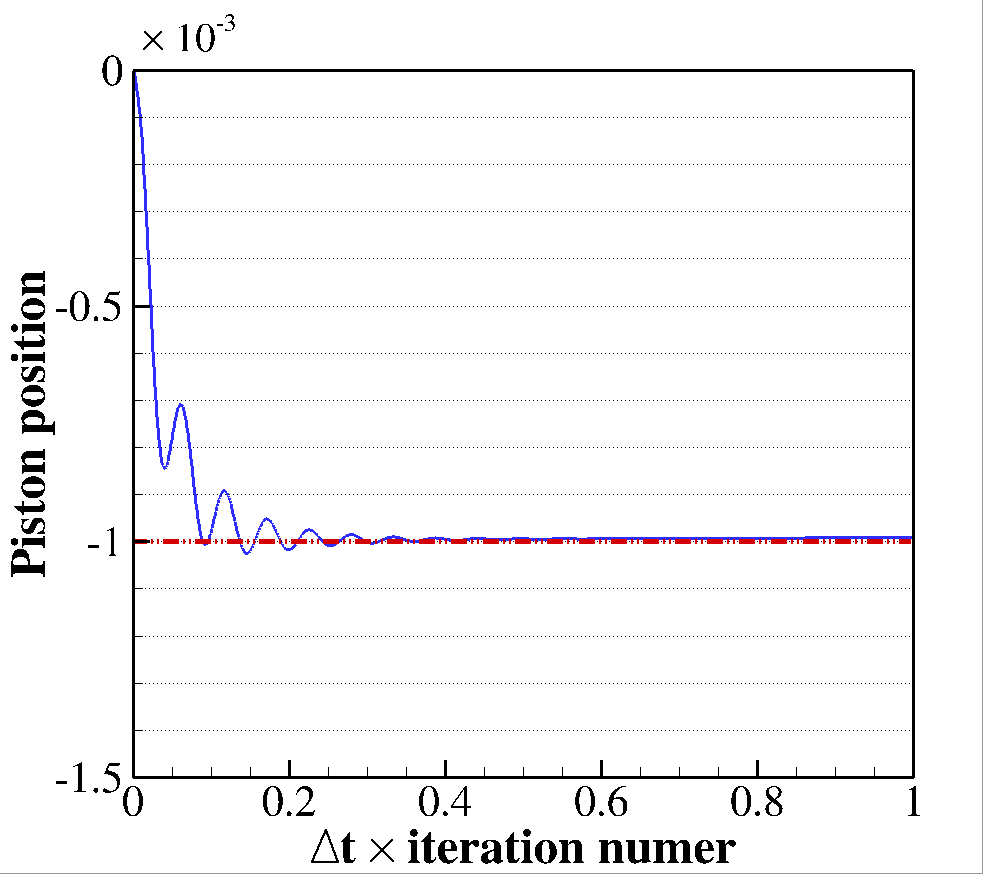}
		\label{piston_loc_kn0p31}
	}
	\caption{\label{piston_loc} The histories of piston position for (a) $Kn=0.031$ and (b) $Kn=0.31$ (The red dash and dot lines represent the theory solutions).}
\end{figure}

\begin{figure}[!htp]
	\centering
	\subfigure[]{
		\includegraphics[width=0.4 \textwidth]{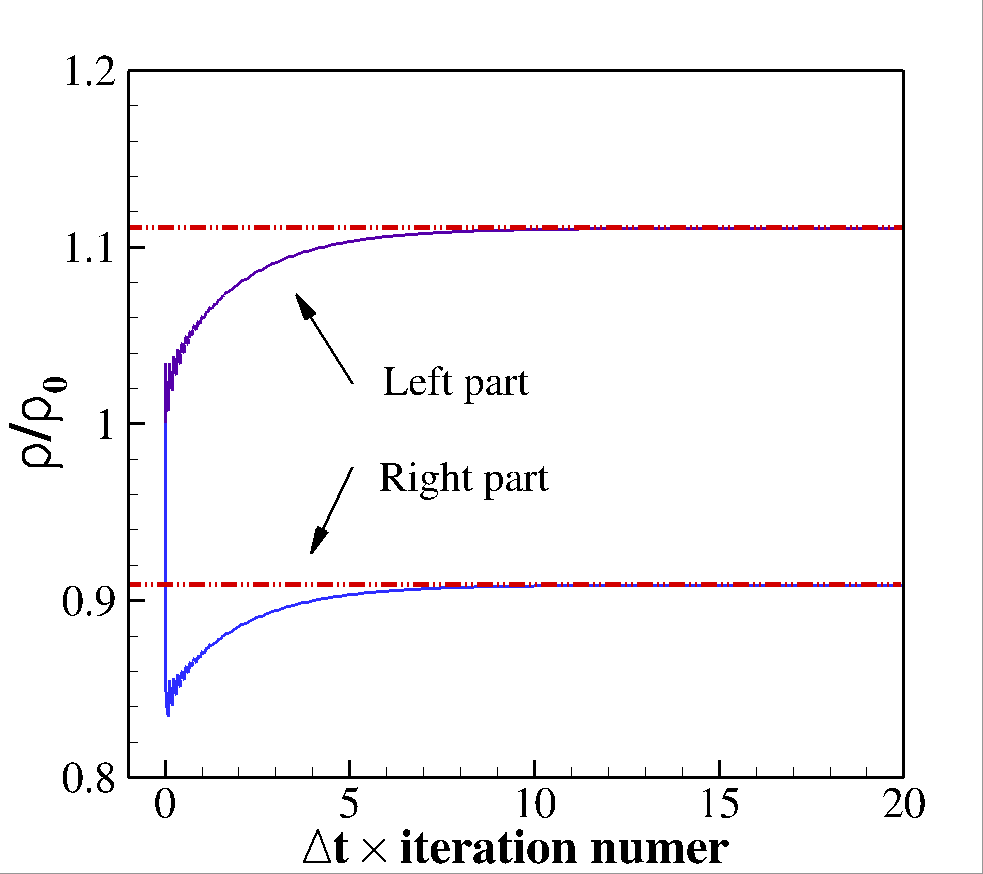}
		\label{piston_rho_kn0p031}
	}
	\subfigure[]{
		\includegraphics[width=0.4 \textwidth]{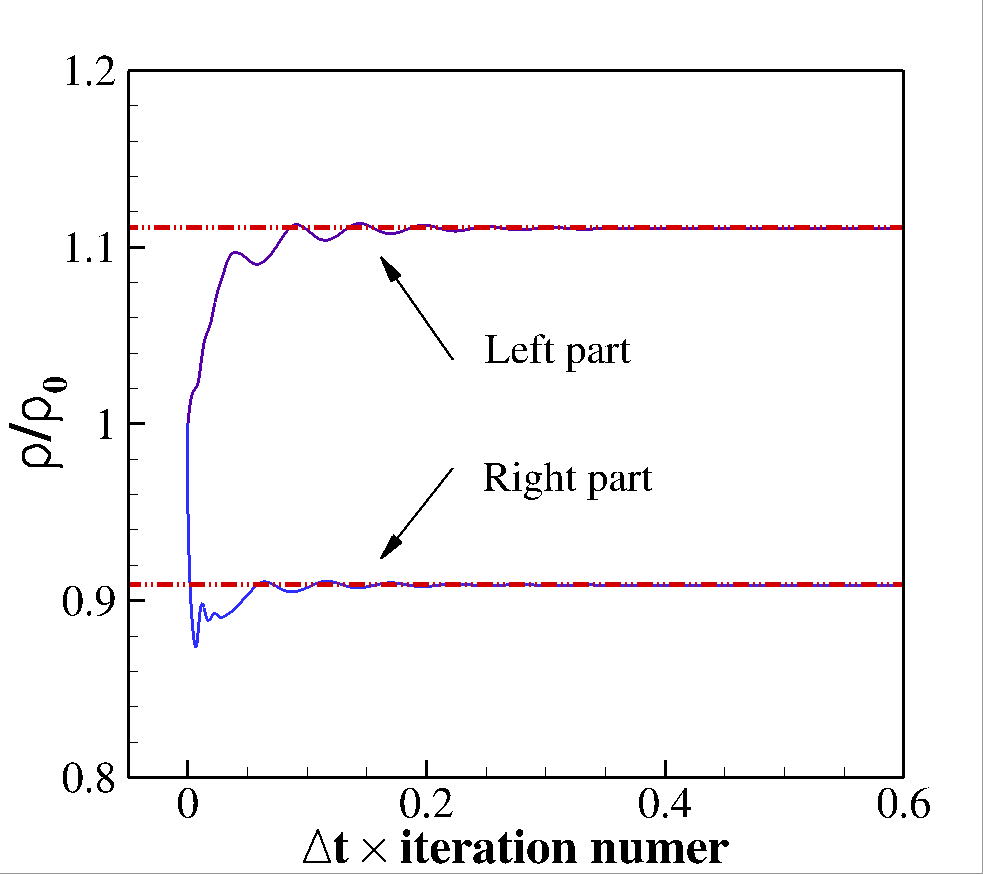}
		\label{piston_rho_kn0p31}
	}
	\caption{\label{piston_rho} The time evolutions of density at left and right subdomain for (a) $Kn=0.031$ and (b) $Kn=0.31$ (The red dash and dot lines represent the theory solutions).}
\end{figure}

\begin{figure}[!htp]
	\centering
	\subfigure[]{
		\includegraphics[width=0.4 \textwidth]{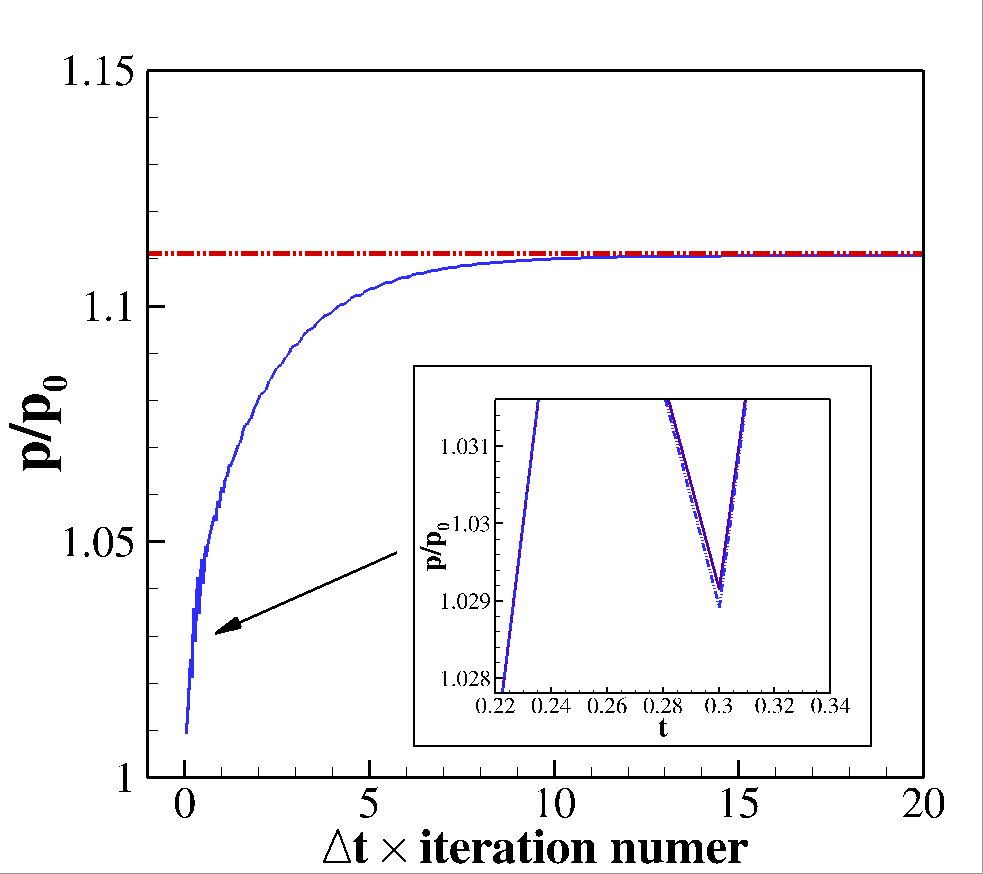}
		\label{piston_pre_kn0p031}
	}
	\subfigure[]{
		\includegraphics[width=0.4 \textwidth]{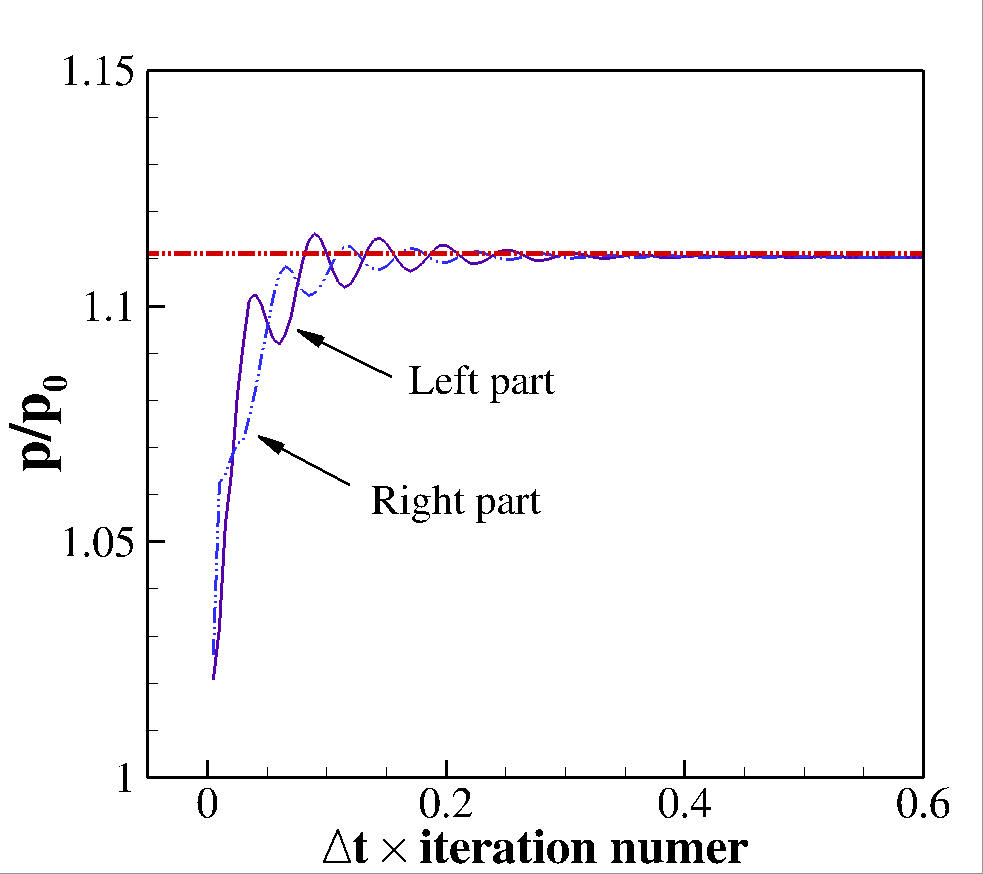}
		\label{piston_pre_kn0p31}
	}
	\caption{\label{piston_pre} The time evolutions of pressure at left and right subdomain for (a) $Kn=0.031$ and (b) $Kn=0.31$ (The red dash and dot lines represent the theory solutions).}
\end{figure}

Fig.~\ref{piston_loc} shows the time evolutions of the piston location at two $Kn$ numbers. The similar profiles also described in Ref.~\cite{shrestha2015numerical}. That is the position of piston for $Kn=0.31$ will much faster converge to its equilibrium position than that for $Kn=0.031$. In addition, DSMC at small value of $Kn$, the results will fluctuating and need perform several independent runs to reduce this stochastic fluctuations in time-dependent problems. On the contrary, our results are much smooth even at small $Kn$. Besides, the red dash and dot lines shown in figure are the theory solutions (Eq.~(\ref{piston_theory})), the errors between the numerical results and theory solutions are both less than $1\%$. Fig.~\ref{piston_rho} and Fig.~\ref{piston_pre} show the time evolutions of density and pressure at two sub-domains, respectively. Our numerical results are also consistency with the theory solutions. During the evolutions, the pressure difference of two walls for piston at $Kn=0.31$ are larger than that of $Kn=0.031$, it may also indicate that the piston at computation condition of large $Kn$ will faster moving to its converged equilibrium position.

\begin{figure}[!htp]
	\centering	
    \includegraphics[width=0.4 \textwidth]{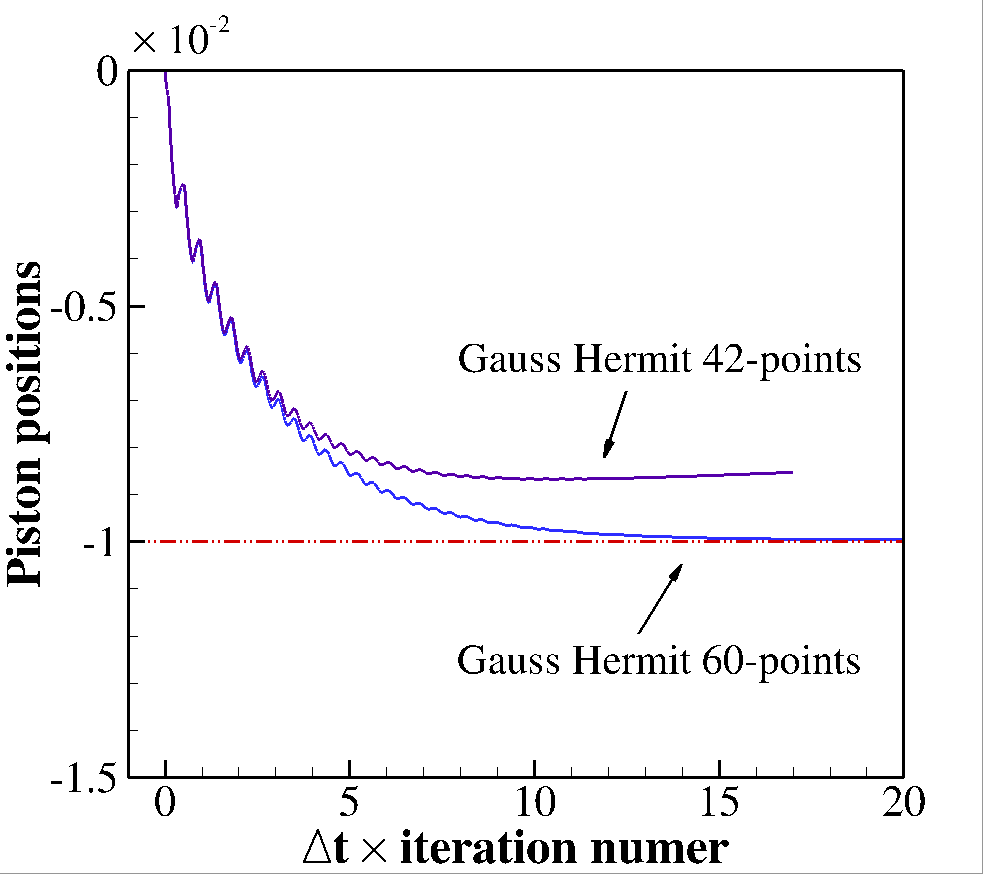}
	\caption{\label{piston_integral} The histories of piston position at different number of Gauss-Hermit quadrature points with $Kn=0.031$ (The red dash and dot line represent the theory solution).}
\end{figure}

In this case, the Gauss-Hermit quadrature rule is used \cite{zhu2016discrete} to integral the macro-quantities, and the codes to calculate the abscissas and weights are shown in Ref.~\cite{press1992numerical}. Fig.~\ref{piston_integral} shows the influence of number of quadrature points to the results. For $Kn=0.031$, from our test, the integral accuracy with 28 quadrature points is good enough for left sub-domain to calculate the macro-quantities. But, for the right side, about two times number of point is needed to get good results. Due to the error of integral, the density in right part will continue decline, and the velocity will not converge to zero. Consequently, even though the pressure difference of two walls for piston converge to a very small value, the piston will move from left to right with a small value of velocity, finally the simulation will blow up.

\subsection{Rarefied flow caused by a plate oscillating in its normal direction}
\begin{figure}[!htp]
	\centering	
	\includegraphics[width=0.4 \textwidth]{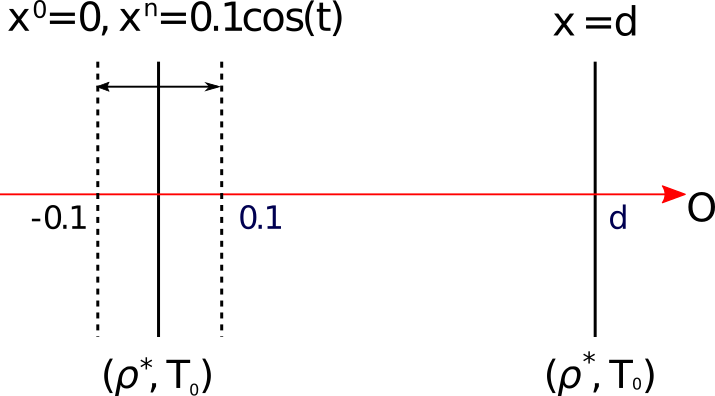}
	\caption{\label{oscillate_plate} Schematic of the moving plate oscillated in its normal direction and the stationary plate.}
\end{figure}

In this section, another rarefied flow test case will be simulated, which has been studies by Tsuji et al. \cite{tsuji2014gas}. Fig.~\ref{oscillate_plate} shows the schematic of this problem. In one-dimensional domain, the right wall is stationary, and left wall is oscillating with cosine function $x(t)=a_wcos(\omega{t})$ ($a_w=0.1$ and $\omega=1$ in this case), the amplitude equal to 0.1 make sure this case is low-speed flow. Following the set up described in Ref.~\cite{tsuji2014gas}, in initial time, the length of computational domain is $d=2\pi\sqrt{5/6}$, which is the wavelength of the sinusoidal acoustic wave with angular frequency $\omega$ in an inviscid (Euler) gas \cite{aoki2017unsteady}. The Knudsen number is given as
\begin{equation}
   Kn=\frac{2}{\sqrt{\pi}}K,
\end{equation}
where $K$ is the special $Kn$ number, and two $K$ numbers, which equal to 0.5 and 1.0, respectively, will be considered in this case. $\rho_0=1.0$, $T_0=1.0$ and $u_0=0.0$ will be used to initial the distribution function $g$ and $h$. 400 cells is used to discretize the computational domain for $K=0.5$, and that of 200 for $K=1.0$. For time step, in each oscillating period ($T=2\pi$), 3,2000 iteration computations will be implemented for $K=0.5$, and 6,4000 for $K=1.0$.

\begin{figure}[!htp]
	\centering
	\subfigure[]{
		\includegraphics[width=0.4 \textwidth]{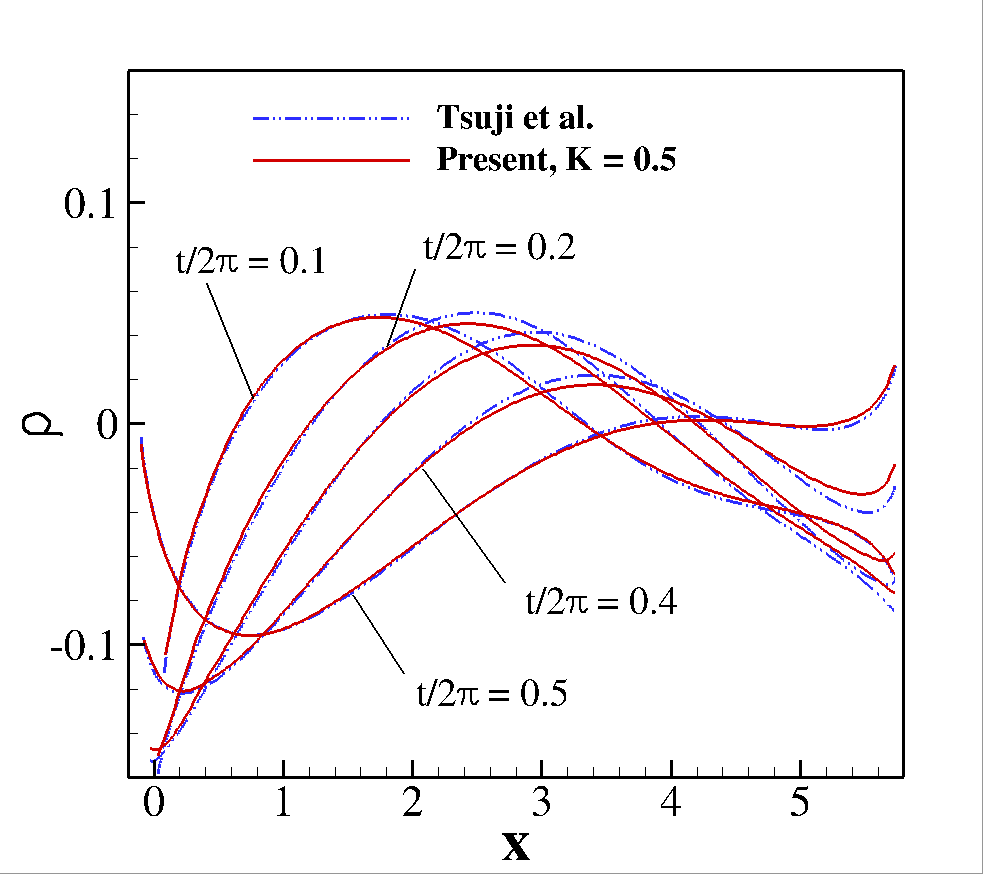}
		\label{plate_rho1_kn0p5}
		\includegraphics[width=0.4 \textwidth]{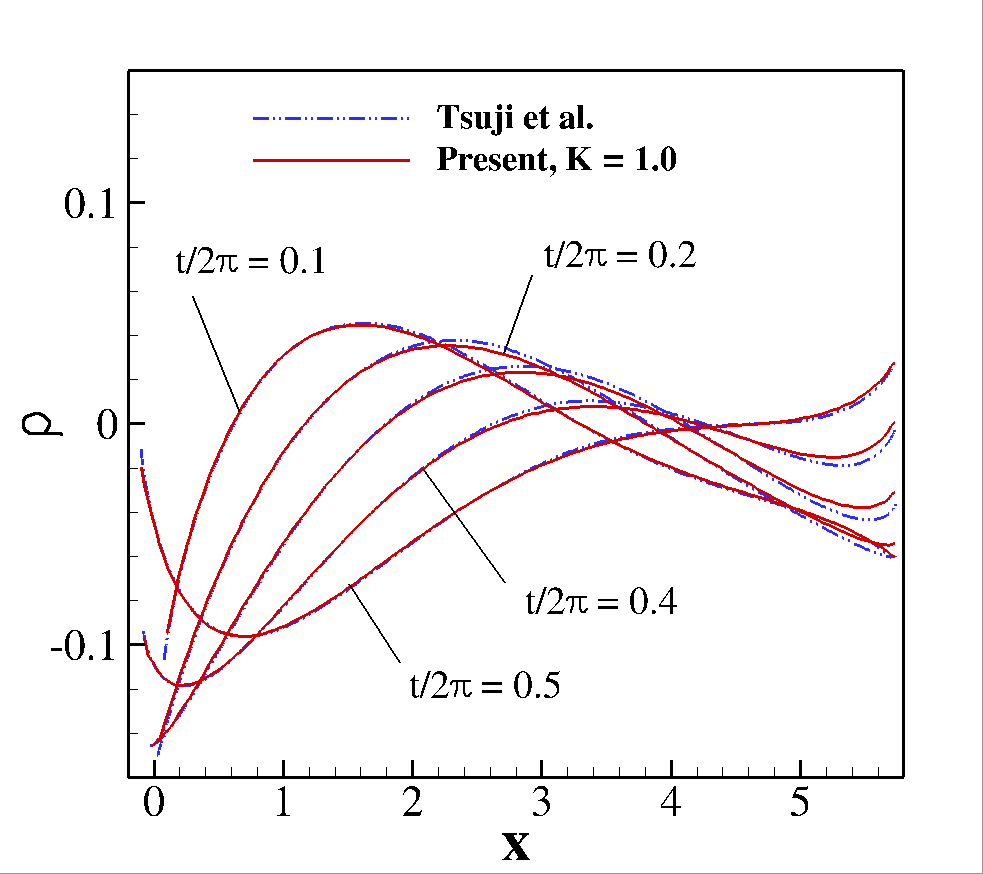}
		\label{plate_rho1_kn1}
	}
	\subfigure[]{
		\includegraphics[width=0.4 \textwidth]{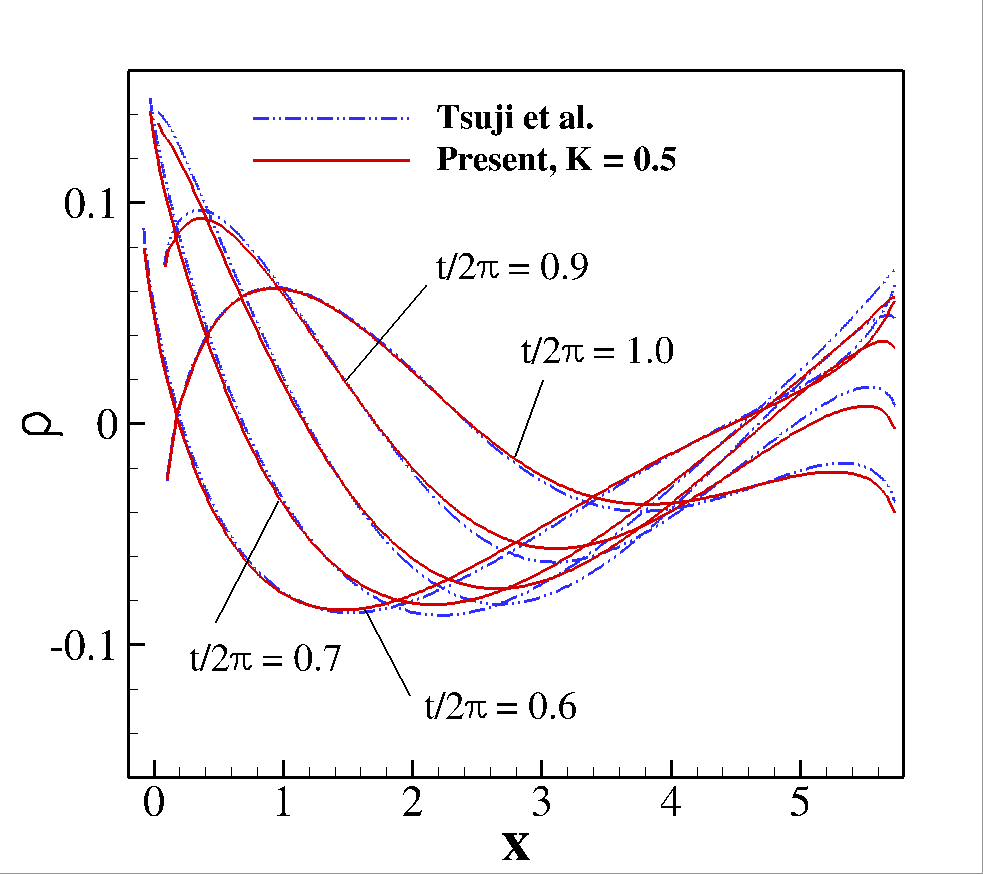}
		\label{plate_rho2_kn0p5}
		\includegraphics[width=0.4 \textwidth]{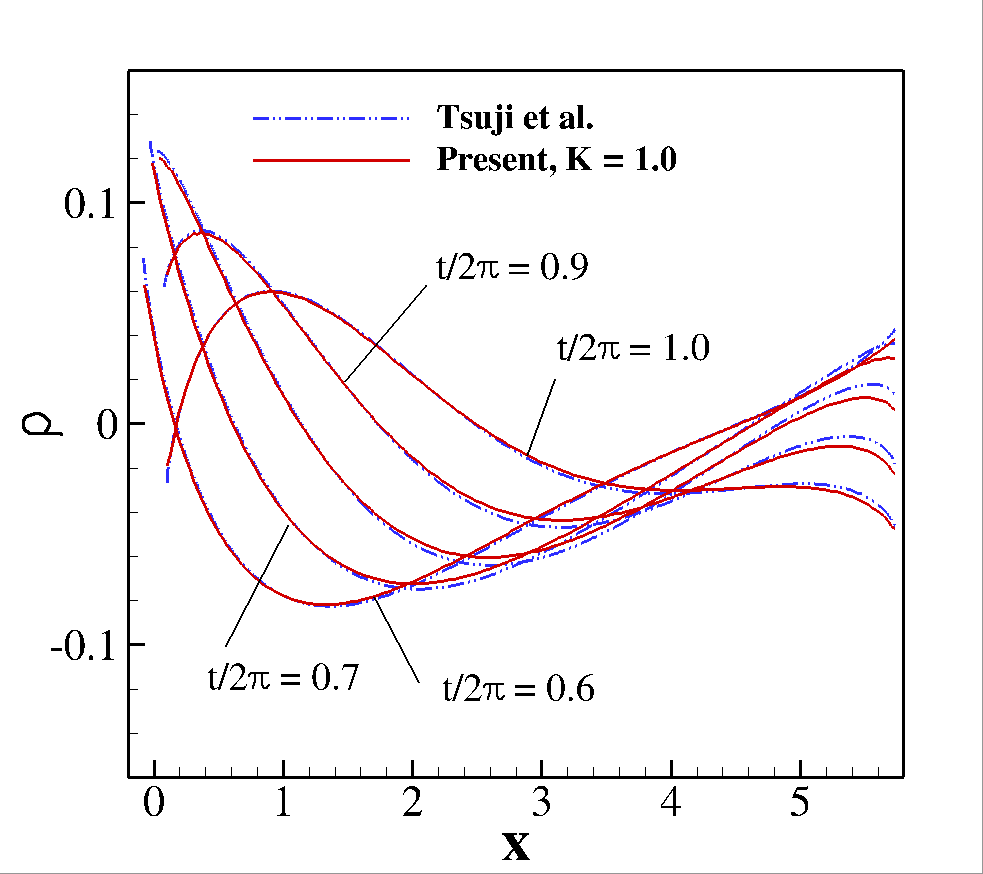}
		\label{plate_rho2_kn1}
	}
	\caption{\label{plate_rho} Profiles of density at (a) $t/2\pi=0.1, \dots, 0.5$ and (b) $t/2\pi=0.6, \dots, 1.0$ at two $K$ numbers, 0.5 and 1.0, respectively.}
\end{figure}

\begin{figure}[!htp]
	\centering
	\subfigure[]{
		\includegraphics[width=0.4 \textwidth]{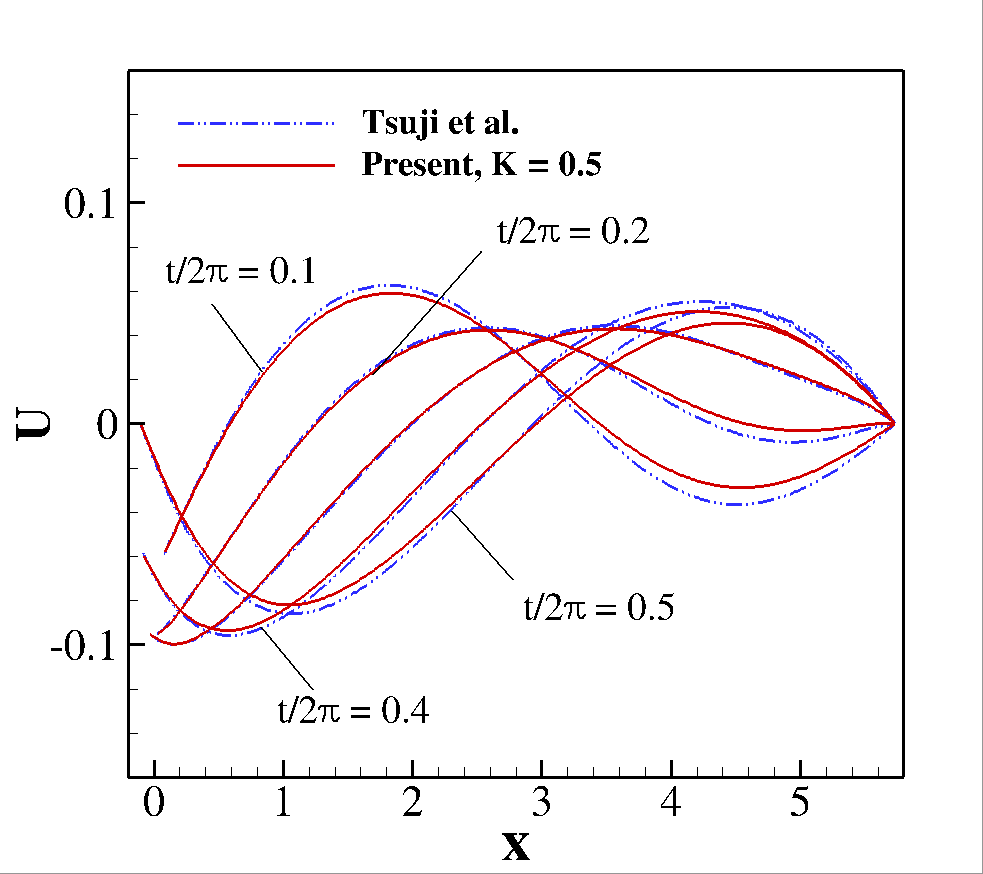}
		\label{plate_vel1_kn0p5}
		\includegraphics[width=0.4 \textwidth]{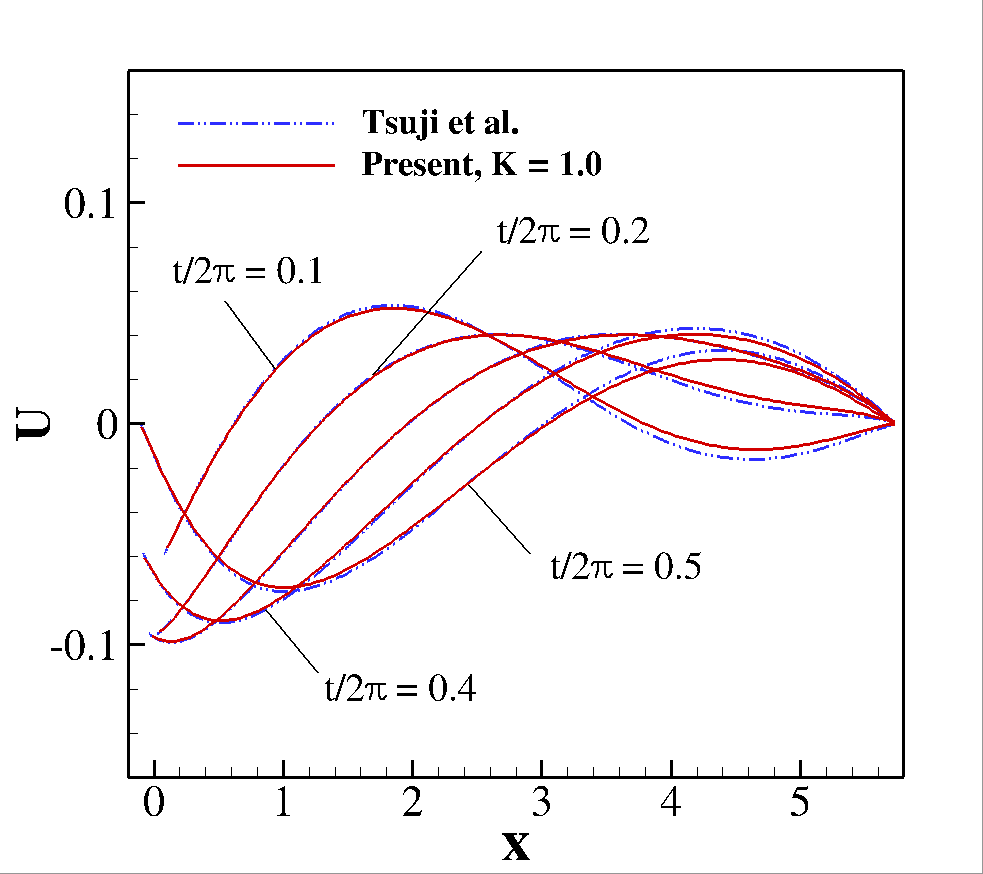}
		\label{plate_vel1_kn1}
	}
	\subfigure[]{
		\includegraphics[width=0.4 \textwidth]{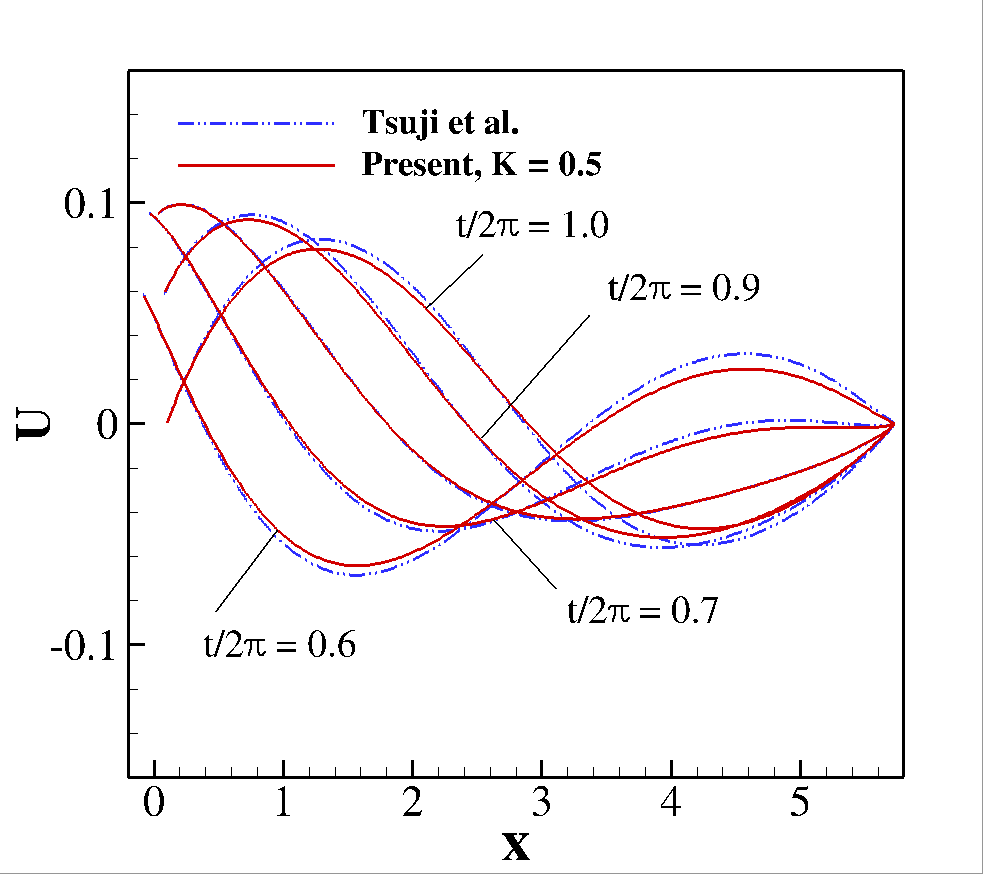}
		\label{plate_vel2_kn0p5}
		\includegraphics[width=0.4 \textwidth]{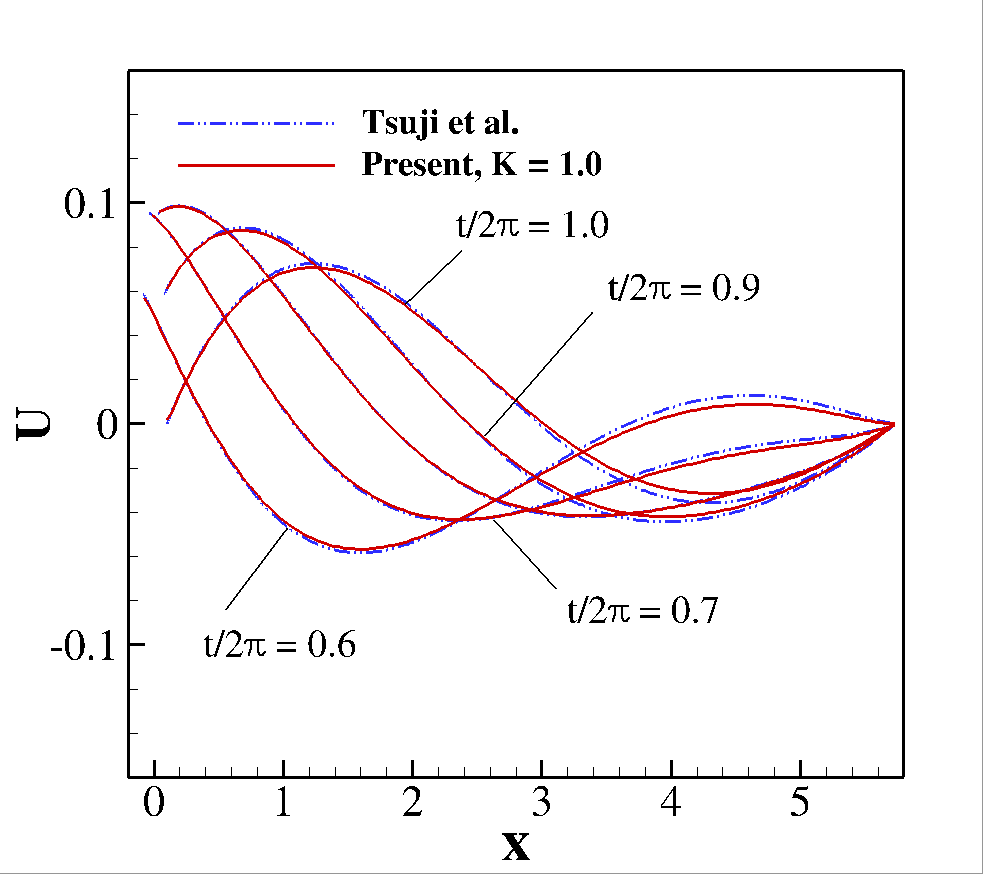}
		\label{plate_vel2_kn1}
	}
	\caption{\label{plate_vel} Profiles of velocity at (a) $t/2\pi=0.1, \dots, 0.5$ and (b) $t/2\pi=0.6, \dots, 1.0$ at two $K$ numbers, 0.5 and 1.0, respectively.}
\end{figure}

\begin{figure}[!htp]
	\centering
	\subfigure[]{
		\includegraphics[width=0.4 \textwidth]{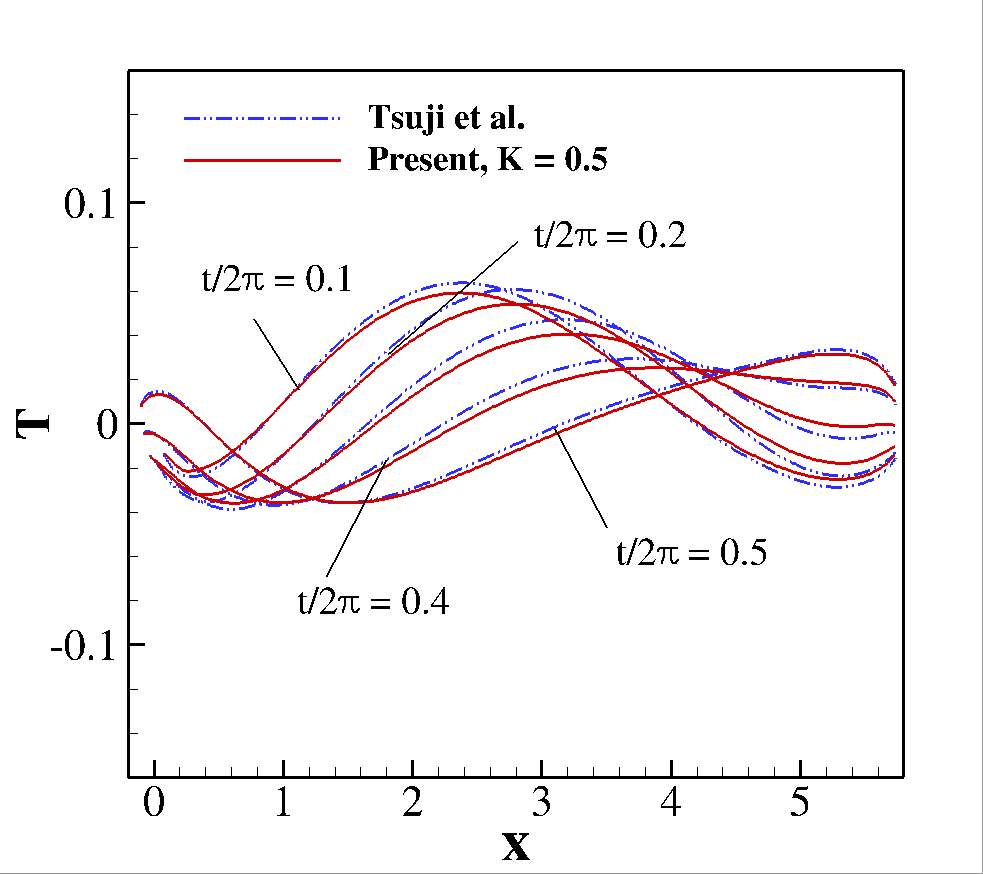}
		\label{plate_tem1_kn05}
		\includegraphics[width=0.4 \textwidth]{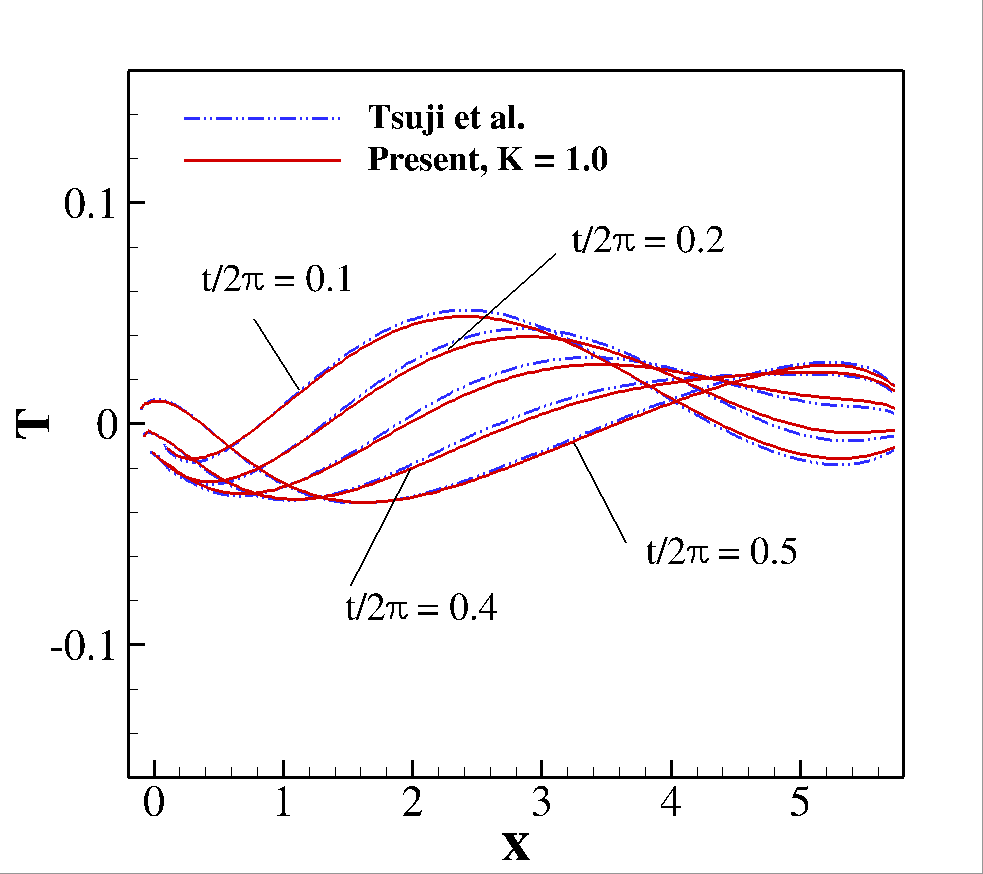}
		\label{plate_tem1_kn1}
	}
	\subfigure[]{
		\includegraphics[width=0.4 \textwidth]{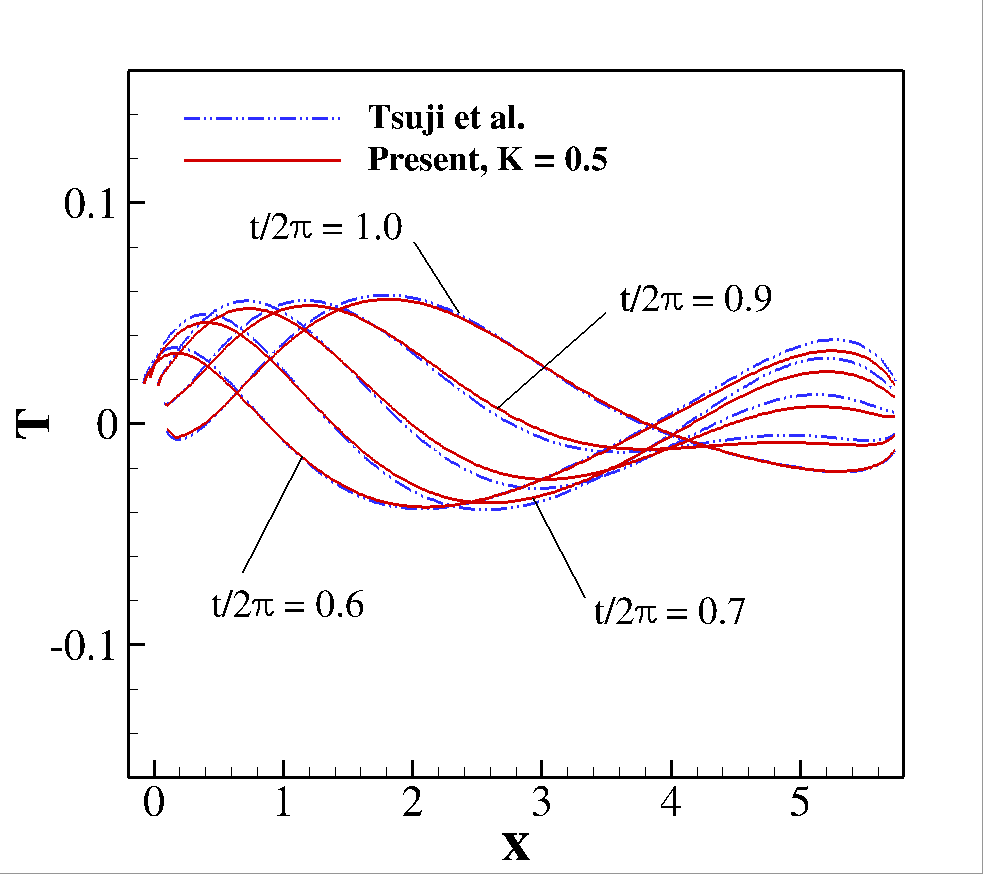}
		\label{plate_tem2_kn0p5}
		\includegraphics[width=0.4 \textwidth]{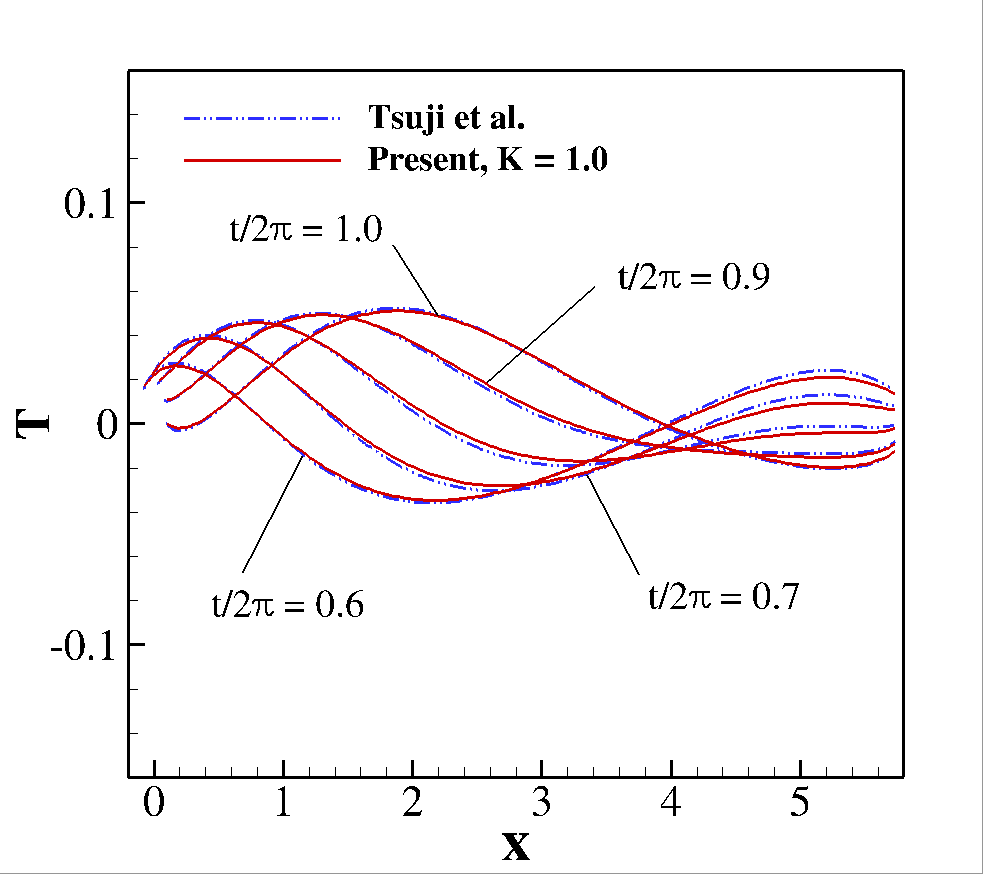}
		\label{plate_tem2_kn1}
	}
	\caption{\label{plate_tem} Profiles of temperature at (a) $t/2\pi=0.1, \dots, 0.5$ and (b) $t/2\pi=0.6, \dots, 1.0$ at two $K$ numbers, 0.5 and 1.0, respectively.}
\end{figure}

Figs.~\ref{plate_rho} - \ref{plate_tem} show the profiles of density, velocity and temperature, respectively, at ten moments in each oscillating period. Generally, our results are agree well with that of Tsuji et al. \cite{tsuji2014gas}, especially at high value of $K$. Without more reference, it¡¯s hard to identify which result is better. From the figures, it is clear that at $t/\pi=1.0$, the velocity profile is close to the sinusoidal shape, but the profiles of density and temperature deviate from it significantly. Furthermore, for velocity profile, this shape will deviates more from the sinusoidal shape and tends to attenuate more rapidly as $K$ increased, especially for the right part of wave.

For the quadrature rule to calculate macro-quantities, the Newton-Cotes rule will be used. For higher value of $K$, due to the number of particles is small, the wave generated from the left moving wall and the reflected wave from the right stationary wall will lead to the singular of distribution function \cite{tsuji2013moving}. So, to get the smooth results, the number of abscissas is much larger for higher $K$. From our test, for $K=0.5$, 50 abscissas is enough to get the convergent and smooth results. But for $K=1.0$, about 200 abscissas can get the smooth results.

\section{Conclusion}\label{Conclusion}
In the present work, the original DUGKS is extend to ALE-type DUGKS. The mesh moving velocity is introduced into the Boltzmann-BGK equation to modify the net flux of cell interface. Consequently, based on the constructed mesh moving velocity, the remapping-free-type ALE method is used to develop the current ALE-type DUGKS. To exclude the GCL error, three DGCL compliance schemes are discussed. As the present DUGKS only the middle time is needed to calculate the flux of cell interface, based on the geometry average, the geometry information at this time level is easy to defined, so the DGCL scheme1 is a good choice. Further improved work such as high-order or multi-time-levels implicit ALE-type DUGKS which the geometry information at these time levels are not easy to define, DGCL scheme2 and scheme3 will be the good choice. From the uniform flow test case, all these three schemes have good performance which no disturbances will be introduced into the computational domain. Four test cases of low-speed flow are simulated, two of them are the continuum flows, and others are the rarefied flows. Results of all the cases are in good agreement with the other numerical and/or experiment results. Therefore, for continuum flow, similar to the macro-methods based on the N-S equations, the present ALE-type DUGKS has the capability to cope with more complex low-speed moving boundary problems. And for rarefied flows, which out of ability for macro-methods, the present method also has the power to deal with them. Further works, such as parallel computing, implicit accelerated method, etc., will be continued, to enhance the ability of present ALE-type DUGKS for simulating the moving boundary problems at different flow regime.

\section*{Acknowledgements}
The current work is supported by the National Natural Science Foundation of China (Grant No. 11472219) and the 111 Project of China (B17037).

\clearpage

\bibliography{ref}

\end{document}